\documentclass[12pt,twoside]{book}
\usepackage{fancyhdr}
\usepackage{parskip}
\usepackage[utf8]{inputenc}
\usepackage{fancyhdr}
\usepackage[font=small,labelfont=bf]{caption}
\usepackage[super]{nth}
\usepackage{subfiles}
\usepackage{enumerate}
\usepackage{graphicx}
\usepackage{sectsty}\usepackage{bbm}
\usepackage[paperwidth=169mm, paperheight=239mm]{geometry}
\usepackage{fancyhdr}
\newcommand{\di}{\text{d}}
\newcommand{\did}{\text{d}^\dagger}
\newcommand{\be}{\begin{equation}}
\newcommand{\ee}{\end{equation}}

\newcommand{\F}{\mathcal{F}}
\newcommand{\A}{\mathcal{A}}
\newcommand{\cF}{\mathcal{F}}
\newcommand{\bLaa}{\boldsymbol{\Lambda}}

\newcommand{\bF}{\mathbf{F}}

\newcommand{\eps}{\Lambda}

\newcommand{\mone}{{\scriptscriptstyle-1}}

\newcommand{\de}{\delta}
\newcommand{\Laa}{\Lambda}
\newcommand{\laa}{\lambda}

\newcommand{\ex}{\text{x}}
\newcommand{\ordr}[1]{\mathcal{O}(\rho^{#1})}

\newcommand{\eqs}[1]{\begin{equation} #1 \end{equation}}

\newcommand{\eq}[2]{\begin{equation} #1 \label{#2} \end{equation}}

\newcommand{\x}{\vec{x}}

\newcommand{\M}{\mathcal{M}}
\newcommand{\B}{\mathcal{B}}
\newcommand{\C}{\mathcal{C}}
\newcommand{\I}{\mathscr{I}}

\newcommand{\pp}{{\prime\prime}}
\newcommand{\p}{\prime}

\newcommand{\blaa}{\boldsymbol{\laa}}
\newcommand{\blist}{\begin{itemize}}

	\newcommand{\pii}{\textbf{p}}
	
	\newcommand{\Hh}{\mathcal{H}}

	\newcommand{\beps}{\boldsymbol{\epsilon}}
	\newcommand{\ord}[1]{\mathcal{O}(#1)}

	\newcommand{\Ti}{{\scriptstyle T}}

	\newcommand{\T}{{\scriptscriptstyle T}}
	\usepackage[T1]{fontenc} 
	
	\usepackage[percent]{overpic}
	\usepackage{wrapfig}
	\usepackage{tabu}
	\usepackage{diagbox}
	
	\usepackage{graphicx}
	\usepackage{tikz}
	
	\usetikzlibrary{arrows,chains,shapes,matrix,positioning,scopes}
	\usetikzlibrary{decorations.markings}
	\tikzstyle arrowstyle=[scale=1]
	\tikzstyle directed=[postaction={decorate,decoration={markings,
			mark=at position .65 with {\arrow[arrowstyle]{stealth}}}}]
	\tikzstyle reverse directed=[postaction={decorate,decoration={markings,
			mark=at position .65 with {\arrowreversed[arrowstyle]{stealth};}}}]
	\usepackage{import}
	
	\usepackage{mathrsfs,amsmath,amssymb,slashed}
	\usepackage{multirow,multicol}
	\usepackage{accents}
	\usepackage[perpage]{footmisc}

	\definecolor{darkgreen}{rgb}{0,0.3,0}
	\definecolor{darkblue}{rgb}{0,0,0.3}
	\definecolor{darkred}{rgb}{0.7,0,0}

	\newcommand{\bA}{\mathbf{A}}
	\newcommand{\beqs}{\begin{equation*}}

	\newcommand{\nn}{\nonumber}
	
	\newcommand{\ii}{\mathtt{i}}
	\newcommand{\jj}{\mathtt{j}}
	\newcommand{\kk}{\mathtt{k}}

	\newcommand{\elist}{\end{itemize}}

	\providecommand{\href}[2]{#2}

	\newcommand{\dl}{\mathbbmss{L}}

	\DeclareFontFamily{OT1}{rsfs}{}
	
	\DeclareFontShape{OT1}{rsfs}{m}{n}{ <-7> rsfs5 <7-10> rsfs7 <10->rsfs10}{} 
	
	\DeclareMathAlphabet{\mycal}{OT1}{rsfs}{m}{n}

	\DeclareMathOperator{\extdm}{d}
	
	\newcommand{\extd}{\extdm \!}

	\usepackage{amsmath,amssymb}
	\usepackage{amsthm}
	\usepackage{thmtools}
	
	\declaretheoremstyle[headfont=\bfseries]{normalhead}
	\declaretheorem[style=normalhead]{example}
	
	\newtheorem{defin}{Definition}
	\newtheorem{theo}{Theorem}
	\newtheorem{pro}{Proposition}
	\newtheorem{defi}{Definition}

	\usepackage[linktocpage=true,pagebackref]{hyperref}
	\hypersetup{
		colorlinks=true,
		linkcolor=blue,
		filecolor=magenta,      
		urlcolor=cyan,
	}
	
	\usepackage{titlesec}
	\titleformat{\chapter}[display]
	{\normalfont\huge\bfseries}
	{\chaptertitlename\ \thechapter}{20pt}{\Huge}
	\titleformat{\section}
	{\normalfont\Large\bfseries}
	{\thesection}{1em}{}
	
	\usepackage[pages=some]{background}
	\backgroundsetup{
		scale=1.2,
		color=black,
		opacity=1,
		angle=0,
		contents={
			\includegraphics[width=\paperwidth,height=\paperheight]{titlepage.pdf}
		}
	}

	\newcommand\summaryname{Abstract}
	\newenvironment{Abstract}%
	{\begin{center}%
		\bfseries{\summaryname} \end{center}}

	\newtheorem{definition}{Definition}

	\fancypagestyle{IHA-fancy-style}{%
		\fancyhf{}
		\fancyhead[LE]{\slshape \leftmark}
		\fancyhead[RO]{\slshape \rightmark}
		\fancyfoot[LE,RO]{\thepage}
	}
	
	\fancypagestyle{plain}{%
		\fancyhf{}%
		\fancyfoot[LE,RO]{\thepage}%
	}
	
	\chapternumberfont{\Large} 
	\chaptertitlefont{\LARGE}
	
	\graphicspath{{Figures/}{../Figures/}}

\begin{document}

	\frontmatter
	\begin{titlepage}
	
	\begin{center}

	\includegraphics[width = 50mm]{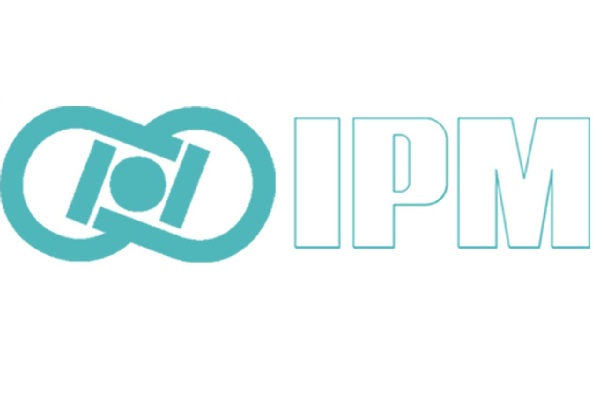}
	
	
	{\scshape  Institute for Research in Fundamental Sciences (IPM)}\\
	\vspace{2mm}
	{\scshape  School of Physics}\\
	\vspace{2 cm}
	
	{\scshape\LARGE\textbf{ $p$-form gauge fields: charges and memories}\par}
	\vspace{1.5 cm}
		{\Large{Erfan Esmaeili Fakhabi}}
	\vspace{2 cm}
	
	\textit{\large Thesis presented  for the degree of Doctor of Science}
	\\
	\vspace{1.5 cm}

	{ Supervisor: {Prof. Mohammad M. Sheikh-Jabbari}}
\\
	
	{Co-advisor: { Dr. Hamid R. Afshar}}
	\vfill
	
	September 2020
	\end{center}

	\end{titlepage}
	\newpage
	\newgeometry{outer=18mm,inner=18mm,top=25mm,bottom=25mm}
	
	\begin{Abstract}
	
	\begin{changemargin}{1.5cm}{1.5cm}
	In this thesis, we study the asymptotic structure of $p$-form theories on flat space. $p$-form theories are generalizations of Maxwell's theory of electrodynamics in which the gauge potential is a higher-rank differential form. As in the Maxwell theory, there is a choice of boundary conditions with an infinite-dimensional asymptotic symmetry
	group. For higher $p$-form theories on $(2p+2)$-dimensional flat spacetime, the surface charges are classified
	by the Hodge decomposition of gauge parameters on $2p$-sphere into exact and co-exact charges. For the exact parameters which are special features of higher-form theories, the surface charges are
	non-commuting.  In presence of $p$-branes, there are non-trivial zero-mode charges analogous to the electric charge for point particles. Although our $p$-form discussion is restricted to $2p+2$ dimensions, we generalize the study of  Maxwell theory not only to general dimensions but also to the anti-de Sitter background. 
	
	Finally, motivated by the IR triangle of gauge theories which relate asymptotic symmetries, memory effects, and soft theorems,  we  introduce
	a new kind of memory effect exerted on the internal modes of
	closed and open strings. On closed string probes, soft 2-form radiation rotates left- and right-moving modes oppositely, which amounts to a unit shift in the spin of probe particles. This provides the first instance of an `internal memory effect' on a non-local object. We conclude by general remarks on the physical significance of soft modes.
	\end{changemargin}
	\end{Abstract}

	\newpage


	\tableofcontents
	
	\frontmatter
	
	\chapter{Preface}
	This thesis encompasses my research on \emph{asymptotic symmetries in $p$-form theories} at the high energy physics group of IPM. It focuses on the problems I collaborated in solving, their immediate conceptual background, and their possible implications. My effort here has been to give a transparent account of the logic behind the calculations and results, which have already been published in detailed versions. The composition has a linear theme, with the earlier chapters providing the basis for the later ones. The last chapter on memory effect could be read independently nonetheless.
	The material of this thesis is mainly based on the following papers
	\begin{itemize}
	\item E.~Esmaeili, V.~Hosseinzadeh, and M.~Sheikh-Jabbari,
	``Source and response soft charges for Maxwell theory on AdS$_{d}$,''
	JHEP \textbf{12} (2019), 071
	\href{https://arxiv.org/abs/1908.10385}{1908.10385}
	
	\item
	E.~Esmaeili,
	``Asymptotic Symmetries of Maxwell Theory in Arbitrary Dimensions at Spatial Infinity,''
	JHEP \textbf{10} (2019), 224
	\href{https://arxiv.org/abs/1902.02769}{1902.02769}
	\item
	H.~Afshar, E.~Esmaeili, and M.~Sheikh-Jabbari,
	``String Memory Effect,''
	JHEP \textbf{02} (2019), 053
	\href{https://arxiv.org/abs/1811.07368}{1811.07368}
	
	\item H.~Afshar, E.~Esmaeili and M.~Sheikh-Jabbari,
	``Asymptotic Symmetries in $p$-Form Theories,''
	JHEP \textbf{05} (2018), 042
	\href{https://arxiv.org/abs/1801.07752}{1801.07752}
	\end{itemize}
	\newpage
	\begin{center}
\textbf{\large Acknowledgement}
	\end{center}

	This thesis—as a social enterprise—is based on thousands of human relations. Nonetheless, I am unable to imagine myself in similar status without my admirable supervisor \textbf{Prof. Shahin Sheikh-Jabbari}. His disciplined, reliable, and canonical practice of science formed my viewpoint of theoretical research and has been the cornerstone of this activity. 
	I have to declare my utter respect to Shahin for his key role in the most of my scientific activity as he created a researcher \emph{ex nihilo}: by defining rich problems in my Master and Ph.D., by giving terrific lectures which constitutes the most of my physics knowledge, and by his absolute promoting act anywhere there is a will to do better physics. His deep insight into basics and his critical eye on the literature and creative spontaneous ideas have been the source of encouragement in my work. I feel fortunate to have been participated in many workshops and scientific events under his support and endorsement.
	
	My progress in the subject of this thesis is fundamentally beholden to \textbf{Dr. Hamid-Reza Afshar}.  It was joyful to prove the statements that Hamid saw on the first day by his striking physical intuition.  I admire his determination and patience in the course of the work and his friendliness and understanding in desperate moments of a Ph.D. student.
	
	I am grateful to \textbf{Prof. Daniel Grumiller} for hosting my Junior Research Fellowship at Erwin Schro\"dinger Institute for Physics and Mathematics. It has been an honor for me to join his active research group at Vienna Technical University. He granted me the opportunity to explore new physical topics (namely, 2D gravity and higher spin theory) in the last year of my Ph.D. program which expanded my knowledge in high energy physics in new areas.
	
	I would like to thank my Ph.D. jury: Aliakbar Abolhasani, Amjad Ashoorioon, Marc Henneaux, Daniel Grumiler, and Farhang Loran for the time they dedicated to this thesis.
	
	I am thankful to the faculty of physics at IPM, especially to \textbf{Prof. Yasaman Farzan} for her elegant lectures and encouragements.  People at High Energy Physics Group of IPM provided a rich web of ideas, questions, arguments, and stimulations, to be acknowledged: Hamed Adami, Sajjad Aghapour, Moslem Ahmadvand, Arash Ardehali,  Kamal Hajian, Vahid Hosseinzadeh, Ghadir Jafari, Reza Javadinezhad, Reza Mohammadi,  Ali Mollabashi, Mojtaba Najafizadeh, Maryam Noorbakhsh, Abasalt Rostami, Saeedeh Sadeghian, Hamid Safari, Ali Seraj, Hesam Soltanpanahi, and Mohammad H. Vahidinia.

	I am deeply grateful to my wife \textbf{Maryam}, as well as our parents and family, who made the world more beautiful and without whom this job was simply impossible.
	\paragraph{Financial Support.}
	I attribute my fortune to focus on research to the Institute for Research in Fundamental Sciences (IPM) for its five-year financial support. In addition, I would like to thank the Erwin Schr\"odinger International Institute for Mathematics and Physics for its generous grant which made a four-month-visit to Vienna possible. I also benefited financial support from  INSF grant No 950124, ICTP network scheme NT-04, and the Iran-Austria scientific IMPULSE project.

	\chapter{Notation and conventions}
	Most of the symbols in this text have the universal definitions explained here. The spacetime dimension is $d$ and the metric is Lorentzian with signature $(-+\cdots+)$. We will work in hyperbolic coordinates where
	\begin{center}
	\begin{tabular}{c|l}
	$\rho$   & is the hyperbolic radial coordinate \\
	$x^a$   & are hyperbolic transverse coordinates.
	\end{tabular}
	\end{center}
	
	Tensors  on different coordinates have particular index notations as follows
	
	\begin{center}
	\begin{tabular}{c|l}
	$\mu,\nu,\alpha$  & spacetime indices in any coordinate system\\
	$a,b,c$ & indices on $(d-1)$-dimensional de Sitter space\\
	$i,j,k$ & indices on $(d-2)$-sphere\\
	$\ii,\jj,\kk$ & Cartesian indices
	\end{tabular}
	\end{center}
	
	Regarding connections,    $D$  denotes the covariant derivative on $(d-1)$-de Sitter space while 
	$\mathcal{D}$ is its counterpart on $(d-2)$-sphere. We presume asymptotic power-law expansions in radial coordinate $\rho$ for gauge fields. We use different fonts for the exact gauge field versus the leading term in the expansion:
	
	\begin{center}
	\begin{tabular}{c|l}
	$\mathcal{A}_{\mu_1\cdots\mu_p}$ & gauge fields in spacetime\\
	$A_{\mu_1\cdots\mu_p}$ & the leading asymptotic coefficient of gauge fields
	\end{tabular}
	\end{center}
	
	In chapter \ref{p chapter} we utilize form notation:
	
	\begin{center}
	\begin{tabular}{c|l}
	$\wedge$& exterior (wedge) product\\
	$\mathbf{A}$ & a differential $p$-form: $\mathbf{A}=\A_{\mu_1\cdots\mu_p}dx^{\mu_1}\wedge\cdots \wedge dx^{\mu_p}$\\
	$\star$ & Hodge star operator defined in Appendix \ref{AppDiff}\\
	$\di$ &exterior derivative\\
	$\did$ & codifferential operator defined in Appendix \ref{AppDiff}
	\end{tabular}
	\end{center}

	When discussing worldsheet theory in chapter \ref{memory chapter}, we use the notation defined below
	\begin{center}
	\begin{tabular}{c|l}
	$\tau,\sigma$ & worldsheet coordinates\\
	$X^\mu$ & target space coordinates\\
	$\gamma_{ab}$& worldsheet metric\\
	$\ell_s$ & string length\\
	$\alpha^{\p}$ & $1/$string tension
	\end{tabular}
	
	\end{center}
	
	Finally, $\approx$ means equality on the mass shell and $\equiv$ shows equality by definition.
	
	\mainmatter
	\chapter{Introduction}
	
	What is the symmetry group of a given gauge theory? Of course, a gauge theory is–by definition–a theory with \emph{gauge symmetry}; it is invariant under the transformation of variables by arbitrary functions of spacetime coordinates. Strictly speaking, this vast amount of freedom is not a symmetry, for it does not transform observables of the theory into each other. For a deterministic theory, all observables are predicted by dynamics in terms of given data in the past. Gauge variables are not observable. Instead, their \emph{relative value} (depending on the specific theory) is observable and is predicted by dynamics. A number of gauge variables serve as clocks, rods, the ground potential, etc. (generally speaking, \emph{reference frames}) with respect to which the other variables are predicted and measured. Hence, gauge freedom conveys nothing but redundancy in the description of the theory. So, why bother with gauge symmetry?  Gauge variables show possible ways that a system can interact with another one, by all possible ways that gauge-invariant quantities can be constructed. They provide \emph{handles} \cite{Rovelli:2013fga} through which systems can interact. For instance,  electromagnetic fields can interact with matter fields and particles through local gauge interactions.

	In Lagrangian theories with a Lie group of global symmetries, there corresponds a conserved current $j^\mu$ to every generator of the  Lie algebra by Noether's first theorem, such that conserved Noether charges are given by volume integrals on $(d-1)$-hypersurfaces.
	
	Similarly, using Noether's second theorem there corresponds a conserved 2-form superpotential $k^{\mu\nu}$ to   {reducibility parameters} \cite{Barnich:2001jy}, also known as global part of gauge symmetries.  This latter set of symmetries have \emph{surface charges}, obtained by integrating $k^{\mu\nu}$ on a $(d-2)$-hypersurface. A familiar example is charge conservation in $U(1)$ abelian gauge theory.

	Gauge theories on open manifolds have particular features due to the presence of \emph{boundaries}. In such cases (which constitute the most physically interesting examples), the classical solution space of the theory is determined by imposing the correct number of gauge conditions and prescribing boundary conditions. The latter are often asymptotic fall-offs at infinity or on the black hole horizon\footnote{The term `boundary condition' can be alternatively used in a general sense as any condition that completely defines the phase-space under consideration. This definition is inclusive of gauge conditions as well as phase-space dependence of the fields. } (c.f.  \cite{Ruzziconi:2019pzd, Ashtekar:2018lor}). \emph{Residual gauge transformations}, by definition preserve gauge and boundary conditions. The control on the asymptotic behavior of gauge fields, allows a large (usually infinite-dimensional) algebra of residual transformations to have well-defined surface charges in the asymptotic region of spacetime. In non-abelian gauge theories such as gravity theories, the explicit form of residual gauge transformations can be quite complicated and in particular field-dependent: the generator depends on the background field explicitly. In such cases, the finite form of transformations is usually elusive. Furthermore, the surface charges might lack one or more of their expected properties; namely, conservation, finitude, and integrability on solution space. For example, the asymptotic symmetry algebra for a gravitational system in four dimensions is \emph{BMS algebra}: a semi-direct product of Lorentz transformations and an infinite-dimensional extension of translations \cite{Bondi:1962,Sachs:1962}, the so-called \emph{supertranslations}. The surface charges corresponding to supertranslations are not integrable on radiative backgrounds.
	
	For analyzing the residual gauge symmetries and associated conserved charges, there are different systematic frameworks and formulations. The simplest one is based on the usual Noether's theorem suitably extended to capture these symmetries e.g. see \cite{Banados:2016zim, Avery:2015rga, Fatibene:1994vc}. There are other approaches based on Hamiltonian formulation e.g. see \cite{ Regge:1974zd,Castellani:1981us,Brown:1986nw, Brown:1986ed} and covariant phase space method \cite{Lee:1990nz,Iyer:1994ys,Barnich:2001jy}.

	Responding to the question raised in the first line of this chapter,  the asymptotic symmetry group is a subgroup of the entire symmetry group of the theory.  If the bulk theory is holographic, its asymptotic symmetry group is isomorphic to the global symmetry group of the dual field theory \cite{Aharony:1999ti}. This fact motivates the extensive study of boundary conditions in various backgrounds and their symmetries to elucidate the basic properties of putative dual theories \cite{Ashtekar:1999jx,Mishra:2017zan, Mishra:2018axf, Compere:2019bua,Anninos:2010zf,Ashtekar:2014zfa,Poole:2018koa,Compere:2008us}.

	In asymptotically flat spacetimes, the observation that \emph{supertranslations} of BMS group in gravity are induced by zero-frequency mode of gravitational radiation, provided a powerful criterion for identifying physical boundary conditions for a large set of gauge theories. It was shown that in many physically relevant theories, an infinite-dimensional algebra is realized simply because of infinite possible directions on the celestial sphere that soft bosons are radiated. The surface charges as quantum operators commute with the Hamiltonian and this leads to a derivation of Weinberg's soft theorems \cite{Weinberg:1965nx}  as Ward identities corresponding to asymptotic symmetries \cite{ Strominger:2013lka,Strominger:2013jfa,Kapec:2014opa,He:2014cra}. The derivation of soft theorems from asymptotic symmetries has been carried out in numerous theories such as  gravity, higher spin theories, anti-symmetric $2$-form theories \cite{Kapec:2014zla,Kapec:2015ena,Campiglia:2015qka,Seraj:2016jxi,Conde:2016csj,DiVecchia:2017gfi,Campoleoni:2017mbt,Campoleoni:2017qot} Yang-mills theory  \cite{Campoleoni:2019ptc} and even scalar theories \cite{Campiglia:2017dpg,Francia:2018jtb}.
	Alongside this mathematical motivation for soft charges, the observability of soft modes is attributed to their memory effects. 	This \emph{IR triangle} of asymptotic symmetries, soft theorems, and memory effects expanded our understanding of the asymptotic structure of gauge theories in asymptotically flat backgrounds  (see e.g. \cite{Heissenberg:2019fbn,Strominger:2017zoo,Gomez:2016hxz,Mitra:2017wkm,Mao:2016bzy} for review).

	The permanent trace of zero-frequency mode of radiation on a probe is called \emph{memory effect}  \cite{Christodoulou:1991cr,Bieri:2013hqa,Strominger:2014pwa,Pasterski:2015tva,Pasterski:2015zua,Susskind:2015hpa,Pate:2017fgt,Hamada:2017atr,Gomez:2017soz,Compere:2019odm,Hollands:2016oma,Pate:2017fgt,Campoleoni:2017qot,Garfinkle:2017fre,Mao:2017wvx,Campiglia:2017xkp,Satishchandran:2017pek,Laddha:2018vbn}. It happens when the initial and the final state of the gauge field or the metric differ by a pure gauge or coordinate transformation, while  there is a non-zero field strength in the middle of the process, due to radiation.  The memory effect is the imprint of this evolution of the gauge field  on a charged probe, which clearly involves low-energy gauge bosons. 	For instance, the gravitational memory is the change in the relative position of two near-by objects when some form of energy passes by at finite duration. The phenomenon was introduced by Zel'dovich and Polnarev\cite{zeldo} in 1974    as the displacements of geodesic congruences imprinted by the passage of gravity waves \cite{Christodoulou:1991cr,Zel'dovic} within general relativity, but never confirmed experimentally so far.

	Not surprisingly, one of the first instances of the IR triangle was  (massless) QED \cite{He:2014cra,Campiglia:2015qka}.  Asymptotic analysis at the null infinity (see e.g. \cite{He:2014cra, Campoleoni:2018uib,Campoleoni:2017qot, Kapec:2014zla,Satishchandran:2019pyc,Hosseinzadeh:2018dkh}) was generalized to subleading orders in asymptotic expansion of electromagnetic potential  \cite{Seraj:2016jxi, Campiglia:2016hvg,Campiglia:2018dyi,Hirai:2018ijc,Campiglia:2014yka}.
	
	In this thesis we explore three paths to generalize the asymptotic symmetry results of four-dimensional QED:
	
	\begin{itemize}
	
	\item Higher dimensions at spatial infinity (chapter \ref{max chapter})
	
	\item Asymptotic symmetry group in anti-de Sitter background (chapter \ref{max chapter})

	\item Higher form gauge potentials (chapter \ref{p chapter})
	
	\end{itemize}

	Our derivation in all cases is a generalized implementation of the calculations first performed by Campiglia and Eyheralde for Maxwell theory in four dimensions \cite{Campiglia:2017mua}, in which boundary conditions are imposed in the hyperbolic coordinates of Minkowski space \cite{Ashtekar:1991vb}. This coordinate system not only makes Lorentz symmetry manifest but covers the whole spatial infinity of flat space. In addition,   the surface charges defined at the null infinity  are recovered if the integration surface approaches the boundaries of the asymptotic de Sitter space. This old approach  has also been used in gravity \cite{Mann:2006bd,deBoer:2003vf,Compere:2011db} and in temporal infinity appropriate for studying massive particles \cite{Campiglia:2015qka}.

	\paragraph{Soft electromagnetic charges in higher dimensional flat space.}
	
	The spatial infinity analysis of Maxwell theory in four dimensions \cite{Seraj:2016jxi, Campiglia:2017mua,Henneaux:2018gfi,Prabhu:2018gzs} generalizes to higher dimensions by applying the canonical methods in a standard way. Notably, the Hamiltonian formulation of the theory in higher dimensions by Henneaux and Troessaert \cite{Henneaux:2019yqq} shows that the boundary structure in higher dimensions is similar to the four-dimensional case (except for parity conditions). Examination of the theory at spatial infinity has its special merits. Derivation of the surface charges in terms of Cauchy data makes their physical significance more transparent and can be helpful in the characterization of quantum states defined on equal-time slices. Moreover, contrary to radiative boundary conditions that are sensitive to the parity of spacetime dimension \cite{Hollands:2004ac,He:2019jjk},  Coulombic modes at spatial infinity have a uniform structure in even and odd dimensions.

	In particular, the 3-dimensional theory has a simple solution space in terms of left- and right-moving modes on $dS_2$ which is covered in section \ref{3d sec}. Asymptotic symmetries of the 3d Einstein-Maxwell theory is studied both at the null infinity and the near-horizon geometries \cite{Donnay:2018ckb,Barnich:2015jua}. We will show how the symmetry algebra at the null infinity  is recovered when taking a limit from spatial infinity. In addition, our hyperbolic setup parallels the study of asymptotically flat gravity with BMS$_3$ symmetry at spatial infinity  \cite{Compere:2017knf}. Thus, we expect that the combined  Einstein-Maxwell analysis will reproduce the symmetry group found in \cite{Barnich:2015jua}.

	In deriving soft theorems from commutation of asymptotic symmetry charges with $S$-matrix 
	\begin{equation}
	\langle out|Q^+_{\epsilon_+} S-SQ^-_{\epsilon_-}|in\rangle=0
	\end{equation}
	one needs a boundary condition that identifies gauge parameters $\epsilon_+$ and  $\epsilon_-$ between the future and the past null infinity respectively. The requirement that the parameters are mapped \emph{antipodally} i.e. $\epsilon_+(\hat{x})=\epsilon_-(-\hat{x})$ is called the \emph{antipodal matching condition} \cite{Strominger:2014pwa}. Classically, the radial electric field of constantly moving electric charges (so-called Li\'enard–Wiechert solution) happens to be antipodally related between past and future null infinities. Therefore, \emph{charge conservation} $Q^+_{\epsilon_+}=Q^-_{\epsilon_-}$ as equality of phase space functions holds classically only if the parameters are identified antipodally as well.
	
	An important merit of using hyperbolic coordinates is the simple formulation of antipodal matching condition as a map between future and past boundaries $\mathscr{I^+},\,\mathscr{I}^-$ of asymptotic de Sitter space. We will show how this condition is required to avoid logarithmic behavior in electromagnetic fields in four dimensions. We also explain its necessity to ensure regularity of field strength at light cone for $d\geq 4$. In the Hamiltonian formulation of the theory, the symplectic form is logarithmically divergent in four dimensions unless certain parity conditions are imposed on the solution space \cite{Henneaux:2018gfi}.

	\paragraph{Soft electromagnetic charges on $AdS_d$.} In section \ref{Maxwell AdS} we discuss the asymptotic structure of the Maxwell theory on anti-de Sitter background.  Study of asymptotic symmetries of anti-de Sitter gravity is well-understood \cite{Brown:1986nw,Balasubramanian:1999re,Ashtekar:1999jx}. 	The timelike conformal boundary of anti-de Sitter space makes its causal structure  different from flat space.  Anti-de Sitter space is not globally hyperbolic: hyperbolic equations do not have globally unique solutions for given initial data.  In a near boundary analysis, one sets Dirichlet conditions on boundary fields (so-called \emph{sources}) and subleading terms are determined by algebraic recursive equations in terms of the sources. This procedure terminates at some finite order and certain components are left undetermined (so-called \emph{responses}). In the geometric approach \cite{Ashtekar:1999jx}, asymptotic symmetries are defined as gauge transformations that preserve boundary conditions (here the sources) modulo those which are trivial on the boundary. In anti-de Sitter gravity, the asymptotic structure is invariant under $AdS_d$ isometry group $O(d-1,2)$ for $d>3$. The corresponding conserved charges have thermodynamic significance (like mass and angular momentum) and they are functions of the responses.
	
	Similarly, the Dirichlet condition is set in $U(1)$ gauge theory by fixing the induced gauge field on the boundary. This structure is preserved by gauge transformations which asymptote to a constant function. For this reason, the asymptotic symmetry algebras of bulk gauge theories are finite-dimensional in the standard boundary conditions advocated in AdS/CFT correspondence, and in $U(1)$ theory, the algebra incorporates only the total electric charge.
	
	This status quo calls for generalized boundary conditions that allow infinite-dimensional asymptotic symmetries for anti-de sitter gravity and gauge theory. This possibility and its holographic consequence has been investigated in gravity  \cite{Compere:2008us,Takayanagi:2011zk,Compere:2019bua,Compere:2020lrt}.  These results can help us understand  the conceptual and technical gap between flat space and $AdS$ holographies.

	In section \ref{Maxwell AdS}, we study Maxwell theory on  AdS$_d$ and construct a conserved symplectic form. Our main result is finding two sets of boundary gauge transformations, which may be called \emph{source} and \emph{response} transformations, akin to AdS/CFT terminology. These two sets commute up to a central charge.

	\paragraph{Soft $p$-form charges.} To explore theories with a more interesting gauge structure, we look at $(p+1)$-form gauge fields $\A_{\mu_0\cdots\mu_p}$  which are natural extensions of the usual Maxwell theory of electrodynamics.
	Higher form field gauge fields are part of supersymmetry multiples, and they are present in supersymmetric theories  \cite{Cremmer:1979up,Salam:1989fm}.  Prominent examples are $p$=even ($p$=odd) forms in 10d type IIb (IIa) supergravity which arise as the low energy limit of type II string theories. 
	Higher form gauge theories in Lagrangian or Hamiltonian description and their Dirac-type quantization conditions and duality properties have been extensively studied  \cite{Deser:1997mz,Henneaux:1997ha,Teitelboim:1985yc,Teitelboim:1985ya,Banados:1997qs,Henneaux:1986ht,Cremmer:1997ct,Cremmer:1998px,Bremer:1997qb, Henneaux:1985kr, Baulieu:1987pz, Brown:1986nw,Henneaux:1999ma,Bekaert:2000qx}. 
	$p$-forms received a renewed attention after the introduction of D$_p$-branes in string theory \cite{Polchinski:1995mt}. $D_p$ branes couple to $(p+1)$-form gauge fields and carry RR-charge.
	Higher form fields come with a $p$-form gauge symmetry, where the gauge parameter is a generic $p$-form. This is a direct generalization of the case of Maxwell theory where the gauge parameter is a scalar.

	We choose to fix the Lorenz gauge which preserves Lorentz symmetry, and we specify the fall-off asymptotic behavior of gauge fields.  We fix the fall-off behavior such that we find finite and well-defined expressions for the conserved surface charges in $d=2p+4$ dimensions. This spacetime dimension is special in some different ways, especially with regards to soft radiation and memory effect.  The radiation flux of a localized source in $d$ dimensions has radial fall-off $r^{-(d-2)}$, which means the  field yielding this radiation should fall off as $r^{-\frac{d-2}{2}}$ \cite{Ortaggio:2014ipa,Campoleoni:2017qot}, usually called radiation fall-off behavior. On the other hand, the fall-off behavior of fields generated by electric sources, the so-called Coulomb fall-off behavior, for a $(p+1)$-form is $r^{-(d-p-3)}$ \cite{Ortaggio:2014ipa}. The simplest instances of memory effect are in examples  (like gravity and QED in four dimensions)  when these two are equal. In $(p+1)$-form theories, this   happens in $d=2p+4$.  Otherwise, one can focus on other special dimensions of interest. For example, a 2-form theory in four dimensions is dual to a scalar theory, and its asymptotic symmetries can account for the scalar IR triangle \cite{Campiglia:2018see, Francia:2018jtb, Henneaux:2018mgn}.

	The central feature in the analysis of $p$-form charges ($p>0$) is that the parameter (pulled back to $S^{2p+2}$)  can be decomposed into exact and coexact parts, leading to two distinct sets of asymptotic surface charges which we call  \emph{exact and coexact asymptotic charges}. In addition, the surface charge algebra shows that the coexact charges commute among themselves and with exact charges. The exact charges, however, do not commute and form an infinite-dimensional Heisenberg algebra. Electric charges of $p$-branes are found by applying Gauss law in orthogonal directions, and their contribution to black brane thermodynamics has been studied  \cite{Spindel:2018cgm,Detournay:2018cbf, Compere:2008cv,Peng:2016qnz,Rogatko:2009th}. These `zero-mode charges'  are relevant to thermodynamics while ignorant of the motion of the brane. However, the higher modes that we derive store the information of brane boost and alignment.

	\paragraph{String Memory}
	
	Chapter \ref{memory chapter} is devoted to the string memory effect.  Depending on the quantity imprinted on the probe we may have different types of memory effects. In gravity, this memory effect in the language of soft charges is associated with supertranslations (see also \cite{Compere:2018ylh,Laddha:2018vbn,OLoughlin:2018ebk,Zhang:2018srn}).
	In electrodynamics, the leading memory effect  is the shift of the velocity of a charged point particle under the 1-form gauge field radiation\footnote{The memory effect in Yang-Mills theory is called `color memory'  \cite{Pate:2017vwa}.} (the kick memory effect \cite{Winicour:2014ska,Mao:2017axa}).

	Generically, $p$-branes (with $p>0$) have internal degrees of freedom which can be degenerate, so  the transition between these internal modes has no energy cost. In particular, there could be transitions between these modes by the exchange of \emph{soft $(p+1)$-forms}. This yields a new type of memory effect which is associated with the change in internal excitations of an extended object, like a $p$-brane. The internal memory effect may be put in contrast with other so-far discussed memories which are associated with the change in a spacetime property of the probe like position, momentum, or spin.

	We study such an internal memory effect by focusing on the $2$-form theory (the $p=1$ case) which is naturally coupled to strings.    $2$-form gauge field backgrounds, as well as the fundamental string probes, are both contents of string theory, thus this effect will henceforth be dubbed \emph{string memory effect}. We also connect the string memory effect to 2-form soft charges to construct one edge of the $2$-form soft triangle.


	\chapter{Charges and Memories}\label{2 chapter}
	
	In this chapter, we briefly review covariant phase space formalism and electromagnetic memory effect. The content of later chapters that contain new results is mainly based on the concepts and tools defined here. Section \ref{ASG def} is devoted to definition of surface charges in covariant phase space method (a detailed review can be found in \cite{Hajian:2015eha}). In section \ref{memory section} we illustrate the relation between surface charges and soft radiation in QED and we derive kick memory effect.
	
	\section{Asymptotic symmetries}\label{ASG def}
	

	\subsection{Poisson and symplectic structures}
	Hamiltonian systems can be covariantly described by symplectic geometry. The Poisson structure quantifies how observables evolve over time. In addition, one can express symmetries in geometric terms in a transparent way. The first definition is motivated by our understanding of phase space as a Poisson manifold.

	\begin{definition}
	Let $M$ be a manifold. A Poisson structure on $M$ is an antisymmetric bilinear map
	\begin{equation}
	\{\cdot,\cdot\}:C^\infty(M)\times C^\infty(M)\to C^\infty(M): F,G\mapsto \{F,G\}
	\end{equation}
	that satisfies Leibniz and Jacobi rules. Endowed with $\{\cdot,\cdot\}$, $M$ is called a \textit{Poisson manifold}.
	\end{definition}
	If we fix one of the functions \textit{e.g. }$ F$, the Poisson bracket performs some kind of derivation\footnote{cause it satisfies Leibniz rule.} on the input functions: 
	$G\mapsto \{F,G\}$. It turns out that  any of these derivations corresponds to exactly one  vector field on $M$:
	\begin{equation}
	\xi_{H}\equiv-\{H,\cdot\}\label{math-hamiltonian vector}
	\end{equation}
	called the \textit{Hamiltonian vector field associated with function $H$}. Definition\eqref{math-hamiltonian vector} relates the derivative of $G$ along the vector field $\xi_{H}$, to its Poisson bracket with $H$:
	\begin{equation}
	\xi_H(G)=-\{H,G\}.\label{math-poisson-vector}
	\end{equation}
	One may find the integral curve $\gamma(s)$ whose tangent vector is $\xi_H$. That is, $d\gamma(s)/ds=\xi_{H}\mid_{\gamma(s)} $. Then \eqref{math-poisson-vector} can be rewritten as\begin{equation}
	\frac{dG}{ds}=-\{H,G\}\,,
	\end{equation}
	that can be recognized as the equations of motion when $H$ is  the Hamiltonian. In particular, if the Poisson bracket of two functions $G$ and $H$ vanishes, each of them is \textit{conserved} along the integral curves of the other.

	Now that we have associated with any function $F$ its Hamiltonian vector field $\xi_F$, it is natural to define an object that acts on a pair of vectors, namely a two form.
	\begin{definition}
	A symplectic manifold is a manifold  $M$ 
	endowed with a closed, non-degenerate two-form $\omega$ on $M$\footnote{non-degeneracy means that $\omega$ has no null eigenvector $v$:\quad $\omega(v,w)=0\quad \forall w$.}.
	\end{definition}
	Since $\omega$ is anti-symmetric, symplectic manifolds are \textbf{even-dimensional}. By \eqref{math-hamiltonian vector} we linked functions and vector fields on a Poisson manifold.  Lacking a Poisson structure on symplectic manifolds, we have to redefine Hamiltonian vector fields. Given $F\in C^{\infty}(M)$ where $M$ is a symplectic manifold, its Hamiltonian vector field is defined to  satisfy
	\begin{equation}
	\omega(\xi_{F},\cdot)=dF.\label{math-symplectic-hamiltonian}
	\end{equation}
	Both sides are one-forms on $M$, and ``$\cdot$'' is replaced by the vector field on which the form acts.

	Starting from a symplectic manifold, it is possible to define a Poisson structure such that
	\begin{equation}
	\{F,G\}\equiv\omega(\xi_{F},\xi_{G}).\label{math-omega-poisson}
	\end{equation}
	Conversely, the same definition allows us to define a symplectic structure on a Poisson manifold\footnote{A Poisson manifold need not be even-dimensional. The symplectic structure can be defined on the subspace spanned by Hamiltonian vector fields, the so-called \textit{symplectic leaves.}}.

	\subsection{Covariant phase space}
	Most theories of interest are Lagrangian theories in which the action functional is given by
	\begin{equation}
	S[\Phi]=\int \mathbf{L}[\Phi,\partial_\mu\Phi,\partial_{\mu}\partial_\nu\Phi,\cdots]
	\end{equation}
	where $\mathbf{L}=L d^dx$ is the Lagrangian and $\Phi$ denotes the dynamical fields. A transformation on the fields depending on arbitrary parameters $f=(f^\alpha)$ and their derivatives is called a gauge transformation
	\begin{align}\label{gauge def}
	\delta_f\Phi=R_\alpha f^\alpha+R_\alpha^\mu \partial_\mu f^\alpha+R^{(\mu\nu)}_\alpha\partial_\mu\partial_\nu f^\alpha+\cdots\,.
	\end{align}
	The functions $R_\alpha^{(\mu_1\cdots\mu_k)}$ are functions of the coordinates, the fields and their derivatives. The Lagrangian has gauge symmetry \eqref{gauge def} if it transforms up to a boundary term
	\begin{eqnarray}
	\delta_f\mathbf{L}=\di\mathbf{B}_f\,.
	\end{eqnarray}
	\begin{example}	
	For $(p+1)$-form theory
	\begin{equation}\label{p action}
	S=-\frac{1}{2}\int \mathbf{F}\wedge\star\mathbf{F}
	\end{equation}
	where $\mathbf{F}=\di\mathbf{A}$ is a $(p+2)$-form in our conventions,
	the gauge transformation $\delta_f\mathbf{A}=\di \bLaa$ is a symmetry of the theory for arbitrary $p$-form $\bLaa$.\hfill$\square$	
	\end{example}	
	
	Gauge symmetry shows a redundancy in the description of the theory, as there is arbitrary freedom in the value of dynamical fields. To eliminate this freedom, one imposes additional algebraic or differential constraints on the fields, the so-called gauge-fixing conditions
	\begin{equation}\label{gauge condition}
	G[\Phi]=0.
	\end{equation}
	The number of gauge-fixing conditions must be equal to the number of free parameters $f_\alpha$ in order to eliminate the freedom completely.   After the imposition of the right number of gauge-fixing conditions, there usually remains a number of residual gauge transformations, i.e. those which preserve the gauge conditions \eqref{gauge condition}, namely
	
	\begin{eqnarray}
	\delta_fG[\Phi]=0
	\end{eqnarray}
	where $f$ is no longer arbitrary.

	Equations of motion of the theory are obtained by the Euler-Lagrange derivative of the Lagrangian $L$
	\begin{eqnarray}\label{EULER}
	\frac{\delta L}{\delta\Phi}\equiv \sum_{k\geq 0}(-1)^k\partial_{\mu_1}\cdots\partial_{\mu_k}\Big(\frac{\partial L}{\partial \Phi_{\mu_1\cdots\mu_k}}\Big)
	\end{eqnarray}
	where $\Phi_{\mu_1\cdots\mu_k}=\partial_{\mu_1}\cdots\partial_{\mu_k}\Phi$.
	The solution space depends on the choice of boundary conditions, which are often fall-off conditions on the fields in the asymptotic region. Therefore, the choice of boundary conditions defines the physical theory under consideration. 
	Variation of Lagrangian can be written as
	\begin{equation}
	\delta{\mathbf{L}}=\frac{\delta\mathbf{L}}{\delta\Phi}\delta\Phi+\di\boldsymbol{\theta}(\Phi,\delta\Phi)\label{LET}\,.
	\end{equation}

	\begin{example}
	For $(p+1)$-form theory,
	\begin{equation}
	\boldsymbol{\theta}(\mathbf{A},\delta\mathbf{A})=-\delta\mathbf{A}\wedge\star\mathbf{F}
	\end{equation}
	$\hfill\square$
	\end{example}

	Given a set of boundary conditions, the solutions space of the theory is specified such that any field configuration $\Phi$ that satisfies the equations of motion is a point in this space. Cotangent bundle of the solution space consists of field variations $\delta\Phi$ and variation of their derivatives $\delta\partial_{\mu_1}\cdots\partial_{\mu_k}\Phi$. For instance, $\boldsymbol{\theta}(\Phi,\delta\Phi)$ is a 1-form in field space. $\delta$ is Grassmann odd $\delta^2=0$.

	From the $(d-1)$-form $\boldsymbol{\theta}(\Phi,\delta\Phi)$ one defines the presymplectic density 
	\begin{equation}
	\boldsymbol{\omega}(\Phi;\delta\Phi,\delta\Phi)=\delta[\boldsymbol{\theta}(\Phi;\delta\Phi)]
	\end{equation}
	
	which is a 2-form on solution space and a $(d-1) $-form on spacetime. This quantity is conserved on-shell  $\di\boldsymbol{\omega}\approx 0$, which is proven by acting by $\delta$ on \eqref{LET}.

	The \emph{symplectic 2-form} of the theory is defined as its integration on a Cauchy surface
	\begin{eqnarray}\label{symp}
	\Omega=\int_\Sigma\boldsymbol{\omega}(\Phi;\delta\Phi,\delta\Phi)\,.
	\end{eqnarray}

	The symplectic form defined as above is generically degenerate: there may exist non-trivial vector fields in solution space $V=\delta_V\Phi$ such that $\Omega(\delta_V\Phi,\cdot)=0$. A typical example of degeneracy is gauge transformation. Except for special classes   which are non-trivial on the boundary, gauge transformations make the symplectic form degenerate.

	Field variations in solution space generated by gauge transformations  have  the following  special property \cite{Wald90, Lee:1990nz}.
	\begin{pro}
	If $\delta_\epsilon\Phi$ is a gauge transformation generated by $\epsilon$, then the presymplectic form $\boldsymbol{\omega}$ acted on  $\delta_\epsilon\Phi$  is exact
	\begin{eqnarray}
	\boldsymbol{\omega}(\Phi,\delta\Phi,\delta_\epsilon\Phi)\approx \normalfont{d}\mathbf{k}_\epsilon(\Phi,\delta\Phi)\,
	\end{eqnarray} 
	provided $\delta\Phi$ satisfies linearized equations of motion.
	\end{pro}

	This proposition shows that Hamiltonian functions associated with $\delta_\epsilon\Phi$ defined in \eqref{math-symplectic-hamiltonian} are \emph{surface integrals}\footnote{For a derivation of  the superpotential $\mathbf{k}_\xi$ for diffeomorphisms $\delta_\xi\Phi$  see e.g. \cite{Hajian:2015eha}. }
	\begin{eqnarray}\label{charge eps}
	\slashed{\delta}Q_\epsilon=\oint\mathbf{k}_\epsilon(\Phi,\delta\Phi)\,.
	\end{eqnarray}
	The charge variation \eqref{charge eps} is not necessarily integrable on solution space: different paths between two solutions must bring about the same charge differences. This happens only if $\mathbf{k}_\epsilon(\Phi,\delta\Phi)$ is a total variation.

	\begin{definition}
	A gauge transformation with generator $\epsilon$ is called an \textbf{ asymptotic symmetry} if its superpotential $k_{\epsilon}^{\mu\nu}$ falls off at infinity in such a way that its surface charge $\eqref{charge eps}$ is  finite and generically non-zero. The corresponding transformation is called a \textbf{large gauge transformation} which forms the \textbf{the asymptotic symmetry group} of the theory.
	\end{definition}

	\begin{example} The symplectic form for a $(p+1)$-form theory is
	\begin{eqnarray}
	\Omega=\int_\Sigma \delta\bA\wedge\star\delta \bF\,.
	\end{eqnarray}
	
	For a gauge transformations $\delta_\Lambda\bA=\di\bLaa$, we have $\boldsymbol{\omega}(\bA;\delta_\Laa\bA,\delta\bA)=\di\bLaa\wedge\star\delta\bF$, so that
	
	\begin{eqnarray}
	\mathbf{k}_\Laa=\bLaa\wedge\star\delta\bF\,.
	\end{eqnarray}
	
	If the gauge parameter is field-independent, that is $\delta\bLaa=0$, the charge is integrable, 
	
	\begin{equation}\label{Naive charge}
	Q_\Laa=\oint\bLaa\wedge\star\bF\,.
	\end{equation}
	
	This quantity is in general neither finite nor conserved. We will discuss in chapter \ref{p chapter} how fixing the Y-ambiguity can lead to a set of conserved charges for specific boundary conditions. $\hfill \square$
	\end{example}

	In gravity theories, interesting boundary conditions often require field-dependent diffeomorphism vector fields and integrated expressions for the charges as in \eqref{Naive charge} are not available \emph{a priori}. Moreover, the charge expressions are often well-defined only asymptotically where the behavior of the fields is controlled by boundary conditions. However, there is usually a subgroup of asymptotic symmetry group which leaves each solution invariant. This \emph{global subgroup} of asymptotic symmetries is usually exact and conserved deep into the bulk as long as the integration surface does not hit a source. In the above example of $(p+1)$-form theories,  closed parameters $\di\bLaa=0$ leave the gauge field invariant. The charge \eqref{Naive charge} is non-vanishing only if the integration surface has non-trivial topology. These \emph{zero-mode charges} are discussed in chapter \ref{p chapter}.

	If the gravitational coupling is weak $G\to0$, the matter theory is effectively living on a fixed background, and it is symmetric under the isometry group of the background and the corresponding charges can be obtained using covariant phase space formalism.

	\begin{example} In $(p+1)$-form theory, Killing vector fields $\xi^\mu$  act on the gauge field by Lie derivative
	\begin{equation}\label{Lie A}
	\delta_\xi\bA=\xi\cdot\bF+\di(\xi\cdot\bA)
	\end{equation}
	The formula \eqref{math-symplectic-hamiltonian} gives
	\begin{eqnarray}
	\slashed{\delta}Q_\xi=\Omega(\bA,\delta_\xi\bA,\delta\bA)=\int_\Sigma\big(\delta_\xi\bA\wedge\star\delta \bF-\delta\bA\wedge\star\delta_\xi\bF\big)\,.
	\end{eqnarray}
	Straightforward algebra leads to the following formula for the isometry charges
	\begin{align}\label{Naive Killing}
	\slashed{\delta}Q_\xi=\delta\int_\Sigma\Big((\xi\cdot\bF)\wedge\star\bF-\frac{1}{2}\xi\cdot(\bF\wedge\star\bF)\Big)+&\delta\oint(\xi\cdot\bA)\wedge\star\bF\nn\\
	&\quad\quad-\oint\xi\cdot(\delta\bA\wedge\star\bF)
	\end{align}
	
	The first integral is the Noether current $j^\mu=\xi_\nu T^{\mu\nu}$ integrated on a Cauchy surface, as could have been obtained by the standard Noether's method. The second term has the same structure of a surface charge \eqref{Naive charge} for a gauge transformation generated by parameter $\bLaa=\xi\cdot\bA$, as in the second term of \eqref{Lie A}\footnote{Note that a field-dependent transformation in general lacks integrable charge as it is the case for $\bLaa=\xi\cdot\bA$.}. The last term is not a total variation and shows that the charge is in general not integrable. Moreover, even the infinitesimal charge variation is not in general conserved, since the flux on the boundary can be non-vanishing. The resolution of these subtleties requires explicit knowledge of the boundary conditions of the theory. $\hfill\square$
	\end{example}

	\subsection{Action principle, conservation and integrability}
	For a  physically interesting boundary condition, the action principle is well-defined and the surface charges of the theory are finite and generically non-vanishing.
	Given  an action functional $S[\Phi]$, and for  fixed values of the dynamical fields on initial and final spacelike surfaces, respectively $\Sigma_1$ and $\Sigma_2$,
	the classical trajectory $\Phi_{cl}$ is defined as the solution of
	\begin{equation}
	\frac{\delta S}{\delta \Phi}=0\,.
	\end{equation}	
	The functional derivative of the action is equal to Euler-Lagrange derivative of the Lagrangian \eqref{EULER} and gives the equations of motion, if the boundary term that appears after variation vanishes
	\begin{equation}\label{boundary term theta}
	\int _\Gamma\boldsymbol{\theta}(\Phi,\delta\Phi)=0\,.
	\end{equation}
	Depending on the fall-off behavior of the fields (as well as gauge-fixing conditions), \eqref{boundary term theta} may or may not hold, as is discussed in subsequent chapters. This criterion is closely related to the conservation of the symplectic form, as the  quantity \eqref{symp}  is not necessarily conserved for general boundary conditions. The difference of symplectic forms integrated on two spacelike surfaces $\Sigma_1$ and $\Sigma_2$ is 
	\begin{equation}\label{Omega conservation}
	\Omega_2-\Omega_1=\int \di\boldsymbol{\omega}+\int_\Gamma\boldsymbol{\omega}\,.
	\end{equation}
	In particular, non-conservation of the infinitesimal charge variation for transformation $\delta_\epsilon\Phi$ is found by
	\begin{equation}
	\slashed{\delta}Q_\epsilon\big|_{t_1}^{t_2}=\int_{t_1}^{t_2} \di\mathbf{k}_\epsilon(\Phi,\delta\Phi)=\int_{t_1}^{t_2} \boldsymbol{\omega}(\Phi,\delta\Phi,\delta_\epsilon\Phi)\,.
	\end{equation}
	
	Yet another potential problem is integrability of surface charges, especially those of diffeomorphisms. For a generally covariant theory, the infinitesimal charge variation for a diffeomorphism is given by \cite{Iyer:1994ys}
	\begin{eqnarray}\label{xi charge}
	\slashed{\delta}Q_\xi=\int_\Sigma\delta J_\xi+\oint\xi\cdot\boldsymbol{\theta}(\Phi,\delta\Phi)\,.
	\end{eqnarray}
	
	If the pull-back of $\boldsymbol{\theta}$ on timelike boundary vanishes, then the action principle is well-defined, the symplectic form is conserved and the charges are integrable.

	Integrability of the surface charges is non-trivial in physical systems. For example, in four-dimensional asymptotically flat gravity, the charges are neither integrable nor conserved and not even finite on the entire solution space  \cite{Barnich:2011mi,Barnich:2010eb,Barnich:2009se}. In any case, non-conservation can be interpreted as the flux through boundary \cite{Compere:2019gft}.
	
	\begin{example} Consider an electrostatic system on $AdS_4$ with boundary conditions
	\begin{eqnarray}\label{Ads4}
	\A_t=\frac{f(x^i)}{r}+\ord{r^{-2}},\qquad \A_r=0,\qquad \A_i=\partial_i\phi(x^j)
	\end{eqnarray}
	The subleading terms in $\A_t$ are determined by field equations. One sees that the boundary term of action vanishes since
	\begin{eqnarray}
	\boldsymbol{\theta}_{tij}(\A,\delta \A)=\frac{f(x^i)\delta f(x^i)}{r} \overset{r\to\infty}{\longrightarrow} 0\,.
	\end{eqnarray}
	The symplectic form is therefore conserved and the charges \eqref{Naive charge} for residual gauge transformations $\A_i\to\A_i+\partial_i\laa(x^j)$ are given by
	\begin{eqnarray}\label{soft ads4}
	Q_\laa=\oint\sqrt{q}d^2\hat{x} \laa(x^i)f(x^i)\,.
	\end{eqnarray}
	
	This phase space enjoys infinitely many conserved and finite soft charges. Moreover, the boundary condition \eqref{Ads4} is preserved by rotations and time translation. The four conserved charges of these global symmetries are integrable cause the last term in \eqref{xi charge} vanishes asymptotically. Therefore, energy and angular momenta given by \eqref{Naive Killing}  are  well-defined too.  Energy becomes the usual integral of the energy-momentum tensor
	\begin{equation}
	E=\int_\Sigma\sqrt{-g}{T_t}^t
	\end{equation}
	and the integral converges at large $r$. On the contrary, the angular momenta have no bulk contribution since $T_{ti}=0$ identically in this phase space. However, there is a boundary contribution to angular momenta from the second term in \eqref{Naive Killing}
	\begin{equation}
	Q_\xi=\oint \sqrt{q}d^2x\xi^i\partial_i\phi(x^j)f(x^j)\,.
	\end{equation}
	Notice that the angular momentum of the system is non-vanishing only if the pure-gauge mode $\phi(i)$ is present. In addition, the angular momentum generically changes by large gauge transformations $\phi\to\phi+\laa$. This can be seen by the fact that soft charges \eqref{soft ads4} are representations of the rotation group $SO(3)$:
	\begin{equation}
	\{Q_\laa,Q_\xi\}=\oint \sqrt{q}d^2x\xi^i\partial_i\laa(x^j)f(x^j)=Q_{\tilde{\laa}},\qquad\qquad\tilde{\laa}=\mathcal{L}_\xi\laa
	\end{equation}
	
	This example exhibits a simple realization  of an infinite-dimensional (Abelian) asymptotic symmetry algebra with conserved charges, for which energy and angular momenta are integrable. Notice the fact that soft charges commute with the energy functional—as the word \emph{soft} suggests. $\hfill\square$
	\end{example}

	\paragraph{$\mathbf{Y}$-ambiguity.}The definition  \eqref{LET} of $\boldsymbol{\theta}$ is ambiguous by arbitrary $(d-2)$-forms
	\begin{equation}\boldsymbol{\theta}(\Phi;\delta\Phi)\sim\boldsymbol{\theta}(\Phi;\delta\Phi)+\di\mathbf{Y}(\Phi;\delta\Phi)\,.
	\end{equation}	
	In consequence, the symplectic form is defined up to a surface term
	\begin{eqnarray}
	\Omega\sim\Omega+\oint\delta\mathbf{Y}\,.
	\end{eqnarray}
	The choice of $\mathbf{Y}$ is important in the conservation of the symplectic form. For instance, suppose that the boundary term in \eqref{Omega conservation} is not strictly vanishing, but its pull-back on the boundary is exact
	
	\begin{equation}\label{c cons}
	\boldsymbol{\omega}\big|_{\text{boundary}}= \di \textbf{c}\,.
	\end{equation}
	By choosing $\mathbf{Y}=-\mathbf{c}$, we obtain a conserved symplectic form
	\begin{eqnarray}
	\Omega=\int_\Sigma\boldsymbol{\omega}-\oint\mathbf{c}\,.
	\end{eqnarray}
	
	This freedom allows wider classes of boundary conditions to have conserved charges, as long as \eqref{c cons} is fulfilled by some boundary equations of motion.
	
	We move on to review the memory effect in the electromagnetic case.

	\section{Memory effect} \label{memory section}
	
\begin{figure}[t]
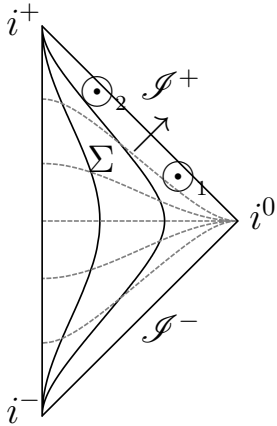

	
	\centering 
	
	\begin{overpic}[width=0.2\textwidth,tics=1]{flat-penrose0.pdf}
	
	\put(23,69) {\large$\nearrow$}
	
	\put(12,62)  {\large$\Sigma$}
	
	\put(10,79)  {\Large$\odot_{\text{\scriptsize2}}$}
	
	\put(30,58){\Large$\odot_{\text{\scriptsize1}}$}
	
	\put (25,20) {\large$\mathscr{I}^-$}
	
	\put (25,80) {\large$\mathscr{I}^+$}
	
	\put (52,48) {\large$i^0$}
	
	\put (-8,0) {\large$\displaystyle i^-$}
	
	\put (-8,96) {\large$\displaystyle i^+$}
	
	
	\end{overpic}
	
	\caption{Observation of memory effect due to the out-going radiation by massive probes. The $\odot_{1,2}$ symbols show detection points, located at constant $r$. One could set the detection times earlier to observe the in-going memory effect. The dashed lines depict some Cauchy surfaces and as we see they all intersect at spatial infinity $i_0$. In the figure, we have implicitly imposed the antipodal matching through $i^0=\mathscr{I}^+_-=\mathscr{I}^-_+$.}
	
	\label{memory}
	
	\end{figure}

	In four dimensional QED, the electromagnetic field $\A_{\mu}(x)$ has mode expansion in terms of photon creation and annihilation operators of photons. The operators, say $a_{\pm}(\omega\hat{q})$, are labeled by  helicity $\pm$, frequency $\omega$ and three-momentum direction $\hat{q}$.  In retarded Bondi coordinates, where the Minkowski metric is
	
	\begin{eqnarray}\label{Bondi ret}
	ds^2=-du^2-2dudr+4r^2\frac{dzd\bar{z}}{(1+z\bar{ z})^2}\,.
	\end{eqnarray}
	By a large-$r$ saddle-point approximation, one can show that field at the future null infinity $A_z(u,z,\bar{ z})\equiv\lim_{r\to\infty}A_z(u,r,z,\bar{ z})$ has the same mode expansion in terms  of the aforementioned operators, so it is the radiation field at infinity
	\begin{equation}
	A_z(u,z,\bar{ z})=\frac{i}{1+z\bar{ z}}\int_0^\infty d\omega\big[a_+^{\text{out}}(\omega\hat{x})e^{-i\omega u}+a_-^{\text{out}}(\omega\hat{x})e^{i\omega u}\big]\,.
	\end{equation}
	The \emph{soft photon mode} is defined as the zero-frequency limit of electromagnetic field at null infinity
	\begin{equation}
	\lim_{\omega\to0}\int du e^{i\omega u}F_{uz}=A_z(z,\bar{z})\big|_{u=-\infty}^{u=+\infty}\equiv\partial_z\alpha
	\end{equation}
	where in the last relation, the   absence of long-range magnetic fields has been assumed. This presumption entails  that the difference of $A_z$ between early and late times is a gauge transformation. This pure gauge difference between the  two ends of the future null infinity while given by the soft factor of the corresponding scattering, can be related to soft charges as follows. Using  equation of motion at the leading order in the absence of massless sources we have
	\begin{eqnarray}
	\partial_u (A^{(1)}_u-\mathcal{D}^zA_z^{(0)} )=0
	\end{eqnarray}
	and   the charge difference is given by
	\begin{align}\label{softchargememory}
	Q_\laa\big|_{u=-\infty}^{u=+\infty}=\int_{\mathscr{I}^+} \sqrt{|g|}\laa\,\partial_u \F^{ur}&=\oint\sqrt{|g|}\laa\mathcal{D}^z A^{(0)}_z\big|_{u=-\infty}^{u=+\infty}\nonumber\\
	&=\oint\sqrt{|g|}\laa\mathcal{D}^z\mathcal{D}_z\alpha\,.
	\end{align}
	Note that the zero-mode charge is invariant in this process (see figure \ref{memory} \cite{Afshar:2018sbq}).

	Our next task is to discuss the electromagnetic memory effect originating from soft photons.
	First, we provide a definition of a memory effect.
	A classical mechanical system can be described by some phase-space coordinates $(q_n,p_n)$ and a Hamiltonian $H$. Suppose that the system interacts with an external gauge field $\A$ through a gauge interaction. The equations of motion are gauge-invariant so that the phase space and the Hamiltonian are invariant under gauge transformations. This is true for all observable quantities $\mathcal{O}[p,q;\A]$ and the statement of gauge invariance takes the following form
	
	\begin{equation}
	\mathcal{O}[p,q;\A]=\mathcal{O}[p,q;\tilde{\A}]
	\end{equation}
	where the external gauge fields $\A$ and $\tilde{\A}$ are related by some gauge transformation. Since gauge transformations of the external field leave all observables invariant, one expects them to corresponds to canonical transformations of the system. The experimental setup for tracking the memory consists of an apparatus located far away from a radiating system. It is a reasonable assumption that the radiation field depends only on the retarded time $\A=\A(u)$ i.e. the apparatus is smaller than spatial fluctuations of the radiation field. We now state the following proposition to be proved in specific examples we discuss.

	\begin{pro}\label{canonic}
	Let a classical system $(p_n,q_n;H)$ interact through an external gauge potential $\A$ with time-dependent field strength $\F(u)$ which is  vanishing at early and late times:
	\eq{	
		\lim_{u\to\pm\infty}\F(u)=0\,,	
	}{}
	which means that the system evolves with free Hamiltonian in the limit $u\to\pm\infty$.\footnote{This is typically the case when $F(u)$ is a radiative field, being strong only in a finite duration.} Defining the pure gauge configurations,	
	\begin{align}
	\A^+\equiv \A(u=+\infty),\qquad
	\A^-\equiv\A(u=-\infty),
	\end{align}	
	the system at very early and late times is governed by free Hamiltonians $H[\A^+]$ and $H[\A^-]$. These two Hamiltonians are related by a canonical transformation $U[q,p;\A^+,\A^-]$.	
	\end{pro}

	As a result, the phase space and the Hamiltonian at $u\to\pm\infty$ are related by a canonical transformation $U[q,p;\A_+,\A_-]$ and free solutions are mapped to free solutions by the same operator.

	In the above setting, we propose the following definition for the memory effect:
	
	\begin{defi}[Memory effect]
	Let $\mathcal{O}[q,p;\A]$ be an observable of the system described above, for which the following quantities are well-defined:
	\begin{align}
	\mathcal{O}^+=\lim_{u\to+\infty}\mathcal{O}[q,p;\A],\qquad 
	\mathcal{O}^-=\lim_{u\to-\infty}\mathcal{O}[q,p;\A].
	\end{align}
	If $\mathcal{O}^+$ and $\mathcal{O}^-$ are {not related} to each other by the canonical transformation defined in proposition \ref{canonic}, we say that gauge field $\A$ has induced a memory effect on the system.
	\end{defi}

	\subsection{Electromagnetic kick memory}
	One may analyze the problem in thhe action or the Hamiltonian formulations. Let us start with the action and equations of motion:
	\begin{equation}
	S=\int d\tau \,\eta\left(\frac{1}{2}\eta^{-2}\partial_\tau X^\mu\partial_\tau X^\nu g_{\mu\nu}+q\,\eta^{-1}\partial_\tau X^\mu\A_\mu-\frac{1}{2}m^2
	\right),
	\end{equation}
	where $\eta(\tau)$ is the einbein of the metric on the worldline $ds^2=\eta^2(\tau)d\tau^2$. We take $X^\mu=(u,r,X^i)$ as Minkowski coordinates  defined in \eqref{Bondi ret}. This action is gauge invariant up to a boundary term and we can fix the radial gauge $\A_r=0$. To fix the reparametrization freedom we use the light-cone gauge which sets $u$ as the clock, i.e. $u=\tau$. The action is then
	
	\begin{equation}\label{worldline}
	S=\int d\tau \left(\frac{1}{2\eta}(-1-2\dot{r}+\dot{X}^i\dot{X}^j\delta_{ij}) +q\left(\dot{X}^i\A_i+\A_u\right)-\frac{1}{2}\eta\, m^2
	\right).
	\end{equation}
	
	
	Assuming that the gauge field components $\A$  have `very mild' $r$ dependence (according to fall-off behavior given below), the equation of motion for $r$ yields $\eta\approx const$ and we must choose this constant equal to $1/m$ to be consistent with the non-relativistic limit, taken below\footnote{\label{constraint-footnote}Variation of the action \eqref{worldline} with respect to $\eta$ gives $\dot{X}^\mu\dot{X}_\mu+\eta^2m^2=0$ as a constraint.}.
	Equations of motion for $X^i$ are
	\begin{align}\label{EMeom}
	\ddot {X}^i=\frac{q}{m} (\cF^i_{\ j} \dot X^j+\cF^i_{\ u})\,. 
	\end{align}
	We now focus on cases where the particle is coupled to the radiation  field through $\A_i$ with  the  boundary conditions at large $r$ \cite{He:2014cra},
	\begin{align}\label{A-falloff}
	\A_i\sim\ord{1/r}\,,\qquad \A_u\sim\ord{1/r}\,,\qquad \A_r=0\,.
	\end{align}
	
	The field strength component $\F_{ui}$ is at order $r^{-1}$, while the other components of the field strength $\F_{ij}$ and $\F_{ur}$ are subleading. 
	By integrating \eqref{EMeom} along the retarded time $u$, we are left with an electric \emph{kick memory effect},
	\begin{align}\label{kickmemory}
	\Delta \dot{X}^i=\dot{X}^i(u=\infty)-&\dot{X}^i(u=-\infty)\simeq \frac{q}{m}\int_{-\infty}^{\infty}\cF^i{}_u\extd u\nn\\
	\qquad&=-\frac{q}{m r}\left(A_i(u\to\infty)-A_i(u\to-\infty)\right)\,,
	\end{align}
	in which $A$ is the leading term in the asymptotic falloff \eqref{A-falloff} and terms of $\ord{r^{-2}}$ are ignored.

	In order to apply and use the results of the proposition \ref{canonic}, we present these results in the Hamiltonian formulation. 
	From \eqref{worldline} the canonical momentum $\tilde{P}_i$ and $\tilde{P}_r$ conjugate to the position coordinates $X^i$ and $r$ are,
	\begin{equation}
	\tilde{P}_i=\frac{1}{\eta}\dot{X}_i+q\,\A_i\,,\qquad \tilde{P}_r=-\frac{1}{\eta}\,.
	\end{equation}
	The Hamiltonian is
	\eq{	
		H=-\frac{m^2+\tilde{P}_r^2+(\tilde{P}_i-q\,\A_i)^2}{2\tilde{P}_r}-q\,\A_u\,.	
	}{}
	The constraint equation (\emph{c.f.} footnote \ref{constraint-footnote}), $\eta^2 m^2-1-2\dot{r}-(\tilde{P}_i-q\A_i)^2=0$ associated with  the reparametrization invariance of the probe action can be solved for $\eta$. Thus, for a  non-relativistic particle, deviation of $\eta$ from $1/m$ is small, suggestive of  defining a shifted radial momentum $P_r\equiv \tilde{P}_r+m\ll1$.  Finally, the Hamiltonian for a non-relativistic charged particle of rest mass $m$ moving in  the  light-cone gauge $\tau=u$ for the particle and in the radial gauge $\A_r=0$ for the  gauge field, takes the familiar form
	
	\eq{	
		H=m+\frac{P^2_r+(\tilde{P}_i-q\A_i)^2}{2m}-q\,\A_u\,,\qquad i=1,2\,.
	}{classical hamilton}
	
	We drop the rest mass constant term $m$ from now on.
	Clearly, the Hamiltonian \eqref{classical hamilton} takes different forms at $u\to\pm\infty$.  
	Recalling the falloff behavior of the gauge field at large $r$ \eqref{A-falloff}, the particle is `free' at order $r^{-1}$ and one expects early and late Hamiltonians to be related by a canonical transformation. To show this,
	consider new dynamical variables $(P,X)$ related to old variables $(\tilde{P},X)$ by the canonical transformation on momenta
	
	\begin{equation}\label{free-elect-canonic}
	P_r=\tilde{P}_r+m\,,\qquad P_i=\tilde{P}_i-q\,\A_i\,,
	\end{equation}
	while the coordinates remain unaltered. The new Hamiltonian $K$ is
	\begin{equation}
	K=\frac{P_r^2+P_i^2}{2m}+q\,X^i\partial_u\A_i\,,\qquad i=1,2\,.
	\end{equation}
	
	To verify this statement, we note that new and old variables in the large $r$ limit are related as
	\begin{equation}\label{genrel}
	\dot{X}^a\tilde{P}_a-H=\dot{X}^aP_a-K+\frac{d{G}}{du}\,,\qquad a=1,2,3,
	\end{equation}
	and the generating function\footnote{The function $G$ as appears in \eqref{genrel} does not generate the transformation. Different kinds of generating functions are obtained only by adding certain combinations of old and new coordinates and momenta (like $XP$) to $G$.} for the transformation is 
	\begin{equation}G= qx^i \A_i+q\int_{u_0}^u\A_u(u^\p)du^\p.
	\end{equation}
	The integral term in $G$ does not appear in transformation relations \eqref{free-elect-canonic}; it becomes subleading in $r$ when Cartesian derivatives are performed.

	The new Hamiltonian $K$ has the same form at $u\to\pm\infty$ since the electromagnetic radiation $\F_{ui}=\partial_u\A_i+\ord{r^{-2}}$ by assumption vanishes at both temporal limits. We are now ready to find an observable in the new coordinate system which has different asymptotic values at early and late times. The   equations of motion in the new basis are
	\eq{	
		\dot{X}^a=[X^a,K]=\frac{P_a}{m}\,,\qquad \dot{P}_i=[P_i,K]=-q\,\dot{\A}_i\,,\qquad\dot{P}_r=0\,.
	}{}
	Integrating on the whole time interval of the process gives:

	\begin{pro}[Memory effect]
	Electromagnetic  Memory effect on a free charged non-relativistic particle is the difference between late and early transverse momenta:
	\begin{equation}\label{kickmemory2}
	\Delta P_i=-q\,\Delta \A_i\,,
	\end{equation}
	where for a generic variable $V$, $\Delta V$ measures the difference between the late and early values, 
	\eq{\Delta V\equiv V(u=+\infty)-V(u=-\infty)\,.}{Deltadef} 
	\end{pro}
	Eq. \eqref{kickmemory2} is the analogue of \eqref{kickmemory} in the Hamiltonian formulation. 
	The above example provides a smooth transition to a quantum treatment of the memory effect. The interested reader may consult Ref. \cite{Afshar:2018sbq}.
	
	After this short review to set the terminology, we 	start focusing more on Maxwell fields in the next chapter.

	\chapter{Soft Charges in Maxwell Theory}\label{max chapter}
	In this chapter, we analyze the asymptotic symmetries of Maxwell theory in three and higher dimensions. 
	The chapter is divided into two parts with similar structure: Minkowski background (section \ref{Minkowski Maxwell}) and anti-de Sitter background (section \ref{Maxwell AdS}). In either case, physical boundary conditions are motivated and the soft charges are computed.

	\section{Minkowski background}\label{Minkowski Maxwell}

\begin{figure}[t]
	\centering
	\begin{overpic}[width=0.2\textwidth,tics=1]{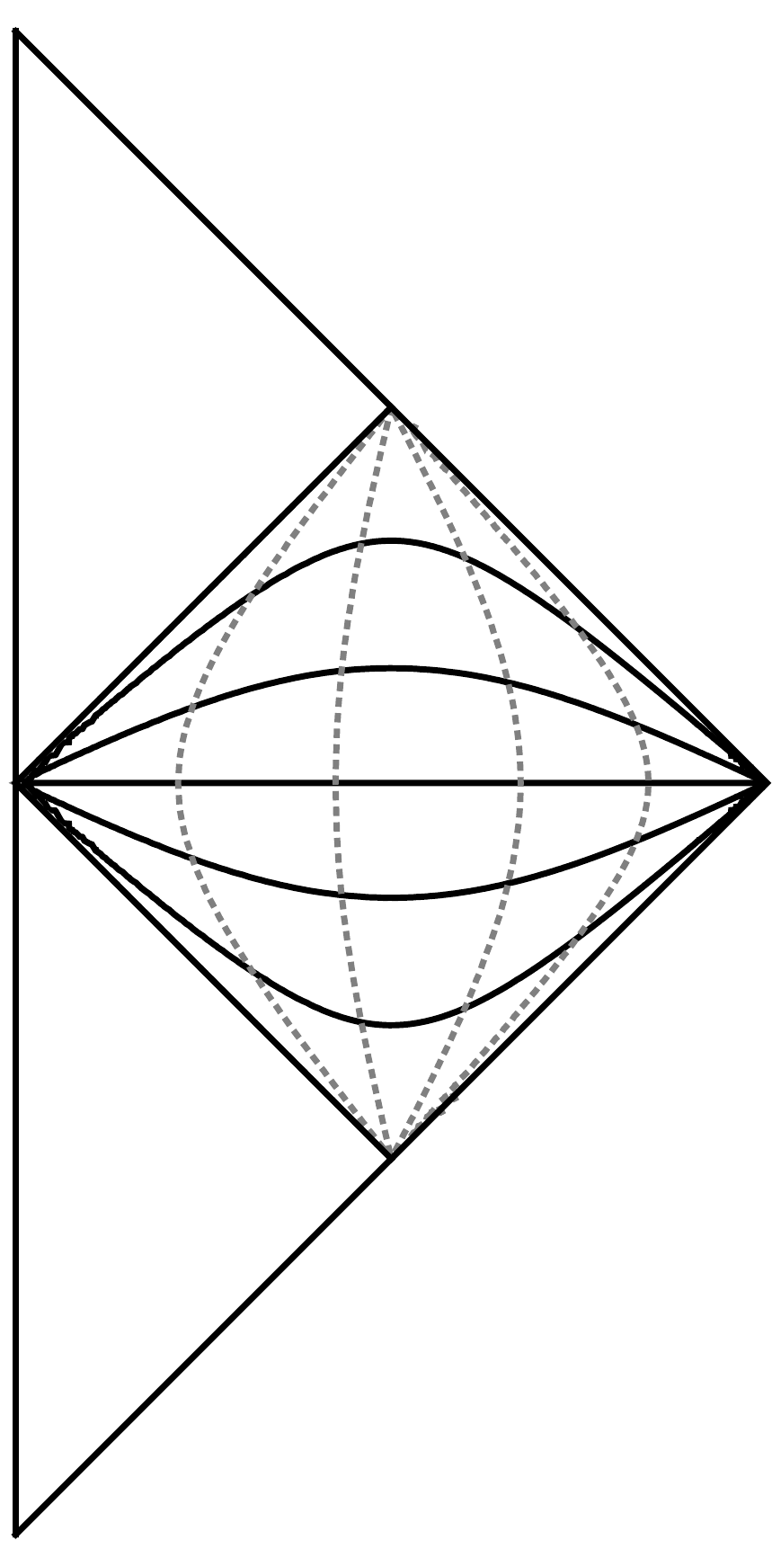}
	\put (10,30) {$\mathscr{L}^-$}
	\put (10,65) {$\mathscr{L}^+$}
	\put (25,75) {$\tau=0$}
	\put (25,20) {$\tau=\pi$}
	\put (52,48) {\large$i^0$}
	\put (-8,0) {\large$\displaystyle i^-$}
	\put (-8,96) {\large$\displaystyle i^+$}
	\end{overpic}
	\caption{Penrose diagrams of Minkowski flat spacetime ${\cal M}_d$ \cite{Esmaeili:2019hom}.  The Rindler patch covers the events outside the light cone. The solid lines are constant $\tau$ slices, while dotted lines are constant $\rho$ hyperboloids. }
	\label{Rind fig}
	\end{figure}
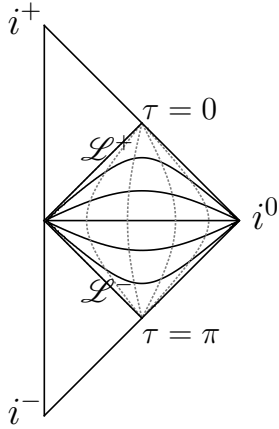

	\paragraph{{Hyperbolic coordinates.}}
	Given an arbitrary point $\mathcal{O}$ in Minkowski space, one can define null coordinates $u=t-r$ and $v=t+r$. The future light cone $\mathscr{L}^+$ of $\mathcal{O}$ is the $u=0$ hypersurface, while the past light cone $\mathscr{L}^-$ is at $v=0$. $\mathscr{L}^+$ and $\mathscr{L}^-$ intersect at the origin $\mathcal{O}$. We call the set of points with space-like distance to $\mathcal{O}$, the \emph{Rindler patch} and denote it by Rind$_{d-1}$ (see figure \ref{Rind fig}). The Rindler patch is conveniently covered by coordinates $(\rho,\tau,x^i),\, i=1,\cdots ,d-2$, such that
	\begin{equation}\label{Hypercoo}
	\left\{\begin{array}{c}
	\rho^2=r^2-t^2\\
	\cos \tau=t/r
	\end{array}\right.
	\qquad\qquad
	\left\{\begin{array}{c}
	t=\rho\cot \tau\\
	r=\rho/\sin \tau
	\end{array}\right.,\qquad 0\leq \tau\leq \pi\,.
	\end{equation}
	and Minkowski metric is
	\begin{equation}
	ds^2=\di \rho^2+\rho^2h_{ab}dx^adx^b=\di \rho^2+\frac{\rho^2}{\sin^2\tau}\left(-\di \tau^2+ q_{ij}\di x^i\di x^j\right)\,.
	\end{equation}
	
	The boundaries of this patch are reached at as follows. \emph{Future light cone} $\mathscr{L}^+$ is at $(\rho=0, \tau=0)$ and \emph{past  light cone} $\mathscr{L}^-$ is at$(\rho=0, \tau=\pi)$.
	For instance, a point at radius $r$ on $\mathscr{L}^+$ is reached   by taking the limit $\tau\to0$ such that $\rho=r\tau$.
	The	\emph{Spatial infinity} $i^0$ defined as the destination of spacelike geodesics is at $(\rho\to\infty, 0<\tau<\pi)$, shown as the intersection of future and past null infinities on the Penrose diagram. The limit $(\rho\to\infty,\tau\to0,\pi)$ covers the portion of the null infinity outside the light cone. The point at retarded time $u$ on future null infinity is reached by taking the limit $\rho\to\infty$ such that $u=-\rho \tau/2$.

	\paragraph{Field equations.} Maxwell's equations 
	\begin{equation}
	\nabla_{[\alpha} \F_{\mu\nu]}=0\qquad\qquad \nabla^\mu\F_{\mu\nu}=0
	\end{equation}
	take the following form in hyperbolic coordinates $(\rho,x^a)$
	\begin{subequations}\label{subs equation}
	\begin{align}
	&  D_a\F^{a\rho}=0\label{eq rho}\\
	&  \rho^{1-d}\partial_\rho(\rho^{d-1}\F^{\rho a})+D_b\F^{ba}=0\label{eq a}\\
	&  \partial_\rho\F_{ab}+2\partial_{[a}\F_{b]\rho}=0\label{Bianchi rho}\\
	& \partial_{[a}\F_{bc]}=0 .\label{Bianchi a}
	\end{align}
	\end{subequations}
	
	Note that $\F_{a\rho}$ and $\F_{ab}$ are different representations of de Sitter isometry group $O(d-1,1)$.

	\subsection{Boundary conditions}\label{boundary}

	The electromagnetic field of a static electric source at $r=0$ is $
	\F_{tr}=r^{2-d}\nn$. 
	In hyperbolic coordinates we have
	\begin{equation}
	\F_{\tau\rho}=-\frac{\rho}{\sin \tau}\F_{tr}=-\rho^{3-d}\sin^{d-3}\tau.
	\end{equation}
	Applying a boost (which belongs to the isometry group of the hyperboloid) will turn on other de Sitter components of $\F_{a\rho}$ with the same fall-off; so one generally has $\F_{a\rho}\propto\rho^{3-d}$. Therefore, we propose the following boundary conditions for $d$-dimensional theory
	\begin{equation}
	\F_{\rho a}\sim\ord{\rho^{3-d}}\,,\qquad \F_{ab}\sim \ord{\rho^{3-d}}\,.\label{maxwell b.c.}
	\end{equation}
	The fall-off for $\F_{ab}$ is implied by Poincar\'e invariance. 
	
	\paragraph{{Solution space}}
	
	\begin{enumerate}
	\item {Electric monopoles:}
	If we set $\F_{ab}=0$, by \eqref{eq a}  we have
	$ \F^{a\rho}\propto\rho^{1-d}.$
	Furthermore, by \eqref{Bianchi rho} and \eqref{eq rho},
	\begin{equation}
	\F_{a\rho}=\rho^{3-d}\partial_a\psi(x^b)\,,\qquad   D^aD_a\psi=0 \label{indp electric}
	\end{equation}
	In this case, the solution consists of a scalar degree of freedom $\psi$ which satisfies the equation
	\begin{equation}
	D^aD_a \psi(x^b)=0 \label{maxwel de sitter eom}
	\end{equation}
	and the solution is obtained by spectral decomposition of Laplace operator on $S^{d-2}$, being $\mathcal{D}^2Y_\ell(\hat{x})=-\ell(\ell+d-3)Y_\ell(\hat{x})$. Then, \eqref{maxwel de sitter eom} will
	simplify to
	\begin{equation}
	(1-y^2)\psi^{\pp}_\ell(y)+(d-4)y\psi_\ell^\p(y)+\ell(\ell+d-3)\psi_\ell(y)=0\qquad y=\cos\tau\,.
	\end{equation}
	The general solution is
	\begin{equation}\label{solution}
	\psi(y,\hat{x})=(1-y^2)^{\frac{d-2}{4}}\sum_{\ell=1}Y_\ell(\hat{x})\left(a_\ell P_{(2l+d-4)/2}^{(d-2)/2}(y)+b_\ell Q_{(2l+d-4)/2}^{(d-2)/2}(y)\right)\,,
	\end{equation}
	where $P_l^m$ and $Q_l^m$ are associated Legendre functions of the first and second kind respectively. For $\ell=0$, the solutions are
	\begin{equation}\label{zero sol}
	a_0+b_0 y\,\,{}_2F_1(\frac{1}{2},\frac{4-d}{2},\frac{3}{2},y^2)\,.
	\end{equation}
	\item  {Non-compact magnetic branes:}
	Switching $\F_{a\rho}$ off, fixes the $\rho$-dependence by \eqref{Bianchi rho} to $ \F_{ab}\propto\rho^0$ and the field equation \eqref{eq a} reduces to
	\begin{equation}
	D^a\F_{ab}=0 \label{indp magnet}\,.
	\end{equation}
	The set of solutions \eqref{indp magnet} covers magnetic charges  moving freely in space and crossing the origin at $t=0$. In dimensions larger than 4,  the magnetic monopoles are replaced by  magnetic branes  extended in all directions. This solutions are necessarily singular as the sources hit the celestial sphere. For simplicity, we are considering boundary conditions which exclude magnetic charges altogether.
	
	\item {Electric multipoles and compact magnetic branes:} Any other solution involves both $\F_{ab}$ and $\F_{a\rho}$. The solutions with power-law fall-off in $\rho$ correspond to multipoles of electric and magnetic branes. Electric monopoles generate the independent solution \eqref{indp electric} for $\F_{a\rho}$, while magnetic mono-poles(-branes) generate the independent solution \eqref{indp magnet}  for$\F_{ab}$. To arrange multipole configurations, one acts on monopole solutions with translation generators, which leads to fields of lower fall-off with both $\F_{a\rho}$ and $\F_{ab}$ components. In the radial gauge $\A_\rho=0$,
	our prescribed boundary condition implies
	\begin{align}\label{asymp expand}
	\A_a=\frac{\rho^{4-d}}{4-d}\partial_a\psi+\rho^{4-d}\sum_{n=1}A_a^{(n)}\rho^{-n}\,
	\end{align}
	and the field equations become second order equations on de Sitter space:
	\begin{subequations}
	\begin{align}
	& D^bF_{ba}^{(n)}+n(n+d-4)A_{a}^{(n)}=0\\
	&  D^aA^{(n)}_{a}=0\\
	&   D^aD_a\psi=0\label{difff}
	\end{align}
	\end{subequations}
	If $d=4$, the leading term in \eqref{asymp expand} will be $\partial_a\psi\log \rho$.
	
	\end{enumerate}

	
	
	\subsection{Soft Charges}
	The asymptotic expansion \eqref{asymp expand} in the radial gauge contains gauge-invariant components. It is possible to add pure-gauge fluctuations independent of $\rho$, as
	\begin{align}
	\A_a=\partial_a\alpha+\frac{\rho^{4-d}}{4-d}\partial_a\psi+\rho^{4-d}\sum_{n=1}A_a^{(n)}\rho^{-n}\,.
	\end{align}
	Under residual gauge transformations, the boundary field shifts as
	\begin{eqnarray}\label{res alp}
	\alpha\to\alpha+\lambda,
	\end{eqnarray}
	while the subleading terms are gauge invariant. Introducing  pure-gauge variables on the boundary is interesting since one takes into account residual gauge freedom of the theory and indeed the asymptotic symmetry group becomes non-trivial. This comes with the price of a symplectic flux and a finite boundary term for the action. As we will see, the both problems are resolved by a proper restriction of $\alpha$.

	The symplectic flux is
	\begin{eqnarray}
	\int_\Gamma d^{d-1}x\sqrt{|g|}\omega^\rho=\int_\Gamma d^{d-1}x\sqrt{|g|}\delta \A_a\delta \F^{\rho a}=
	\int_\Gamma d^{d-1}x\sqrt{|g|}\delta \partial_a\alpha\delta \F^{\rho a}
	\end{eqnarray}
	

	For the specific boundary conditions \eqref{asymp expand}, we have $\F_{\rho a}=\partial_a\psi$. In consequence, by imposing   a gauge condition on $\alpha$ such as
	\begin{equation}\label{alpha gauge}
	D_aD^a\alpha=0
	\end{equation}
	we have
	\begin{eqnarray}
	\int_\Gamma d^{d-1}x\sqrt{|h|}\delta \partial_a\alpha\delta \partial^a\psi=
	\int_\Gamma d^{d-1}x\sqrt{|h|} D_a\big(\delta D^a\alpha\delta \psi\big)\,.
	\end{eqnarray}
	We can subtract this boundary term from the symplectic form
	\begin{eqnarray}\label{symp2}
	\Omega=\int \delta\A\wedge\star\F-\oint\sqrt{h} n_aD^a\delta\alpha\delta \psi
	\end{eqnarray}
	This is a conserved symplectic form for the theory.

	\begin{pro}
	The symplectic form \eqref{symp2} has non vanishing charges for residual transformations \eqref{res alp} subject to gauge condition \eqref{alpha gauge}
	\begin{equation}\label{charge maxwell 4}
	Q_\laa=\oint d^{d-2}x\sqrt{|h|}\left(\laa \F^{ \tau\rho}-\partial^{\tau}\laa\psi\right)
	\end{equation}
	\end{pro}
	
	To see the conservation explicitly, one may take a time-derivative directly. In addition, the $\tau$-dependence of the charges is given by the Wronskian of the differential equation \eqref{difff} which is a constant. So the charge is independent of $\tau$.
	
	\paragraph{{Physical significance of the charges}}
	Denote the set of solutions for electric monopoles given in \eqref{indp electric} by $\mathcal{E}$. This space covers moving electric charges in space, which are passing the origin simultaneously at  $t=0$, hence their worldlines cross $\mathcal{O}$. The field strength is $\F_{\rho a}\propto \rho^{3-d}$ with no subleading terms. For an arbitrary configuration of freely moving charges, the leading component of the asymptotic field is an element of $\mathcal{E}$, but subleading terms are generally present. In other words, the definition of $\mathcal{E}$ is Lorentz invariant, but not Poincar\'e  invariant. 
	\emph{$\mathcal{E}$ encodes the information of charge values $q_n$ and their velocities $\vec{\beta}_n$.}
	The space $\mathcal{E}$ is isomorphic to the space of boost vectors $\vec{\beta}$, that is $\mathbb{R}^{d-1}$. The space of conserved electric charges we will construct is also isomorphic to $\mathbb{R}^{d-1}$; each point of this space with coordinate vector $\vec{\beta}$ is a conserved charge and gives the total electric charge in space, moving with that specific boost.


	\subsection{Killing Charges}
	
	In previous sections, we computed the conserved charges for large gauge transformations. We can also apply the same method to obtain the charges for
	the background $ISO(d-1,1)$ isometries.
	The Minkowski Killing vectors include Lorentz transformations $L_{\mu\nu}$  and Translations $P_\mu$. Denoting the Minkowski coordinates by $X^\mu$, we have $X^\mu X_\mu=\rho^2$ and we define
	\begin{equation}\label{XMhat}
	\bar{X}^\mu=\frac{X^\mu}{\rho},\qquad\qquad\bar{M}^{\mu}_a	 =\frac{\partial \bar{X}^\mu}{\partial x^a}\,.
	\end{equation}
	The isometries in hyperbolic coordinates are the following.
	\begin{align}
	\text{boosts and rotations}\, \qquad&\qquad L^{\mu\nu}= 2\bar{X}^{[\mu} \bar{M}^{\nu]}_a h^{ab}\partial_b\label{xi-L} \\
	\text{translations}\,\qquad&\qquad P^{\mu}= -\bar{X}^\mu\ \partial_\rho - \frac{1}{\rho} \bar{M}_a^\mu h^{ab}\partial_b\label{xi-T}\,.
	\end{align}
	
	For the translation generators, the surface terms in \eqref{Naive Killing} and \eqref{symp2} are subleading and the charge is given only by the bulk energy-momentum tensor.
	\begin{align}\label{Naive Killing2}
	Q_\xi=\int_\Sigma[(\xi\cdot\F)\wedge\star\F-\frac{1}{2}\xi\cdot(\F\wedge\star\F)]=\int_{\Sigma} \sqrt{\gamma}n_{\alpha}\xi_\beta T^{\beta\alpha}
	\end{align}
	In particular, 	the non-integrable term in \eqref{Naive Killing} vanishes at large $\rho$.	For the Lorentz generators, 
	\begin{equation}\label{final}
	Q_\xi= \int_{\Sigma} \sqrt{\gamma}n_{\alpha}\xi_\beta T^{\beta\alpha}+
	\oint_{\Sigma} \sqrt{h}n_a\Big[D^a(\xi^b\partial_b\phi)\psi-(\xi^b\partial_b\phi) D^a\psi\Big]
	\end{equation}
	This completes our derivation of conserved charges for Minkowski Killing vector fields. The surface terms in \eqref{final} are present only for Lorentz transformations.
	
	Notice that Lorentz transformations lack integrable canonical charges via the original symplectic form of the theory (see a Hamiltonian analysis in \cite{Henneaux:2018gfi}). The modified conserved symplectic form \eqref{symp2}, however, provides integrable charges equal to \eqref{final}.

	\subsection{Light cone regularity and antipodal identification}\label{antip sec}
	
	As far as the field equations are concerned, the whole set of solutions in \eqref{solution} with the two sets of coefficients $a_\ell$ and $b_\ell$ are admissible both for $\psi$ and $\laa$. The two solutions behave differently when de Sitter boundaries at $\tau=0,\pi$ are approached. Imposing the antipodal matching conditions amounts to eliminating half of the solutions in either of $\psi$ and $\laa$ in the right way. We proceed by explaining the four-dimensional theory first.

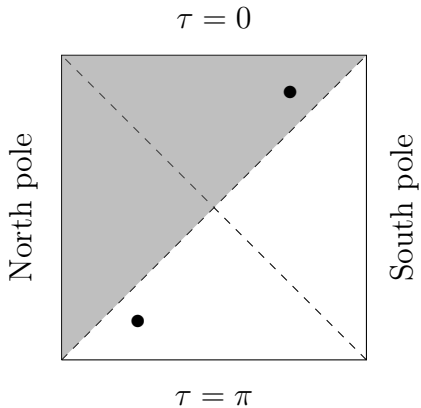
\begin{figure}
	\centering
	\begin{tikzpicture}
	\draw (0,0)--(0,4)--(4,4)--(4,0)--(0,0);
	\draw[dashed](0,0)--(4,4);
	\draw[dashed](4,0)--(0,4);
	\node [rotate=90] at (-.5,2) {North pole};
	\node [rotate=90] at (4.5,2) {South pole};
	\path[fill=gray,opacity=0.5](0,0)--(0,4)--(4,4)--(0,0);
	\node [] at (2,-.5) {$\tau=\pi$};
	\node [] at (2,4.5) {$\tau=0$};
	\node [] at (3,3.5){$\bullet$};
	\node [] at (1,.5){$\bullet$};
	\end{tikzpicture}
	\caption{In the hyperbolic coordinates \eqref{Hypercoo}, the constant $\rho$ surfaces are co-dimension-1 de Sitter spaces.  Penrose diagram of de Sitter space is shown above. The shaded region is the causal future of the north pole, and its dashed $45^{\circ}$ boundary is the future horizon of the south pole. The antipodal map amounts to  a couple of horizontal and vertical flips. Especially, the north pole at far past is mapped to the south pole at far future.}
	\label{dS-Penrose}
	\end{figure}

	\paragraph{Four dimensions.}
	In the general solution \eqref{solution}, let us drop $a_{lm}$ coefficients in $\laa$ and $b_{lm}$ coefficients in $\psi$
	\begin{align}\label{four dim antip}
	\psi(\tau,\hat{x})&=\sin\tau\sum_{l=1}a_{lm} Y_l^m(\hat{x})  P_{l}^{1}(\cos\tau)+a_{0}\cos\tau\,,\\
	\laa(\tau,\hat{x})&=\sin\tau\sum_{l=1}b_{lm} Y_l^m(\hat{x})  Q_{l}^{1}(\cos\tau)+b_{0}
	\end{align}
	To see how these functions behave under antipodal map, recall that
	\begin{equation}\label{parity transform}
	\begin{array}{l}
	P^1_l(-y)=(-1)^{l+1}P^1_l(y)\\
	Q_l^1(-y)=(-1)^{l}Q_l^1(y)
	\end{array}
	\qquad\qquad Y^m_l(-\hat{x})=(-1)^lY_l^m(\hat{x})\,.
	\end{equation} 
	We have (see figure \ref{dS-Penrose} \cite{Afshar:2018apx})
	\begin{equation}\label{4 antip}
	\laa(\pi-\tau,-\hat{x})=\laa(y,\hat{x}), \qquad \qquad  \psi(\pi-\tau,-\hat{x})=-\psi(\tau,\hat{x})\,.
	\end{equation}
	
	Since the soft charges \eqref{charge maxwell 4} are independent of integration surface, we can evaluate them in two limiting cases where the surface is pushed to future or past boundary of de Sitter space at $\tau=0,\pi$. Expanding the exact solutions around $\tau=0,$ we have
	\begin{equation}
	\psi(\tau,\hat{x})= \frac{1}{2}{\tau}^{2}\psi_2(\hat{x})+\ord{ \tau^{4}}\,,\qquad
	\laa(\tau,\hat{x})=\laa_0(\hat{x})+\ord{\tau^2\log\tau}\,,
	\end{equation}
	and the soft charge \eqref{charge maxwell 4} becomes
	\begin{equation}
	Q^+_\laa=\oint d^{2}\hat{x}\sqrt{q}\laa_0(\hat{x}) \psi_2(\hat{x})\,.
	\end{equation}
	In order to write this expression in terms of Bondi coordinates, notice that around $\tau=0$,
	\begin{equation}
	\F_{ur}=\frac{\sin\tau}{\rho}\F_{\rho\tau}=\frac{\sin\tau}{\rho^2}(\partial_\tau\psi+\ord{\rho^{-1}})=\frac{ {\psi}_2(\hat{x})}{r^2}+\ord{1/r^{3}},
	\end{equation}
	and the charge recovers its covariant expression indeed
	\begin{equation}\label{null charge max}
	Q_\laa=\oint d^{2}\hat{x}\sqrt{-g}\laa_0(\hat{x}) \F_{ur}(\hat{x})
	\end{equation}
	For the moment, let us denote expansion coefficients around $\tau=0$ and around $\tau=\pi$ by superscripts $+$ and $-$ respectively. Antipodal properties \eqref{4 antip} reveal expansion coefficients around past boundary at $\tau=\pi$. We have $\laa_0^+(\hat{x})=\laa_0^-(-\hat{x})$ and
	\begin{equation}\label{ra e ant}
	\psi^+_2(\hat{x})=-\psi^-_2(-\hat{x})\quad\Rightarrow\quad F^+_{ur}(\hat{x})=F_{ur}^-(-\hat{x})\,.
	\end{equation}
	
	These are two independent piece of information:
	\begin{itemize}
	\item The soft charge with $\tau$-dependent parameter $\laa(\tau,\hat{x})$ \emph{appears} at past of future null infinity $\mathscr{I}^+_-$ as a charge with parameter $\laa_0(\hat{x})$ and \emph{appears} at future of past null infinity $\mathscr{I}^-_+$ as a charge with parameter $\laa_0(-\hat{x})$.
	\item Radial electric fields  are also antipodally related according to \eqref{ra e ant}. As a result, the charge value at  $\mathscr{I}^+_-$ with parameter $\laa_0(\hat{x})$ is actually equal to the charge value at $\mathscr{I}^-_+$ with parameter $\laa_0(-\hat{x})$. 
	\end{itemize}
	
	What if we exchanged $\laa$ and $\psi$ in \eqref{four dim antip}? In that case we would have $\psi(\tau,\hat{x})=\psi_0(\hat{x})+\tilde{\psi}_2\tau^2\log\tau+\ord{\tau^2}$ and in consequence,
	\begin{equation}
	\F_{ur}=\frac{ \tilde{\psi}_2(\hat{x})\log(u/r)}{r^2}+\ord{1/r^{2}}.
	\end{equation}
	The leading component would be odd under antipodal map $F^+_{ur}(\hat{x})=-F_{ur}^-(-\hat{x})$. This choice is not inconsistent by itself, and has infinitely many soft charges, as was the case for the antipodal-even choice. However, we do not expect logarithmic Coulomb fields from localized sources in four dimensions. Another problem is the logarithmic divergence of boost charges if both antipodal-even and antipodal-odd solutions for $\psi$ are allowed. To see this, let us examine the  boost charges \eqref{final} at $\tau=\pi/2$ (i.e. $t=0$) surface. On the `equator' of asymptotic de Sitter space, $\rho=r$ and $r\F_{tr}=\F_{\rho\tau}$,
	\begin{equation}\label{boost div}
	Q_{\xi}=\text{finite}+\int \frac{dr}{r^2} d^2\hat{x}\sqrt{q}\xi^t(\partial_t\psi(t,\hat{x}))^2|_{t=0}
	\end{equation}
	where the boost vector in $x^{\ii}$ direction is $\xi_{(\ii)}=t\partial_\ii+x^\ii\partial_t$. The expression \eqref{boost div} is logarithmically divergent, unless $\partial_t\psi$ has specific parity (even or odd) under $\hat{x}\to-\hat{x}$. If only one parity is allowed, the integration on 2-sphere vanishes since $\xi^t_{(\ii)}$ is an odd function on 2-sphere. By inspection of Li\'enard–Wiechert solution, one picks \eqref{four dim antip}, for which $F_{tr}(t,\hat{x})|_{t=0}=F_{tr}(t,-\hat{x})|_{t=0}$, and at the same time, avoids logarithmic behaviour at null infinity.

	\paragraph{Higher dimensions $d>4$.} In even dimensions, choose
	\begin{align} \label{higher antip}
	\psi(\tau,\hat{x})&=\sin^{\frac{d-2}{2}}\tau\sum_{\ell=1}Y^m_l(\hat{x}) a_{lm} P_{(2l+d-4)/2}^{(d-2)/2}(\cos\tau) \\
	\laa(\tau,\hat{x})&=\sin^{\frac{d-2}{2}}\tau\sum_{\ell=1}Y^m_l(\hat{x}) b_{lm} Q_{(2l+d-4)/2}^{(d-2)/2}(\cos\tau) 
	\end{align}
	and make the opposite choice in odd dimensions, so that the antipodal conditions \eqref{4 antip} are satisfied. For $l=0$ component \eqref{zero sol}, take $b_0$ for $\psi$ and $a_0$ for $\laa$ in all 
	dimensions. Behavior of solutions around $\tau=0$ is
	\begin{equation}
	\psi= {\tau}^{d-2}{\psi}_{d-2}(\hat{x})+\ord{ \tau^{d}}\,,\qquad
	\laa= \laa_0(\hat{x})+\ord{\tau^2}\,.
	\end{equation}
	The charge expression at $\mathscr{I}^+_-$ will be similar to \eqref{null charge max}. If the opposite choice had been taken, radial electric fields at $\mathscr{I}^+_-$ would have been
	\begin{eqnarray}
	\F_{ur}=r^{-\frac{d}{2}}(-2u)^{\frac{4-d}{2}}\psi_2(\hat{x})+\ord{r^{-\frac{d}{2}-1}}
	\end{eqnarray}
	For these solutions, boost charges are well-defined ($d>4$). But noticing their behavior at $u\to0$, we can eliminate this set of solutions by requiring that:
	\emph{The field strength tensor $\F_{\mu\nu}$ being a physical field must be regular at light cone $\mathscr{L}^\pm$ (i.e. $u=0$ and $v=0$ surfaces in advanced/retarded Bondi coordinates).}
	
	The set of solutions \eqref{higher antip} lead to regular fields at light cone and satisfy antipodal matching condition $F^+_{ur}(\hat{x})=F^-_{ur}(-\hat{x})$ in all dimensions.

	\subsection{Three dimensional theory}\label{3d sec}
	This section is devoted to the three dimensional Maxwell theory at spatial infinity\footnote{The asymptotic symmetry at null infinity was discussed in \cite{Barnich:2015jua}.}. The reason for the separate consideration of the three-dimensional case is mainly the particular form of solutions:  dS$_2$ is conformally flat and the solution space consists of the left- and right-moving scalar modes. For this simplest case, we will also translate the boundary conditions to the retarded Bondi coordinates $(u,r,\varphi)$. The unit  $dS_2$ metric is
	simpler
	in coordinates $x^\pm=\varphi\pm\tau$
	\begin{equation}
	d\tilde{s}^2=\frac{-d\tau^2+d\varphi^2}{\sin^2 \tau}=\frac{\di x^+\di x^-}{\sin^2 \tau}.
	\end{equation}
	\paragraph{Boundary conditions and the solution space.} 
	The static Coulomb solution is
	$
	\F_{tr}=q/r\,.    
	$ 
	The electric field in hyperbolic coordinates becomes $\F_{\T\rho}=-q\,$. The boundary conditions \eqref{maxwell b.c.} for $d=3$ become
	\begin{equation}
	\F_{a\rho
	}\sim \ord{\rho^0}\,,\qquad \F_{bc}\sim \ord{\rho^{0}}\,.
	\end{equation}
	This boundary condition is realized by following power-law asymptotic expansion on the gauge field
	\begin{equation}\label{3d falaf}
	\A_{\rho}=\sum_{n=0}A_{\rho}^{(n)}\rho^{-n}\,,\qquad \A_{a}=\sum_{n=0}A_{a}^{(n)}\rho^{-n}\,.
	\end{equation}
	The asymptotic behavior adopted here allows moving electric charges in $2+1$ dimensions. One can fix a radial gauge $\A_\rho(\rho,x^a)=0$, by an appropriate gauge transformation on \eqref{3d falaf}. The price is the introduction of divergent terms in the new $\tilde{\A}_a$ expansion:
	\begin{equation}\label{radial falaf}
	\tilde{\A}_{a}=\partial_a\psi \rho+\partial_a\chi \log\rho+\sum_{n=0}\tilde{A}_{a}^{(n)}\rho^{-n}\,.
	\end{equation}
	In order to determine a solution for $\A_a(\rho,x^a)$, fixing initial data at (say) future boundary of de Sitter space at $(\rho,\tau\to 0)$ is demanded. Hence, \eqref{radial falaf} would be our prescribed asymptotic expansion in radial gauge.  However, in the present discussion, we will not fix any gauge and work with \eqref{3d falaf}, except at the leading order $A^{0}$.
	
	Analyzing the free solutions is simpler for the Hodge dualized degrees of freedom. In three dimensions, Maxwell theory is dual to a scalar field:
	\begin{equation*}
	\di\Phi=\star\F\,.
	\end{equation*}
	In hyperbolic coordinates the relation is
	\begin{equation}\label{dual 3d}
	\partial_a\Phi={\epsilon^{\rho b}}_{a}\F_{\rho b}\qquad \partial_\rho\Phi={\epsilon^{ab}}_{\rho}\F_{ab}\,
	\end{equation}
	so that $\Phi\sim\ord{1}$. Maxwell's  field equations  are trivially satisfied $\di^2\Phi=0$, while the Bianchi Identity $\di \F=0$ becomes $\di\star\di\Phi=0$ :
	\begin{equation}\label{phi 3d}
	D_aD^a\Phi+\partial_\rho(\rho^2\partial_\rho\Phi)=0\,.
	\end{equation}
	The general solution for the electromagnetic field strength is found by \eqref{phi 3d} and \eqref{dual 3d}.

	At leading and subleading order, $F^{(0)}_{a\rho}=\partial_a A^{(0)}_\rho$ and $F^{(1)}_{a\rho}=\partial_a A^{(1)}_\rho$.
	The relation with the dual scalar field is 
	\begin{equation}\label{duality}
	\partial_aA^{(0)}_\rho=\bar{\epsilon}_{ab}\partial^b\Phi^{(0)}\qquad F_{ab}^{(0)}=-\frac{1}{2}\bar{\epsilon}_{ab}\Phi^{(1)}\qquad     \partial_aA^{(1)}_\rho=\bar{\epsilon}_{ab}\partial^b\Phi^{(1)}
	\end{equation}
	where $\bar{\epsilon}$ is the volume-form on unit $dS_2$.
	The equation of motion for $A^{(0)}_\rho$ and leading Bianchi identity give
	\begin{equation}\label{3d eq}
	D_aD^aA^{(0)}_\rho=0\,,\qquad D_aD^a\Phi^{(1)}=0\,.
	\end{equation}
	The differential operator is the Laplacian on $dS_2$, which takes a nicer form 
	\begin{equation}
	\partial_+\partial_-\psi=0\,.
	\end{equation}
	The general solution  with periodic boundary condition $\psi(\Ti,\varphi)=\psi(\Ti,\varphi+2\pi)$ is the following.
	\begin{equation}\label{3d Maxwell sol}
	\psi(\Ti,\varphi)=a_0+b_0\Ti+\sum_{n\neq 0}\left(a_n e^{inx^+}+b_ne^{inx^-}\right)
	\end{equation}
	
	\paragraph{Action principle and charges.} 
	The boundary term with fall-off \eqref{3d falaf} is finite
	\begin{eqnarray}\label{3d bound}
	\int_{\Gamma}\sqrt{h}\delta A_{a}^{(0)}\partial^aA^{(0)}_\rho=\int_{\Gamma}\sqrt{h}\left[D^a\left(\delta A_{a}^{(0)}A^{(0)}_\rho\right)-D^a\delta A_{a}^{(0)}A^{(0)}_\rho\right]
	\end{eqnarray}
	Integration by parts and fixing the asymptotic gauge  $D^aA^{(0)}_a=0$ makes the integrand a total divergence. In contrast to higher dimensions, fixing the Lorenz gauge $\nabla_\mu\A^\mu$ is not consistent with our power-law asymptotic expansion: The Lorentz gauge in hyperbolic coordinates is
	\begin{equation}
	D^a\A_a+\partial_\rho(\rho^2\A_\rho)=0,
	\end{equation}
	hence at leading order it implies either $A^{(0)}_\rho=0$ or $\A_a\sim\ord{\rho}$. The latter violates the boundary conditions \eqref{3d falaf}, while the former eliminates physical solutions and charges (note: the index on $D^a$ is raised by unit $dS_2$ metric). In the asymptotic gauge $D^aA^{(0)}_a=0$, the boundary term \eqref{3d bound} becomes an integral on the boundaries of $\Gamma$, which give no contribution when the initial and final data are fixed.
	
	The asymptotic gauge fixing leaves residual gauge transformations satisfying $D_aD^a\laa=0$. The conserved charges are obtained by the same method explained before.
	\begin{equation}\label{3d charge}
	Q_\laa=\int_{S^1}d\varphi\left(\partial_\tau\laa A^{(0)}_\rho-\laa \partial_\tau A^{(0)}_\rho\right)=
	\int_{S^1}d\varphi\left(\partial_+\laa A^{(0)}_\rho-\laa \partial_+ A^{(0)}_\rho\right)-\quad (+\leftrightarrow-)
	\end{equation}
	
	With solutions 
	\begin{align}
	A_{\rho}^{(0)}(\tau,\varphi)&=\psi^+_0+\psi^-_0\tau+\sum_{n\neq 0}\left(\psi^+_n e^{inx^+}+\psi^-_ne^{inx^-}\right)\\
	\laa(\tau,\varphi)&=\laa^+_0+\laa^-_0\tau+\sum_{n\neq 0}\left(\laa^+_n e^{inx^+}+\laa^-_ne^{inx^-}\right)
	\end{align}
	the charge becomes
	\begin{equation}
	Q_\laa[A]=\psi^+_0\laa^-_0-\psi^-_0\laa^+_0+2i\sum_{n>0}n(\psi^+_n\laa^+_{-n}-\psi^-_n\laa^-_{-n})
	\end{equation}
	The whole set of charges form two copies of angle-dependent $U(1)$ symmetries, labeled by two functions
	\begin{equation}
	\laa^\pm=\sum_n\laa^\pm_n e^{in\varphi}\,.
	\end{equation}
	We will cut off one copy of the charges by an antipodal condition, explained below.

	\paragraph{Antipodal condition.} 
	The whole set of solutions \eqref{3d Maxwell sol} are regular at light cone. Nevertheless, we opt to impose conditions \eqref{4 antip} which include physical solutions.
	\begin{equation}
	A_\rho^{(0)}(\tau,\varphi)=-A_\rho^{(0)}(\pi-\tau,\varphi+\pi)
	\end{equation}
	The antipodal map $(\tau,\varphi)\to(\pi-\tau,\varphi+\pi)$ is equivalent to $x^+\leftrightarrow x^-$. As a result,  \eqref{3d Maxwell sol} is divided into even and odd parts
	\begin{subequations}
		\begin{align}
		A_\rho^{(0)}(\tau,\varphi)&=c_0 \tau+\sum_{n\neq 0} \frac{c_n}{n} e^{in\varphi}\sin n\tau\,,\qquad c_n=c_{-n}^\ast \qquad \text{odd}\label{odd} \\  
		\laa(\tau,\varphi)&=d_0 +\sum_{n\neq 0} d_n e^{in\varphi}\cos n\tau\,,\qquad d_n=d_{-n}^\ast    \qquad \text{even}\label{even}
		\end{align}
	\end{subequations}
	One can explicitly check that for a boosted electric charge, the gauge field lies in \eqref{odd}. The set of charges consists of one copy of angle-dependent $U(1)$ transformations on circle
	\begin{equation}
	Q_\laa[A]=-\sum_n d^\ast_nc_n\,.
	\end{equation}

	In the next section, we exhibit a similar analysis for the theory on anti-de Sitter space.

	\section{Anti-de Sitter background}\label{Maxwell AdS}
	In this section, we study asymptotic symmetries of Maxwell theory on AdS$_d$, $d\geq 4$.   Boundary conditions in three-dimensional theory are different, as potential grows logarithmically on the boundary. On the other hand, higher-dimensional theories share similar features that we discuss.
	
	We keep on using hyperbolic coordinates, in which the anti-de Sitter metric becomes
	\begin{equation}\label{adshyper}
	ds^2=\frac{\ell^2 d\rho^2}{\rho^2+\ell^2}+\rho^2h_{ab}dx^a dx^b,\qquad a,b=0,1,\cdots,d-2\,,
	\end{equation}
	where $h_{ab}$ is the $(d-1)$-dimensional Lorentzian metric on unit radius de Sitter space
	\begin{equation}\label{AdS-metric}
	h_{ab}dx^a dx^b=\frac{1}{\sin^2\tau}(-d\tau^2+ d\Omega^2_{d-2}),\
	\end{equation}
	see appendix \ref{rindler} for more detail of the AdS Killing vectors in this coordinate system.
	\paragraph{{Field equations.}} In anti-de Sitter space,  field equations determine how boundary data propagate into the bulk. To setup the  AdS radial evolution and describe the boundary data for the evolution in $\rho$ direction, we note that equations of motion $\nabla^\mu {\cal F}_{\mu\nu}=0$ involve two classes of equations: those which are first order in $\rho$ and those which are second order. The former  \begin{equation}
	D_a\F^{a\rho}=0\label{heq1}
	\end{equation}
	may be viewed as constraints among the
	boundary data set at $\rho=\infty$. These initial data (together with the constraint among them) then propagate in the radial direction $\rho$ with the latter set of equations of motion
	\begin{align} (\rho^2+1)^{1/2}\rho^{1-d}\partial_\rho\left(\rho^{d-1}(\rho^2+1)^{-1/2}\F^{\rho a}\right)+D_b\F^{ba}=0\,.\label{heq2}
	\end{align}

\begin{figure}
		\centering
		\begin{tikzpicture}[scale=.4]
		\draw[very thick] (-3,5)--(-3,-5);
		\draw[very thick] (3,5)--(3,-5);
		\draw[dashed] (-3,3)--(3,-3);
		\draw[dashed] (-3,-3)--(3,3);
		
		\draw [thick,blue, directed] (0,0)..controls (2,3)..(1,5);
		\draw [thick,blue, reverse directed] (0,0)..controls (-2,-3)..(-1,-5);
		\draw [thick,blue, directed] (0,0)..controls (-1,3)..(-.5,5);
		\draw [thick,blue, reverse directed] (0,0)..controls (1,-3)..(.5,-5);
		\draw [directed, blue,thick](0,0)--(0,5);
		\draw [directed, blue,thick](0,-5)--(0,0);
		\draw [thick,red, directed] (-.8,-5)--(0,-4)..controls (2.5,-1)..(0,2)..controls (-1.6,3.8)..(-1.8,5);
		%
		\node[] at( 3.5,4) {$\mathscr{I}$};
		\node[] at( -3.5,4) {$\mathscr{I}$};
		%
		%

		\end{tikzpicture}
		\caption{Displaced and boosted charges. Blue curves show geodesics that cross the origin at $t=0$. The paths correspond to different Lorentz boosts applied on a static electric charge at the origin. The red curves show displaced charges, which  cross the origin at a later time.}
		\label{fig ADS}
	\end{figure}
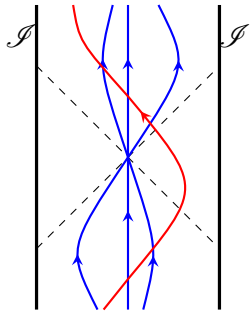

	Any physically relevant boundary condition is expected to allow for moving electric charges. So, we analyze some simple solutions of the field equations before choosing and setting the boundary conditions (see figure \ref{fig ADS} \cite{Esmaeili:2019mbw}).

	\paragraph{Boosted electric charges.}
	The simplest family of solutions is found by assuming $\A_a=\partial_a\mathcal{C}$ for which $\F_{ab}=0$. By  $\di\F=0$ we have
	\begin{equation}
	\partial_a\F_{b\rho}-\partial_b\F_{a\rho}=0.
	\end{equation}
	Then, $\F_{a\rho}=\partial_a\psi$ for a gauge-invariant scalar $\psi=\A_\rho-\partial_\rho \mathcal{C}$. 
	Equation \eqref{heq2} reduces to
	\begin{equation}
	\partial_\rho\left(\rho^{d-1}(\rho^2+1)^{-1/2}\F^{\rho a}\right)=0
	\end{equation}
	and the solution is
	\begin{equation}\label{boosted psi}
	\psi=\rho^{3-d}(1+\rho^2)^{-1/2}\bar{\psi}(x^b),\qquad D^a D_a\bar{\psi}(x^b)=0
	\end{equation}
	
	This is the exact solution for an electric charge with an arbitrary boost, which crosses the event at the origin of AdS at $\rho=0$. To see this, consider a static electric charge at the origin, given by 
	\begin{equation}
	\F_{\tau\rho}=\frac{\rho^{3-d}\sin^{{d-3}}\tau}{\sqrt{\rho^2+1}}=\partial_\tau\psi(\rho,\tau)
	\end{equation}
	in hyperbolic coordinates. 
	$\psi$ is a Lorentz scalar and under boosts,  it acquires angle-dependence so that other components of $\F_{a\rho}$ are also turned on. The phase space of all superpositions of boosted electric charges consists of all functions $\psi(\rho,x^a)$ satisfying conditions \eqref{boosted psi}. 
	Finally, note that near the origin, $\rho\ll 1$, the behavior is $\psi\propto \rho^{3-d}$ which matches flat space solutions \eqref{indp electric}. Near the boundary, however, it falls more slowly as $\psi\propto \rho^{2-d}$.

	\paragraph{Displaced electric charge.}
	Let us now relax the condition $\A_a=\partial_a\mathcal{C}$. Fixing the $\A_\rho=0$ gauge, the constraint equation \eqref{heq1} implies that $\A_a$ is a divergenceless de Sitter vector $D^a\A_a=0$. The spectrum of the Laplace operator on  $\A_a$ is given by representation theory of $SO(d-1,1)$
	\begin{equation}\label{eigen}
	D^b\F^{[\beta]}_{ba}=-\beta(\beta+d-4)\A^{[\beta]}_a\,.
	\end{equation}
	Eq.~\eqref{heq2} then becomes an equation for the eigen-vector $\A^{[\beta]}_a$:
	\begin{equation}\label{eigen eom}
	\frac{\sqrt{\rho^2+1}}{\rho^{d-5}}\partial_\rho\left(\rho^{d-3}(\rho^2+1)^{1/2}\partial_\rho\A^{[\beta]}_a\right)-\beta(\beta+d-4)\A^{[\beta]}_a=0.
	\end{equation}
	Exact expressions  can be expressed in terms of hypergeometric functions. We are only interested in two asymptotic limits $\rho\ll 1$ and $\rho\gg1$.
	\paragraph{Flat region $\rho\ll1$.} The two solutions for eigen-vector fields \eqref{eigen} are
	\begin{equation}
	\A_a^{[\beta]}\propto \rho^\beta,\qquad \qquad \A_a^{[\beta]}\propto\rho^{4-d-\beta}\,.
	\end{equation}
	For integer $\beta$, the solutions with $\rho^{4-d-\beta}$ falloff are of course recognized as the multipole moments of the electromagnetic field \cite{Jackson:1998nia,Esmaeili:2019hom,Herdeiro:2015vaa}. Note that \eqref{eigen} is Lorentz covariant. Therefore, the eigen-vector field $\A_a^{[\beta]}$ describes a system of $2^\beta$-poles moving freely in flat space\footnote{While non-integer in general, for static  flat space configurations (i.e. electric multipoles) $\beta$ becomes a non-negative integer, and by Lorentz covariance of \eqref{eigen}, this remains true for freely moving multipoles. To see this, recall that an electric $2^\beta$-pole configuration in flat space is described by
		\begin{equation}
		\A_t=\frac{\mathcal{Y}_{\beta,m_i}}{r^{\beta+d-3}},\nn\end{equation}
		where $\mathcal{Y}_{\beta,m_i}$ are harmonics on S$^{d-2}$. In hyperbolic coordinates (which is related to global coordinates by \eqref{glo hyper}), and in radial gauge $\A_\rho=0$ this potential is given by
		\begin{align}
		\A_\tau&=\frac{3-\beta-d}{4-\beta-d}\sin^{d+\beta-3}\tau
		\rho^{4-d-\beta}\mathcal{Y}_{\beta,m_i}\\
		\A_i&=-\frac{\tan\tau\sin^{d+\beta-3}\tau}
		{4-\beta-d}\rho^{4-\beta-d}\partial_i\mathcal{Y}_{\beta,m_i}
		\end{align}Straightforward algebra shows that
		\begin{equation}
		D^i\F_{i\tau}=-\beta(\beta+d-4)\A_\tau\,.\nn
		\end{equation}
		
		The above explicit expressions for static fields show that $\beta$ is quantized. This equality holds also in a boosted frame by Lorentz covariance. Thus, $\A_{a}^{[\beta]}$ eigenvectors with integral values for $\beta$
		correspond to electric $2^\beta$-poles. 
	}.
	
	\paragraph{Boundary region $\rho\gg 1$.}
	By a  near boundary expansion
	\begin{equation}
	\A_a=\sum\rho^{s}A_a^{(s)}\,,
	\end{equation}
	equation \eqref{eigen eom} leads to a recursive relation
	\begin{equation}
	s(d+s-3)A^{(s)}_a=-D^bF^{(s+2)}_{ba}\,.
	\end{equation}
	There are two independent solutions that behave near the boundary
	\begin{align}
	& \A_a\sim\ord{1},\label{Nonono}\\
	&\A_a\sim\ord{\rho^{3-d}}\label{normo}\,.
	\end{align}
	Although the recursive relation depends on the moment $\beta$ (and the exact solution is a hypergeometric function 
	depending on $\beta$ and $d$), the leading term is universal in $\beta$.  In other words, the leading term in the asymptotic series is either $s=0$ or $s=d-3$ for all moments $\beta$. This result reveals the drastic difference between asymptotic behavior of multipole moments in flat and anti-de Sitter spaces. In flat space, higher-poles fall successively faster at far regions. In anti-de Sitter space, however, all multipoles have the same order of magnitude near the boundary. In other words, all the data inside are `accessible' to a boundary observer.

	To this end, we need to make sure that the flow of bulk energy momentum tensor $T^{\mu\nu}=\F^{\mu\sigma}{\F^\nu}_{\sigma}-\frac14 g^{\mu\nu}\F \cdot \F$,  vanishes at the boundary $\Gamma$.   
	\begin{align}
	\int_{\Gamma}\sqrt{-g} T^{\rho \mu}\xi_\mu=\int_{\Gamma} \sqrt{-g}T^{\rho\rho}\xi_\rho+\int_{\Gamma} \sqrt{-g}T^{\rho a}\xi_a.
	\end{align}
	Demanding a vanishing flux rules out the solution \eqref{Nonono} for $d>4$. In contrary, for \eqref{normo} we have
	\begin{align}
	T^{\rho\rho}\sim\ord{\rho^{6-2d}},\,\,\,\,\, T^{\rho a}\sim\ord{\rho^{3-2d}},
	\end{align}
	so that
	\begin{align}
	\int_{\Gamma} \sqrt{g} T^{\rho \mu}\xi_\mu\sim\ord{\rho^{3-d}}.
	\end{align}
	We conclude that if $d\geq5$, only the solutions \eqref{normo} are admissible. For $d=4$ there are two possible quantizations, depending on the choice of normalizable modes.     Solutions \eqref{normo} are the usual normalizable modes in the bulk \cite{Skenderis:2002wp,Aharony:1999ti}. 
	

	\paragraph{Boundary conditions.}\label{sec:3.2} As in the usual AdS/CFT \cite{Skenderis:2002wp}, we fix the radial gauge by setting $\A_\rho=0$. For $d>3$, the boundary conditions must be relaxed to allow large gauge transformations which have nontrivial charges. The radial gauge leaves boundary gauge transformations with no subleading terms:
	\begin{equation}\label{bc 3}
	\delta_\lambda\A_a=\partial_a\lambda\sim\ord{1},\qquad \partial_\rho\lambda=0.
	\end{equation}
	The asymptotic behaviour of the field strength components are,
	\begin{equation}\label{F bc}
	\F_{a\rho}\sim\ord{\rho^{2-d}},\qquad \F_{ab}\sim\ord{\rho^{3-d}}.
	\end{equation}
	in 3d case, the electric charge yields a logarithmic function of $\rho$; the 3d should be studied separately. Here we focus on $d>3$ case.
	\begin{equation}
	\begin{aligned}
	&\A_{a}=\partial_a \phi+\frac{A_a}{(3-d)\rho^{d-3}}+\frac{A^{(4-d)}_a}{\rho^{d-4}}+\cdots
	\end{aligned}
	\label{Falloffs}
	\end{equation}
	In a similar manner, the gauge parameter $\lambda$ can be expanded asymptotically,
	\begin{align}\label{gauge-trans-expansion}
	\lambda=\lambda_S+\frac{\hat\lambda}{\rho^{d-3}}+\cdots\,.
	\end{align}
	
	{Let us now focus on the $\A_a$ component. The `constraint equation' \eqref{heq1} then yields $D^a A_a=0$ while $\phi$ remains unconstrained.  Under $\lambda_S$ gauge transformation, $\phi$ shifts while $A_a$ remains intact. The value of $A_a$ is fixed by the reflective boundary conditions on the electromagnetic wave in the bulk. Explicitly, the leading term in the bulk electric field at the boundary is proportional to $A_a$ and the leading magnetic field is $D_a A_b-D_b A_a$.  Nonetheless, one can decompose $A_a$ in \eqref{Falloffs} into an exact part and a transverse part $\hat A_a$
		\begin{equation}\label{A-Psi-hatA}
		A_a=(\partial_a\psi+\hat A_a), \qquad D^a\hat A_a=0.    
		\end{equation}
		The constraint equation $D^a A_a=0$   yields 
		\begin{equation}\label{Psi-eom}
		D_a D^a\psi=0.
		\end{equation}

		Being on the dS$_{d-1}$, which allows for harmonic forms (here a 0-form, a scalar), there is a freedom/ambiguity in making the above decomposition. We quantify this ambiguity through $\lambda_R$ `boundary gauge transformations'. Explicitly, we have two sets of gauge transformations $\lambda=(\lambda_S, \lambda_R)$ such that
		\begin{align}\label{Boundary-Data-variation}
		\delta_\lambda \phi=\lambda_S,\qquad
		\delta_\lambda \psi=\lambda_{R},\qquad
		\delta_\lambda \hat A_a=-\partial_a\lambda_R,\qquad D_a D^a\lambda_R=0,
		\end{align}
		and hence $\delta_\lambda A_a=0$. So, our boundary data are given by $(\phi, \psi; \hat A_a)$ subject to the above boundary gauge transformations.  At this stage, the separation of $A_a$ into $\psi$ and $\hat A_a$ parts seems arbitrary. Its virtue will however become clear in the next section when we study the symplectic structure and the boundary charges.}

	\subsection{Action principle and conserved symplectic form}\label{sec:4}
	
	In this section we supplement the Maxwell action  by appropriate boundary terms, so that with the falloffs \eqref{Falloffs} we have a well-defined variation principle. Starting from \eqref{p action}, we vary the action to get the equations of motion and the symplectic potential,
	\begin{align}
	\delta S= \int_{AdS} \sqrt{g} \nabla_\mu \F^{\mu\nu} \delta \A_\nu+\int_{AdS} \partial_\mu \theta^\mu,
	\end{align}
	with the symplectic potential  $\theta^\mu=-\sqrt{g}\F^{\mu\nu}\delta \A_\nu$. On the solutions , $\delta S \approx \int_{\Gamma} \theta^\rho$. With the falloffs \eqref{Falloffs} we have,
	\begin{align}\label{ActionPrinciple}
	\delta S &\approx -\int_{\Gamma} \sqrt{g}\F^{\rho a}\delta \A_a=-\int_{\Gamma}\sqrt{h}\big(D_a\psi D^a\delta \phi+\hat{A}_a  D^a \delta\phi\big)\\
	&= -\int_{\Gamma}\sqrt{h}D^a\Big(\psi D_a\delta \phi+\hat{A}_a \delta\phi\Big)
	+\int_{\Gamma}\sqrt{h}\Big(\psi D^a D_a\delta \phi+D^a \hat{A}_a \delta\phi\Big)\nonumber
	\end{align}
	where $D_a$ is the covariant derivative with respect to $h_{ab}$ and  $D^a=h^{ab}D_b$. Imposing the boundary gauge condition $D^a D_a\phi=0$,  the last term vanishes (recall that $\hat A_a$ is transverse $D_a\hat{A}^a=0$ by  definition) and we have 
	\begin{equation}\label{kappa-i}
	\theta^\rho=\sqrt{h}D^a \kappa_a,\qquad     \kappa_a=-\Big(\psi D_a\delta \phi+\hat{A}_a \delta\phi\Big)
	\end{equation}
	which guarantees having a well-defined  action principle. 
	
	We conclude that  for a well-posed action principle with the falloff conditions \eqref{Falloffs} we only need to have the following constraints on the boundary data,
	\begin{align}
	D^a D_a\phi=0,
	\end{align}
	which in turn constrains $\lambda_S$ to $D^a D_a\lambda_S=0$. That is, both of our boundary gauge transformations $\lambda_R, \lambda_S$ satisfy the Laplace equation on the boundary.

	
	\paragraph{Conserved symplectic form.}\label{conserve syplfrm}
	Here we choose the constant $\tau$ slices as the foliation slices and define the symplectic structure as, 
	\begin{align}\label{Symplectic Structure}
	\Omega^{\text{\tiny bulk}}_\tau
	=\int_{\Sigma_\tau} \sqrt{g}\tau_\mu \delta\F^{\mu\nu}\delta\A_{\nu}
	\end{align}
	where $\tau_a=\partial_a \tau$ is the 1-form defining constant $\tau$ slices \cite{Wald:1984rg}.  
	The symplectic form is not conserved:
	\begin{align}\label{omflux def}
	\Omega^{\text{\tiny bulk}}_{\tau_2}-\Omega^{\text{\tiny bulk}}_{\tau_1}=\int_{\Gamma} \omega^\rho=\int_{\Gamma} \sqrt{h} D_a \delta\kappa^a
	\end{align}
	where $\kappa^a$ is defined in \eqref{kappa-i}. Since the first term is a total divergence, we can take $\theta_{\text{b'dry}}=-\sqrt{h}\tau_a\kappa^a$ and construct the conserved symplectic form as,
	\begin{align}\label{Symplectic Structure conserved}
	\Omega=\int_{\Sigma_\tau} \sqrt{g} \tau_\mu\delta\F^{\mu\nu}\delta\A_{\nu}+\oint_{\partial\Sigma_\tau}\sqrt{h}\tau_a\Big(\delta\psi D^a\delta \phi+ \delta\hat{A}^a \delta\phi\Big)
	\end{align}
	with the conditions $D^2\phi=0=D^a\hat{A}_a$. 
	Before the computation of the charges, let us pause and discuss our result \eqref{Symplectic Structure conserved}: While the symplectic form of the original Maxwell theory was not conserved, we made it conserved by adding a boundary piece. This boundary piece was extracted through the decomposition \eqref{Boundary-Data-variation} and in itself may be viewed as a symplectic form of a compliment gauge theory with dynamical fields $(\phi,\psi)$ and the `external gauge field' $\hat A_a$.
	Explicitly, substituting the decomposition $\A_\mu=\partial_\mu\phi+\bar{\A}_\mu$ into \eqref{Symplectic Structure conserved} and using the equations of motion, we arrive at,
	\begin{equation}\label{Omega-total}
	\Omega=\int_{\Sigma_\tau} \sqrt{g} \tau_\mu\delta\F^{\mu\nu}\delta\bar{\A}_{\nu}+\oint_{\partial\Sigma_\tau}\sqrt{h}\ \tau_a\Big(\delta\psi D^a\delta \phi+ \delta \phi D^a\delta\psi \Big).
	\end{equation}
	The surface term can be inferred from a boundary action
	\begin{equation}\label{bdry-action}
	S_{\text{\tiny b'dry}}=\int_{\Gamma} \sqrt{h}\ \partial_a\psi\partial^a\phi,\qquad d\neq 3.
	\end{equation}
	This action is invariant under $SO(d-1,1)$ isometry group of the boundary. The corresponding energy-momentum tensor is given by
	\begin{equation}\label{BEM Tensor}
	T^{\text{\tiny b'dry}}_{ab}=\frac{-2}{\sqrt{h}}\frac{\delta S^{\text{\tiny b'dry}}}{\delta h^{ab}}=-\Big[2\partial_{(a}\psi\partial_{b)}\phi-h_{ab}\partial\psi\cdot\partial\phi\Big].
	\end{equation}
	The role of this tensor will be clarified when we compute the Lorentz charge of the bulk theory.

	\subsection{Soft Charges}
	
	We have two classes of gauge transformations: the ``source charge''  denoted by $Q^S_\lambda$ when $\lambda_R=0$ and  the  ``response charge'' denoted by $Q^R_\lambda$ when $\lambda_S=0$. Using \eqref{Boundary-Data-variation} it turns out that,
	\begin{align}\label{source-response-charges}
	\begin{split}
	Q^S_\lambda[\psi]&=-\oint\sqrt{h}\tau_a\Big(\lambda_S D^a\psi-\psi D^a\lambda_S\Big),\\
	Q^R_\lambda[\phi]&=-\oint\sqrt{h}\tau_a\Big(\lambda_R D^a\phi-\phi D^a\lambda_R\Big),
	\end{split}
	\end{align}
	$Q^S_\lambda[\psi],\ Q^R_\lambda[\phi]$ are both conserved and independent of $\tau$ as a result of our earlier discussions on having a conserved symplectic structure. This latter may also be directly verified given the expressions of the charges above.  To see this, recall that the boundary gauge parameter $\lambda$ and the boundary fields $\phi, \psi$ satisfy Laplace equation. This implies that  $dQ^S_\lambda[\psi]/d\tau,\ dQ^R_\lambda[\phi]/d\tau$ become exact forms on the S$^{d-2}$ sphere at $\partial\Sigma_\tau$, and hence vanish.

	One can then readily compute the algebra of charges:
	\begin{align}\label{charges-algebra}
	\begin{split}
	\{Q^S_\lambda, Q^S_\chi\} =0,&\qquad \{Q^R_\lambda, Q^R_\chi\}=0,\\
	\{Q^S_\lambda, Q^R_\chi\}=-\Omega(\delta_{\laa_S},\delta_{\chi_R})&=\oint \sqrt{h}\ \tau_a(\lambda D^a\chi-\chi D^a\lambda).
	\end{split}
	\end{align}
	As we can see the source and response charges do not commute, which justifies the names.

	\subsection{Killing Charges}
	\label{sec:6}
	In this section, we study the conserved charges associated with the isometry transformations of the AdS background.
	
	The isometry algebra consists of Lorentz generators:  $L_{\mu\nu}=X_\mu\partial_\nu-X_\nu\partial_\mu$ 
	and ``AdS-Translations'':
	$L_{\mone \mu}=X_{\mone}\partial_\mu-X_\mu\partial_{\mone}$. 
	In hyperbolic coordinates 
	the associated vectors  are (see \eqref{XMhat})
	\begin{align}
	\text{{Lorentz:}}\, \qquad&\qquad \xi_L= 2\bar{X}^{[\mu} \bar{M}^{\nu]}_a h^{ab}\partial_b\label{xi-L2} \\
	\text{AdS-{translation:}}\, \qquad&\qquad \xi_T = -X_{\mone}\bar{X}^\mu \partial_\rho - \frac{X_{\mone}}{\rho} \bar{M}_a h^{ab}\partial_b\label{xi-T2}.
	\end{align}
	
	From \eqref{transl form}, if $\rho\ll 1$, we have $L_{\mone a}\cong \partial_a$. Hence, $L_{\mone a}$ are actually translation generators near the origin.
	On the other hand if  $\rho\gg1$, i.e. near the boundary of AdS, the AdS-translation generators are
	\begin{equation}
	L^{\mone \mu}\approx  \bar{X}^\mu\rho\partial_\rho + \bar{M}^\mu_a h^{ab}\partial_b\,.
	\end{equation}
	Infinitesimal AdS-translations act on $\rho$ as $\rho\to \rho(1+\epsilon \bar{X}^\mu)$. One can then observe that these
	AdS-translations are conformal Killing vectors of the boundary metric in hyperbolic slicing $h_{ab}$.

	The consistency of our phase space relies  on the constraints $D^2\phi=0=D^a\hat{A}_a$. These are all manifestly invariant under Lorentz part of the AdS isometries. The Lorentz charges are hence straightforward to construct. However, the gauge conditions are not covariant under AdS-translations and so the phase space is not consistent with AdS-translations. 
	A resolution to this problem is that the behavior of the gauge field under AdS-translations must be supplemented by an appropriate (field dependent) gauge transformation, which leaves the boundary constraints invariant.

	For Lorentz transformations labeled by $L_{ab}$ we have (cf. \eqref{xi-L2}), 
	\begin{align}
	\xi_L=(\xi_L^\rho,\xi_L^a)=(0,2\bar{X}^{[\mu} \bar{M}^{\nu]}_b h^{ab})
	\end{align}
	
	The definition $\delta_\xi \A=\mathcal{L}_\xi \A$ implies
	\begin{subequations}\label{Loren Lie}
		\begin{align}
		&\delta_{\xi_L} \psi=\xi_L^b \partial_b \psi\\
		&\delta_{\xi_L} \phi=\xi_L^b \partial_b \phi\\
		&\delta_{\xi_L} \hat{A}_a=\xi_L^b D^{}_b \hat{A}^{}_a+\hat{A}^{}_b D^{}_a\xi_L^b
		\end{align}
	\end{subequations}
	which reflects the fact that they are covariant Lorentz tensors. 
	For AdS-translations labeled by $T^\mu=L^{\mone \mu}$ we have (\emph{cf.} \eqref{xi-T2}),
	\begin{align}
	\xi_T=(\xi_T^\rho,\xi_T^a)=(\bar{X}^{\mone}\bar{X}^\mu\rho,\bar{X}^{\mone} \bar{M}^{\mu}_b h^{ab}).
	\end{align}	
	The boundary fields under AdS-translations transform as\footnote{$\psi$ appears in the asymptotic expansion of $\A_a$ at order $\rho^{3-d}$, \emph{cf.} \eqref{Falloffs} and \eqref{A-Psi-hatA}. Transformation of $\psi$ is defined such that the spacetime function
		$    \psi^\ast\equiv\frac{\psi}{\rho^{d-3}}+\cdots$
		transforms like a scalar field.
	}
	\begin{subequations}\label{Translations}
		\begin{align}
		\delta_{\xi_T} \psi&=\bar{\xi}_T^b \partial_b \psi+(3-d)\psi\bar{\xi}_T^\rho,\label{Translations-a}\\
		\delta_{\xi_T} \phi&=\bar{\xi}_T^b \partial_b \phi,\\
		\delta_{\xi_T} \hat{A}_a&
		=\bar{\xi}_T^b D_b \hat{A}_a+ \hat{A}_b D_a\bar{\xi}_T^b+(3-d)\bar{\xi}_T^\rho\hat{A}_a+(d-3)\psi\partial_a\bar{\xi}_T^\rho.
		\end{align}
	\end{subequations}


	\paragraph{Gauge fixing and AdS isometries.} The gauge fixing $\A_\rho=0$ and the expansion 
	\begin{align}
	\A_a&=\partial_a\phi+\frac{A_a}{(3-d)\rho^{d-3}}+\ord{\rho^{2-d}},\qquad \partial_\rho\phi=0,\label{rho exp}
	\end{align}
	are invariant under the Lorentz $SO(d-1,1)$ subgroup of the AdS$_d$ isometries. The AdS-translations, however,  clearly do not  respect these conditions. For example, starting from $\A_\rho=0$,
	\begin{equation}
	\mathcal{L}_\xi \A_\rho=\A_a\partial_\rho\xi^a,
	\end{equation}
	which is non-vanishing for AdS-translations. In addition, the action of AdS-translations on $\phi$ produces subleading terms which are possibly larger than $\rho^{3-d}$ (actually they are $\ord{\rho^{-2}}$). Both problems are overcome if we supplement AdS-translations by a field-dependent gauge transformation
	\begin{equation}
	\A_\mu\to\A_\mu+\partial_\mu\gamma,\qquad \partial_\rho\gamma=-\A_a\partial_\rho\xi^a.
	\end{equation}
	This transformation is necessary and sufficient to preserve $\partial_\rho\phi=0$. As a result, the structure \eqref{rho exp} is respected by all isometry transformations. For AdS-transformations, $\partial_\rho\xi^a\sim\ord{\rho^{-3}}$, therefore, $\gamma\sim\ord{\rho^{-2}}$ and hence do not contribute to the  canonical charges we computed in the previous section or to the AdS-isometry charges to be discussed in the next subsection. 
	
	The above $\gamma$ gauge transformation, however, is not sufficient since our other gauge conditions $D^2\phi=0$ and $D_b\hat{A}^b=0$ are not covariant under AdS-translations either. While our gauge conditions are Lorentz scalars,
	\begin{subequations}
		\begin{align}
		&\delta_{\xi_L}(D_b D^b\phi)=\mathcal{L}_{\xi_L}(D_b D^b\phi)\\
		&\delta_{\xi_L}(D_b\hat{A}^b)=\mathcal{L}_{\xi_L}(D_b\hat{A}^b),
		\end{align}
	\end{subequations}
	under AdS-translations they transform as,
	\begin{subequations}\label{Translation of gauge}
		\begin{align}
		&\delta_{\xi_T}(D_b D^b\phi)=D_b[\bar{\xi}_T^b D_a D^a \phi+(d-3)\bar{\xi}_T^\rho D^b \phi]\\
		&\delta_{\xi_T}(D_b\hat{A}^b)=D_b[\bar{\xi}_T^b D_a\hat{A}^a+(d-3)\psi D^b\bar{\xi}_T^\rho].
		\end{align}
	\end{subequations}
	
	We will introduce other field-dependent boundary gauge transformations, which we call $\alpha, \beta$ boundary gauge transformations (BGT),  to preserve AdS-translations too. As we will see, the $\gamma$ transformation above and the $\alpha,\beta$ BGT are of different order in $\rho$ and that the latter contribute to the AdS isometry charges, rendering them integrable.
	
	Equations \eqref{Translation of gauge} show how AdS-translations violate gauge condition on $\phi$ and are inconsistent with $D_b\hat{A}^b=0$. This problem can be resolved by supplementing AdS-translations  by an appropriate gauge transformation. The resolution may be provided by {field-dependent} gauge transformations which undo the action of AdS-translations on $\phi$ and $\psi$:
	\begin{align} \label{alpha beta def}
	\delta_\alpha\phi=-\alpha[\phi;\xi]\,,\qquad 
	\delta_\beta \psi=-\beta[\psi;\xi]\,,\qquad 
	\delta_\beta\hat{A}_a=D_a\beta[\psi;\xi], 
	\end{align}
	where $\alpha$ and $\beta$ specify two classes of boundary gauge transformations  such that
	\begin{align}
	(\delta_\alpha+\delta_{\xi_T}) \phi =0,\qquad 
	(\delta_\beta +\delta_{\xi_T}) \psi=0.\nn
	\end{align}
	We denote this phase space action as $\hat\delta_{\xi_T} \equiv \delta_{\alpha\beta}+\delta_{\xi_T}$. 
	One can check that the BGCs are invariant under $\hat\delta_{\xi_T}$, \begin{align}
	\hat\delta_{\xi_T} D^b D_b\phi=&\delta_{\xi_T} D^b D_b\phi-D^2\alpha=0,\nn\\
	\hat\delta_{\xi_T} D^b\hat{A}_b=&\delta_{{\xi_T}}D^b\hat{A}_b+D^2 \beta=D_b\left[\bar\xi_T^b(D^a\hat{A}_a+D^a D_a \psi)\right]\approx 0,\label{divergence variation}
	\end{align}
	where in the second line we have used the equations of motion.  
	
	The improved AdS-translation of $\hat{A}^a$ also would be,
	\begin{align}
	\hat\delta_{\xi_T} \hat{A}_a &=\delta_{\xi_T} \hat{A}_a+D_a \beta\\nn\
	&\approx \bar{\xi}_T^b D_b \hat{A}_a+ \hat{A}_b
	D_a\bar{\xi}_T^b+(3-d)\bar{\xi}_T^\rho\hat{A}_a+D_b\left(\bar{\xi}_T^b D_a \psi-\bar{\xi}^T_a D^b \psi\right)\nn\\
	&= D_b (\bar{\xi}_T^b A_a-A^b \bar{\xi}^T_a)
	\end{align}

	
	
	\paragraph{Integrability of isometry charges.}    One can see that by computing phase space Lie derivative of the symplectic form
	\begin{align}
	\dl_{\xi_L}\Omega&=0\\
	\dl_{\xi_T}\Omega&=(d-3)\oint \sqrt{h} \ \tau_b {\xi}_T^b\  \delta\psi \delta\phi.
	\end{align}
	Therefore, $\delta_{\xi_L}$ has an integrable charge
	This proves non-integrability of AdS-translations. In the discussion above we introduced a modification of the translations by  replacing $\delta_{\xi_T}$ by $\hat\delta_{\xi_T}=\delta_{\xi_T}+\delta_{\alpha\beta}$ which leaves the boundary gauge conditions invariant. 
	Detailed computations \cite{Esmaeili:2019mbw} show that the transformation $\hat\delta_{\xi_T}$ defined above leaves the boundary data $\psi$ and $\Phi$ invariant and importantly, has integrable well-defined canonical charge. In other words, using the equations of motion we can show that $\hat\delta_{\xi_T}$ leaves our phase space invariant and is also a canonical transformation on it, 
	\begin{align}
	\dl_{\hat{\xi}}\Omega=(\dl_{{\xi_T}}+\dl_{{\alpha,\beta}})\Omega \approx 0.
	\end{align}

	Using our gauge conditions and Killing equations, one can easily show that the boundary term is nothing but the boundary energy momentum tensor introduced in \eqref{BEM Tensor} contracted by $\xi_L$. The final expressions are
	\begin{align}\label{T-bdry}
	I_{\xi_L}&=\int_{\Sigma_\tau} \sqrt{g} {\tau}_\mu\xi^L_\nu T^{\mu\nu}+\oint \sqrt{h} \tau_b \bar{\xi}^L_a T_{\text{\tiny b'dry}}^{ab}.\\
	I_{\hat\xi_T}&=\int_{\Sigma_\tau} \sqrt{g} \tau_\mu \xi^T_\nu T^{\mu\nu}.
	\end{align}
	This is a plausible conclusion since the  boundary data are invariant under the improved AdS-translations.
	\subsection{Charge algebra}
	One can compute the algebra of canonical charges $I_{\xi_L}$, $I_{\hat\xi_T}$, $Q^S_\lambda$ and $Q^R_\lambda$. Firstly, for AdS-translations we have
	\begin{align}\label{Lie-T}
	\{I_{\hat\zeta_T},I_{\hat\xi_T}\}=-\hat\delta_{\zeta_T} I_{\hat\xi_T}&=-\int_{\Sigma_\tau} \sqrt{g}\tau^\nu\xi_T^\mu \delta_{\zeta_T} T_{\mu\nu}\nn\\
	&= -\int_{\Sigma_\tau} \sqrt{g}\tau^\nu\left([\xi_T,\zeta_T]^\mu T_{\mu\nu}+\mathcal{L}_\zeta( \xi_T^\mu T_{\mu\nu})\right)\,.
	\end{align}
	Notice that the second term in \eqref{Lie-T} is a boundary term and indeed vanishes by the prescribed boundary conditions and using  the relation
	\begin{align}\label{LIE EM TENSOR}
	\mathcal{L}_\zeta( \xi^\mu T_{\mu\nu})=&\zeta^\sigma\nabla_\sigma (\xi^\mu T_{\mu\nu})+ \xi^\mu T_{\mu\sigma}\nabla_\nu \zeta^\sigma
	=\nabla_\sigma(\zeta^\sigma \xi^\mu T_{\mu\nu}-\zeta_\nu \xi_\mu T^{\mu\sigma})
	\end{align}
	For the Lorentz case we have the similar situation, 
	\begin{align}\label{Lie-L}
	\{I_{\zeta_L},I_{\xi_L}\}&=-\int_{\Sigma_\tau} \sqrt{g}\tau^\nu\xi_L^\mu \delta_{\zeta_L} T_{\mu\nu}-\oint\sqrt{h} \tau_b \bar{\xi}^L_a \delta_{\zeta_L} T_{\text{\tiny b'dry}}^{ab}\\
	&=I_{[\xi_L,\zeta_L]}
	\end{align}
	where we used the  relation  \eqref{LIE EM TENSOR} on de Sitter space. 
	The rest of Poisson brackets can also easily be computed and the result is,
	\begin{subequations}\begin{align}
		\{I_{\hat\xi_T},I_{\hat\zeta_T}\}=I_{\widehat{[\xi_T,\zeta_T]}},\qquad &
		\{I_{\zeta_L},I_{\xi_L}\}=I_{[\xi_L,\zeta_L]},\qquad \{I_{\xi_L},I_{\hat\zeta_T}\}=I_{\widehat{[\xi_L,\zeta_T]}}\\
		\{I_{\hat\xi_T},Q^S_\lambda\}&=0, \qquad \qquad
		\{I_{\hat\xi_T},Q^R_\lambda\}=0\\ 
		\{I_{\xi_L},Q^S_\lambda\}&=Q^S_{\mathcal{L}_{\xi_L}\lambda}, \qquad 
		\{I_{\xi_L},Q^R_\lambda\}=Q^R_{\mathcal{L}_{\xi_L}\lambda}. 
		\end{align}
	\end{subequations}
	Among other things, the commutators above show that all our $Q^R, Q^S$ charges commute with the AdS-translation charges, especially with the Hamiltonian and have zero bulk energy.

	\section{Discussion}
	We conclude this chapter by a brief summary of the main results.
	\begin{itemize}
		\item Both in flat and anti-de Sitter space, the surface charges are not conserved with the original symplectic form of the theory. The addition of surface terms (basically fixing $\mathbf{Y}$-ambiguity) renders the charges conserved.
		\item Our flat space boundary conditions exclude magnetic charges. In higher dimensions $d>4$, magnetic charges are extended $(d-4)$-dimensional objects and their asymptotic behavior depends on their compactification. Sources compactified on $k$ dimensions fall off as $\ord{\rho^{-k}}$. Thus, compact magnetic branes give rise to fields behaving as $\ord{\rho^{4-d}}$ which needs a slight extension of boundary conditions, without much refinement in the analysis.
		\item Conservation of soft charges demands an asymptotic gauge fixing. The momentum conjugate to the soft mode $\laa$ is the gauge-invariant de Sitter scalar $\psi$. Asymptotic gauge fixing enforces $\laa$ to obey the same equation as its momentum; demanded by charge conservation. We also showed that the regularity of field strength at the light-cone implies the antipodal matching condition.

		\item On Anti-de Sitter background we found two sets of charges. `Source charges' belong to the residual part of the bulk gauge transformations after fixing the radial gauge. `Response charges' appear as the ambiguity in the decomposition of the induced gauge field $\A_\mu$ into transverse and longitudinal parts. In a flat space limit, the transverse part becomes subleading, so the longitudinal part can be extracted unambiguously.   Source and Response charges are non-commuting. 
		
		\item  The presence of memory effects in flat space results from the passage of electromagnetic and gravitational radiation. It is well known that there is no Bondi news in AdS space. The natural question will be to identify an $AdS$ observable which encodes the conserved charges, analogous to memory effects in flat space (see e.g. \cite{Chu:2019ssw}).

		\item We computed the action of Minkowski and Anti-de Sitter isometries on soft charges. In Minkowski space, the whole Poincar\'e algebra is canonically realized on phase space. In contrast, the charges for $AdS$-translations in $AdS$ space are not well-defined.

	\end{itemize}

	\chapter{Soft Charges in $p$-Form Theories}\label{p chapter}
	In this chapter, we will consider the asymptotic structure of $(p+1)$-form gauge theories with the Maxwell-like action \ref{p action} on a $(2p+4)$-dimensional Minkowski background. This includes Maxwell theory in four dimensions. Higher-form theories are quite similar to Maxwell theory in this regard, except for the existence of a new set of charges.
	
	In section \ref{gauge reduc}, we discuss the gauge structure of the theory in a Hamiltonian formulation, point out its reducibility, and count the degrees of freedom. To set up the relevant solutions and the boundary conditions, we provide a discussion of $p$-branes in section \ref{p-brane}. The soft charges are computed in section \ref{soft sect} according to boundary conditions set up in its previous section. The content is finally embodied in the example of $2$-form theory in section \ref{2-form theo}.

	We start with the action of the $(p+1)$-form gauge theory in the $d$-dimensional Minkowski spacetime $\M_d$, coupled to electric sources $J_e$\footnote{
		One may of course consider other gauge invariant actions e.g. $p$-form  Chern-Simons theory \cite{Edelstein:2008ry,Bekaert:2002eq,Bunster:2011qp} or Born-Infeld theory \cite{Kalb:1974yc,Chruscinski:2000zm}.}
	
	\eq{
		S=-\tfrac12\int \bF\wedge\star \bF +(-1)^{p} \int \bA\wedge\star J_e  +I_K
	}{pformaction}
	where $I_K$ contains kinetic terms of the branes. 
	Variation with respect to $\bA$ leads to $d\star \bF=\star J_e$. The $(p+1)$-form gauge field minimally couples to the electric branes by the interaction term
	\begin{equation}
	\int \star J_e\wedge \bA=\sum_i q_i\int_{\mathcal{W}_i}\bA\,,
	\end{equation}
	in which $\mathcal{W}_i$ is the world-volume of the $i{\textsuperscript{th}}$ brane.	Charged \emph{particles} being the sources in ordinary electrodynamics are replaced by \emph{extended objects}, which are spatially $p$-dimensional, the so-called \emph{p-branes}. 0-branes are relativistic particles, 1-branes are strings, 2-branes are called membranes etc. 
	
	Suppose that a $p$-brane with world-volume $\mathcal{W}_{p+1}$ is embedded in a $d$-dimensional manifold $M$ with embedding functions $y^\mu$
	\begin{equation}
	x^\mu=y^\mu(\sigma)
	\end{equation}
	where $\sigma_I,\, I=0,\cdots,p$ are world-volume coordinates. To any brane world-volume $\mathcal{W}_{p+1}(\sigma)$ with charge density $q$    one can associate a \emph{de Rham current}
	\begin{equation}
	J^{\mu_{1}\cdots \mu_{p+1}}(x):=q\int_{\mathcal{W}_{p+1}}\delta^{d}(x-y(\sigma))dy^{\mu_1}\wedge\cdots\wedge dy^{\mu_{p+1}}.
	\end{equation}
	If there are more than one charged branes, we define the current as the sum $J=\sum_i J_i$.
	The kinetic terms are usually Nambu-Goto actions
	\begin{equation}
	I_K=S_{NG}[J_e]\,.
	\end{equation}
	The NG action\footnote{To find out more about $p$-form actions and electric-magnetic duaity, see \cite{Bekaert:2002cz}.} for a $p$-brane of world-volume $\mathcal{W}_{p+1}$ is
	\begin{equation}
	S_{NG}=-T_p\int_{\mathcal{W}_{p+1}} \star 1=-T_p\int\sqrt{-\text{det}(g_{ab})}d^{p+1}\sigma
	\end{equation}
	where $T_p$ is the tension and $g_{ab}$ is the induced metric on world-volume.

	\paragraph{The Coulomb and the radiation asymptotic fall-offs.} There are usually two physically relevant fall-off behaviors, the Coulomb and radiation fall-offs.
	For the Coulomb fall-off behavior, let us consider field strengths $\F$ that represent the `electric charges' of the theory. In $p=0$ case, they are electric monopoles moving freely in space, while in generic $(p+1)$-form theories, the sources are extended $p$-branes. For $p=0$, this is the familiar $1/r^{d-3}$ behavior of the electric field and is hence called Coulomb fall-off behavior. Boosting the brane gives rise to purely spatial magnetic components of the field strength with the same fall-off. 
	
	The radiation fall-off behavior corresponds to the intensity of the radiation ${\cal E}$ of  `$(p+1)$-form photons' in $d$ dimensions. This is given by ${\cal E}\propto 1/r^{d-2}$. This energy is carried by the radiation $(p+1)$-form field with temporal component ${\cal A}_u$ (in the standard Bondi frame at null infinity) such that ${\cal E}\propto (\partial {\cal A}_u)^2$, yielding the radiation fall-off behavior ${\cal A}_u\sim 1/r^{\frac{d-2}{2}}$ \cite{Ortaggio:2014ipa, Campoleoni:2017qot}.
	
	For $d>2p+4$, Coulomb field falls off faster than radiation and the converse is true for $d<2p+4$. In $d=2p+4$, which we will be interested in, the both radiation and the Coulomb fields fall off in the same rate and hence the traces of the passing $(p+1)$-form radiation can be recorded in the associated $(p+1)$-form charges, leading to $p$-form memory effect. In addition, $d=2p+4$ is  the dimension in which the theory exhibits conformal symmetry \cite{Deser:1994ca,Raj:2016zjp}. The scaling part of this symmetry may be readily seen from the energy-momentum tensor of the theory
	\begin{equation}
	T_{\mu\nu}=\frac{1}{(p+1)!}\F_{\mu\alpha_1\cdots\alpha_{p+1}}{\F_{\nu}}^{\alpha_1\cdots\alpha_{p+1}}-\frac{1}{2(p+2)!}g_{\mu\nu}\F_{\alpha_1\cdots\alpha_{p+2}}\F^{\alpha_1\cdots\alpha_{p+2}},
	\end{equation}
	which is traceless if $d=2p+4$.
	Demanding scale invariance  hence yields $d=2p+4$. This scale invariance for our theory is enhanced to the $2p+4$ dimensional conformal symmetry \cite{Deser:1994ca}.

	\section{Gauge symmetry and reducibility}\label{gauge reduc}
	The $(p+1)$-form theory has  gauge symmetry  $\bA\to\bA+\di\bLaa$. In contrast to Maxwell theory, this symmetry is \emph{reducible}: There are arbitrary gauge parameters of the form $\bLaa=\di\boldsymbol{\eta}$ which leave the gauge field invariant. This can be best understood in a Hamiltonian formulation as exhibited below.

	Consider the canonical momenta for the theory \eqref{pformaction} 
	\begin{subequations}
		\begin{align}\label{canmumenta1}
		\pi^{0\ii_1\cdots \ii_p}&\equiv\frac{\partial L}{\partial \dot{\A}_{0 \ii_1\cdots \ii_p}}=0\,,\\
		\pi^{\ii_0\cdots \ii_p}&\equiv\frac{\partial L}{\partial \dot{\A}_{\ii_0\cdots \ii_p}}=\dot{\A}_{\ii_0\cdots \ii_p}-(p+1)\partial_{[\ii_0}\A_{|0|\ii_1\cdots \ii_p]}\,.\label{canmumenta2}
		\end{align}
	\end{subequations}
	
	Equation \eqref{canmumenta1}  shows that $A_{0 \ii_1\cdots \ii_p}$ is not a dynamical field since there is no term in the Lagrangian with its time derivative. Vanishing of the associated momenta make them the primary constraints of the theory
	\eqs{
		\phi_1^{\ii_1\cdots \ii_p}\equiv \pi^{0\ii_1\cdots \ii_p}=0\,.
	}
	
	The canonical commutation relations among the fields on the phase space are
	\eqs{
		\{A_{\mu_0\cdots \mu_p}(\ex),\pi^{\nu_0\cdots \nu_p}(\text{y})\}=\delta^{d-1}(\ex-\text{y})
		\delta_{\mu_0\cdots \mu_p}^{\nu_0\cdots \nu_p}\,,
	}
	where the generalized Kronecker delta is equal to $+1$ (respectively $-1$) if the lower indices are even (respectively odd) permutations of upper indices, and zero otherwise. 
	Given the canonical Hamiltonian
	\begin{align}
	H_C&=\frac{1}{p!}\int\di^{d-1}\ex\,\pi^{\ii_1\cdots \ii_p}\dot{\A}_{\ii_1\cdots \ii_p}-L\nn\\
	&=\frac{1}{2}\int\di^{d-1}\ex\Big[
	\frac{1}{(p+1)!}\pi_{\ii_0\cdots \ii_p}\pi^{\ii_0\cdots \ii_p}+
	\frac{1}{(p+2)!}\F_{\ii_0\cdots \ii_{p+1}}\F^{\ii_0\cdots \ii_{p+1}}\,,\nn\\
	&\qquad\qquad\qquad-\frac{1}{p!}A_{0\ii_1\cdots \ii_p}\partial_\kk\pi^{\kk\ii_1\cdots \ii_p}
	\Big]
	\end{align}
	one demands the primary constraints to be preserved in  time. This gives rise to the secondary constraints
	\eqs{
		\{\pi^{0\ii_1\cdots \ii_p},H_C\}=\partial_\kk\pi^{\kk\ii_1\cdots \ii_p}\equiv 
		\phi_2^{\ii_1\cdots \ii_p}\,.
	}
	The $p$-form constraints $\phi_1^{\ii_1\cdots \ii_p}$ and $\phi^{\ii_1\cdots \ii_p}_2$ are first-class and the generator of gauge symmetry will be built out of them. There are no further constraints since $\{\phi_2^{\ii_1\cdots \ii_p},H_C\}=0\,$.
	
	Generators of gauge transformations on phase space are constructed using the procedure \emph{a la} Castellani \cite{Castellani:1981us};
	\eq{
		G[\eps]=\frac{1}{p!}\eps^1_{\ii_1\cdots \ii_p}\phi_1^{\ii_1\cdots \ii_p}
		+\frac{1}{p!}\eps^2_{\ii_1\cdots \ii_p}\phi_2^{\ii_1\cdots \ii_p}
		\,.}{gentrans}
	where gauge parameters $\Laa^1$ and $\Laa^2$ are arbitrary anti-symmetric tensors. 
	The gauge transformations $\de_\epsilon F=\{F,G[\epsilon]\}$ read
	\begin{align}
	&\de \A_{0\ii_1\cdots \ii_p}=\eps^1_{\ii_1\cdots \ii_p}, &
	\de \A_{\ii_0\cdots \ii_p}=(p+1)\partial^{}_{[\ii_0}\eps^2_{\ii_1\cdots \ii_p]}\,, \\&\de \pi^{0\ii_1\cdots \ii_p}=0\,.
	\end{align}

	The extended action;
	\eqs{
		S_E=\int \di ^d\ex \Big[
		\pi^{\ii_0\cdots \ii_p}\dot{\A}_{\ii_0\cdots \ii_p}+\pi^{0\ii_1\cdots \ii_p}\dot{\A}_{0\ii_1\cdots \ii_p}-H_C-u^1_{\ii_1\cdots \ii_p}\phi_1^{\ii_1\cdots \ii_p}
		-u^2_{\ii_1\cdots \ii_p}\phi_2^{\ii_1\cdots \ii_p}
		\Big]\,,
	}
	with Lagrange multipliers $u^1$ and $u^2$  is invariant under transformations generated by \eqref{gentrans} if the constraints are set to vanish.
	The Lagrange multipliers must transform accordingly to retain invariance of the action off-shell \cite{Henneaux:1992ig}:
	\eqs{\label{lagrange mult. transformation}
		\delta u^1_{\ii_1\cdots \ii_p}=\dot{\eps}^1_{\ii_1\cdots \ii_p},\qquad \delta u^2_{\ii_1\cdots \ii_p}=\dot{\eps}^2_{\ii_1\cdots \ii_p}-\eps^1_{\ii_1\cdots \ii_p}\,.
	}
	One usually fixes the Lagrange multipliers corresponding to the secondary and higher generation constraints to zero, reverting to the total action $S_T$ which includes primary constraints only. So we may set $\de u^2=\dot{\eps}^2-\eps^1=0$ in the above,
	\begin{align}
	&\de \A_{0\ii_1\cdots \ii_p}=\dot{\eps}_{\ii_1\cdots \ii_p}, &
	\de \A_{\ii_0\cdots \ii_p}=(p+1)\partial_{[\ii_0}\eps_{\ii_1\cdots \ii_p]}.\label{canonical gauge transf.}
	\end{align}

	The gauge symmetry of the theory, $\bA\to \bA+\di\bLaa$ involves arbitrary $p$-form gauge parameter $\eps_{\mu_1\cdots\mu_p}$. However this generating set is reducible, since any gauge parameter of the form $\eps=\di\eta$ leaves the fields intact \cite{Henneaux:1986ht, Barnich:2001jy} (in \cite{Hajian:2015xlp} these were called exact symmetries). The reducibility manifests itself in identities among second-class constraints $\phi_2$:
	\eqs{
		I_{(1)}^{\ii_2\cdots \ii_p}=\partial_\kk \phi_2^{\kk\ii_2\cdots \ii_p}=0.\label{red identity 1}
	}
	The identities $I_{(1)}$ are not independent either. Taking more divergences produces a chain of identities:
	\eqs{
		I_{(n-1)}^{\ii_n\cdots \ii_p}=\partial_{\kk} I_{(n-2)}^{\kk \ii_n\cdots \ii_p},\qquad n=3,\cdots ,p+1\,.
	}
	
	Yet another way of spotting the reducibility is through the fact that Noether identities corresponding to the gauge symmetries are not independent,
	\eqs{
		\nabla_\alpha\nabla_\beta \F^{\alpha\beta \mu_1\cdots\mu_p}=0\,.
	}
	These are not independent identities, which is evident by taking further derivatives. We can identify the redundant gauge parameters among $\eps_{\ii_1\cdots \ii_p}$ in \eqref{canonical gauge transf.}. If \eqs{\eps_{\ii_1\cdots \ii_p}=p\partial_{[\ii_1}\eta_{\ii_2\cdots \ii_p]}
		\label{eta defined}
	}
	then the corresponding gauge transformation vanishes
	\eqs{
		\eps_{\ii_1\cdots \ii_p}(\text{x})[K(\text{y}),\phi_2^{\ii_1\cdots \ii_p}(\text{x})]=0
	}
	for all $K$ due to reducibility identity $I_1$ \eqref{red identity 1}. However, Lagrange multipliers are left invariant under \eqref{lagrange mult. transformation} only if $\eps_{i_1\cdots i_p}$ is time independent.  The usual procedure in dealing with reducible symmetries, however, is to  introduce new set of parameters $\eps_{0i_2\cdots i_p}$ acting only on Lagrange multipliers as
	\eqs{
		\de u^1_{\ii_1\cdots \ii_p}=-p\partial_{[\ii_1}\eps_{|0|\ii_2\cdots \ii_p]}\,.
	}
	Now the redundant transformations are characterized by $\eps_{0\ii_2\cdots \ii_p}=\dot{\eta}_{\ii_2\cdots \ii_p}$ where $\eta$ is defined in \eqref{eta defined} and is \emph{arbitrary}. With these considerations, the full set of gauge parameters of the theory contain arbitrary $p$-forms $\eps_{\mu_1\cdots \mu_p}$ in spacetime.

	At this stage we can count the propagating degrees of freedom of the theory. The number of conjugate pairs is $C(d,p+1)$. There are $C(d-1,p)$ primary and $C(d-1,p)$ secondary constraints. The $n\textsuperscript{th}$ generation of the reducibility identities $ I_{(n)}$ consists of $C(d-1,p-n)$ number of relations which should be enumerated by alternating signs. Using Pascal's identity $C(d,p+1)=C(d-1,p+1)+C(d-1,p)$ we can write the whole number as an alternating sum:
	\eqs{
		\#\text{degrees of freedom}=\sum_{k=0}^{p+1}C(d-1,k)(-1)^{p-k+1}=C(d-2,p+1)\,.
	}
	For instance, Maxwell theory $(p=0)$ in $d$ dimensions has $d-2$ degrees of freedom.
	One could make a simpler arguments using the residual gauge freedom: 
	The $(p+1)$-form gauge field in $d$ spacetime dimensions has $C(d,p+1)$ independent components. However, gauge symmetry implies that only transverse modes of the form field are propagating. The $p$-form gauge parameter enables us to remove $C(d-1,p)$ components. The residual $p$-form gauge parameters  gauge away another $C(d-2,p)$ components. The total number of degrees of freedom turns out to be,
	\eq{C(d,p+1)-C(d-1,p)-C(d-2,p)=C(d-2,p+1),}{number-of-dof}
	where the last term on the left-hand-side in \eqref{number-of-dof} is the contribution of the residual degrees of freedom.

	\section{$p$-brane solutions}
	
	\label{p-brane}
	
	The solution space we consider consists of all possible configurations of  electric $p$-branes in $d$-dimensional Minkowski space, with $d=2p+4$. 
	
	A static $p$-brane in Minkowski spacetime, extended in $x^1$ to $x^p$ directions can be described by embedding functions $x^\mu(\sigma)$ :
	\begin{equation}
	\left\{\begin{array}{l}
	x^I=\sigma^I,\qquad I=0,\cdots,p\qquad\\
	x^\mathtt{m}=0,\qquad \mathtt{m}=p+1,\cdots d-1\,.
	\end{array}	\right.
	\end{equation}
	
	It is clear that the electromagnetic field generetad by the brane, is the same as a point-like electric charge, living in the orthogonal space $\mathbb{R}^{d-p-1}$, that is
	\begin{equation}
	{\A}_{01\cdots p}=(x^\mathtt{m}x_\mathtt{m})^{\frac{3+p-d}{2}}
	\end{equation}
	up to gauge transformations. From the field strength $\F_{\mathtt{m}01\cdots p }=\partial_\mathtt{m} {\A}_{01\cdots p}$ one can define brane charges integrated on the $(d-p-2)$-sphere enclosing $p$-branes \cite{Teitelboim:1985ya,Henneaux:1986ht}:
	\begin{eqnarray}\label{gauss p-form}
	Q=\int_{S^{d-p-2}}\star\bF\,.
	\end{eqnarray}
	The brane charge flips sign if the orientation of the brane is reversed.  If there are several parallel $p$-branes in space, $\M_{d-p-1}$ intersects each one at a point, and the charge will be the total number of branes with sign $+$ or $-$ for each brane according to its orientation. For example, a couple of parallel branes with opposite orientations have zero total charge.

	The quantity \eqref{gauss p-form} counts the net electric charge of the parallel branes by integration on orthogonal space\footnote{\eqref{gauss p-form} differs from that in \cite{Henneaux:1986ht} due to different  sign conventions at the action level.
	}. The conserved charge \eqref{gauss p-form} may be directly related to the global part of gauge transformations (those with  $\extd\bLaa=0$) via the standard Noether's theorem, 
	\eq{Q_{\Lambda}=\int_{S^{d-2}}\bLaa\wedge\star\bF
		=\int_{T}\star\bF
	}{Noether-charge}
	where $T$ is the non-trivial cycle dual to $\bLaa$ \cite{Compere:2007vx}. These are generalizations of the electric charge to $p$-forms and we call them \emph{zero-mode charges.} If the branes are not parallel, there can be various non-trivial cycles, depending on how the sources are enclosed. The set of all zero-mode charges, hence not only has the information of the brane charges but also their directions.

	\section{Boundary conditions}

	\label{se:boundary conditions}
	
	In order to fully introduce the theory, we have to specify the boundary conditions on the dynamical fields. This together with the equations of motion determines the space of field configurations that defines the theory. In particular one imposes a set of boundary conditions that determine how the fields decline at infinity. Sometimes, specific \emph{gauge} conditions are also set to narrow the space of functions under consideration further.

	\paragraph{Fall-off behavior in de Sitter slicing.} According to the previous discussion, one can verify that
	the fall-off behavior of the field strength associated with a $p$-brane source for any $p$ and $d$  is the following
	\begin{subequations}
		\begin{align}
		\F_{\rho a_0\cdots a_p}=&F_{\rho a_0\cdots a_p}(x^b)\,\rho^{2p+3-d}+\ordr{2p+2-d},\\
		\F_{a_0\cdots a_{p+1}}=&F_{a_0\cdots a_{p+1}}(x^b)\,\rho^{2p+3-d}+\ordr{2p+2-d}\,.  
		\end{align}
	\end{subequations}
	where  the hyperbolic coordinates $(\rho,x^a)$ are introduced in section \ref{boundary}.
	In the presence of extended magnetic $p$-branes, the fall-offs are
	\begin{subequations}
		\begin{align}
		\F_{\rho a_0\cdots a_p}=&F_{\rho a_0\cdots a_p}(x^b)\,\rho^{-2}+\ordr{-3},\\
		\F_{a_0\cdots a_{p+1}}=&F_{a_0\cdots a_{p+1}}(x^b)+\ordr{-1}\,.
		\end{align}
	\end{subequations}
	Therefore, the fall-off that encompasses both sort of sources in $d=2p+4$ is
	\begin{subequations}\label{bnyc1}
		\begin{align}
		\F_{\rho a_0\cdots a_p}=&F_{\rho a_0\cdots a_p}(x^b)\,\rho^{-1}+\ordr{-2},\\
		\F_{a_0\cdots a_{p+1}}=&F_{a_0\cdots a_{p+1}}(x^b)+\ordr{-1}\,.  \label{bnyc0}
		\end{align}
	\end{subequations}
	The boundary conditions on gauge fields compatible with the above are
	\begin{subequations}\label{bnyc2}
		\begin{align}
		\A_{\rho a_1\cdots a_p}&=A_{\rho a_1\cdots a_p}(x^b)\,\rho^{-1}+\ordr{-2},\\
		\A_{a_0\cdots a_{p}}&=A_{a_0\cdots a_{p}}(x^b)+\ordr{-1}\,.
		\end{align}\label{bnyc23}
	\end{subequations}
	We note that \eqref{bnyc2} is not the only possibility following from \eqref{bnyc1}. In principle, we could have chosen a weaker fall-off behavior for allowed gauge transformations than the background gauge fields $\A_{\mu_0\cdots\mu_p}$. Such `leading' gauge transformations do not yield a finite charge.
	
	\subsection{Conserved symplectic form}\label{variational-section}

	First, note that the boundary conditions imply that $A_{\rho a_1\cdots a_p}$  completely determines the $\rho$-component of the field strength tensor
	\eqs{
		F_{a_0\cdots a_p \rho }=(p+1)\partial^{}_{[a_0}A_{a_1\cdots a_p]\rho}\,.\label{ap-coex-Field}
	}
	Next, consider the boundary flux of the symplectic form
	\begin{align}\label{boundary remain}
	\int_\Gamma\omega=-\frac{1}{(p+1)!}\int_{\Gamma}\sqrt{-g}\,n_{\alpha}\, &\delta \A_{\mu_0\cdots \mu_p} \delta\F^{\alpha\, \mu_0\cdots \mu_p}\\
	&\qquad\qquad=\frac{1}{p!}\int_{\Gamma}\sqrt{-h}\, \delta A^{a_0\cdots a_p} \delta\partial_{a_0} A_{\rho\,a_1\cdots a_p}\nn.
	\end{align}
	where $\mathbf{n}=\partial_\rho$ is the unit vector normal to the $\rho=\rho_0$  hypersurface. This boundary term is generically non-zero for our boundary conditions and spoils our initial-value problem. However, we could get rid of this term by fixing a gauge condition and introducing the boundary term. Consider the Lorenz gauge condition
	\begin{equation}\label{Lorenz}
	\nabla^\mu\A_{\mu\nu_1\cdots\nu_p}\,.
	\end{equation}
	Given boundary conditions \eqref{bnyc2}, the leading components are constrained as
	\begin{subequations}
		\begin{align}
		D^{a_0}A_{a_0a_1\cdots a_p}+2A_{\rho a_1\cdots a_p}&=0\,,\label{Lorentz gauge split2}\\
		D^{a_1}A_{\rho a_1\cdots a_p}&=0\,.\label{LG-22}
		\end{align}
	\end{subequations}
	Fixing the gauge only at leading order will allow an improved conserved symplectic form.
	\begin{pro}
		A conserved symplectic form for $(p+1)$-form theory in $d=2p+4$ dimensions is
		\begin{eqnarray}\label{sympP}
		\Omega=\int_\Sigma \delta\bA\wedge\delta\star\bF-\oint\sqrt{g} \delta\A^{\tau i_1\cdots i_p}\delta \A_{\rho i_1\cdots i_p}
		\end{eqnarray}
	\end{pro}

	\subsection{Action principle}
	In a well-defined initial value problem, one fixes the initial and final values of the variable. We need to check that the boundary terms arising in variation of the action on the timelike boundary $\Gamma$ vanishes. That is
	\begin{equation}
	\delta S\big|_\Gamma=  \frac{1}{p!}\int_{\Gamma}\sqrt{-h}\, \delta A^{a_0\cdots a_p} \partial_{a_0} A_{\rho\,a_1\cdots a_p}
	\end{equation}
	By fixing the \textbf{Lorenz gauge} condition asymptotically (i.e. at the leading order in$\rho$), this is a total variation
	and it is sufficient to add the boundary term $S_b$ to the original action \eqref{pformaction} as,
	\begin{equation}\label{Bterm0}
	S_b =-\frac{1}{p!}\int_{\Gamma}\sqrt{g} (\mathbf{n}\cdot\bA)_{\mu_1\cdots \mu_p}(\mathbf{n}\cdot\bA)^{\mu_1\cdots \mu_p}\,.
	\end{equation}
	to ensure  having a well-defined action principle for the $(2p+4, p)$-form theory. Note that  boundary conditions \eqref{bnyc2} and the Lorenz gauge \eqref{Lorenz} are necessary for this conclusion.\footnote{Such a boundary term appears also in the context of $AdS_2$ holography \cite{Castro:2008ms}.} 
	The improved action of $(2p+4,p)$-form theory is thus,
	\begin{equation}\label{improvedaction}
	S=-\frac{1}{2(p+2)!}\int\sqrt{-g}\,\F_{\mu_0\cdots\mu_{p+1}}^2-\frac{1}{p!}\int_{\partial {\Gamma}}\sqrt{-g}\,A^{\rho a_1\cdots a_p }\cdot A_{\rho a_1\cdots a_p}\,.
	\end{equation}

	Although we have fixed the boundary fall-off behavior of our fields, we may need further restrictions on the fields to make sure that the on-shell action (i.e. the boundary term) is a finite quantity for any consistent initial and final data. This will be discussed in section \ref{2-form theo}.

	\section{Soft charges}\label{soft sect}
	In this section, we compute surface charges for large gauge transformations. There are two sets of them: exact charges and coexact charges, based on the geometrical character of the gauge parameter. Exact charges are a feature of higher-form theories $p>0$,
	
	The boundary conditions are preserved by the following fall-off for the gauge parameters
	\begin{subequations}
		\begin{align}
		\Lambda_{\rho a_2\cdots a_{p}}&=\chi_{a_2\cdots a_{p}}(x^b)\,\rho^{-1}+\ordr{-2},\label{bpgt2}\\
		\Lambda_{a_1\cdots a_{p}}&=\lambda_{a_1\cdots a_{p}}(x^b)+\ordr{-1}\,.\label{bpgt3}
		\end{align}
	\end{subequations}
	and Lorenz gauge demands
	\begin{eqnarray}
	\nabla^\mu(\nabla_\mu\Lambda_{\nu_1\cdots\nu_p}- p\nabla_{[\nu_1}\Lambda_{|\mu|\nu_2\cdots \nu_p]})=0\,.
	\end{eqnarray}
	Fix the reducibility by a temporal gauge for the parameter  $\Lambda_{\tau \nu_1\cdots \nu_{p-1}}=0
	$. The remaining components are $\Laa_{i_1\cdots i_p}$ and $\Laa_{\rho i_1\cdots i_{p-1}}$. At the leading order,
	they satisfy
	\begin{subequations}\label{resideq}
		\begin{align}
		D^{a_0}D_{[a_0}\laa_{a_1\cdots a_p]}+2D_{[a_1}\chi_{a_2\cdots a_p]}&=0\,,\label{Lorentz gauge split}\\
		D^{a_1}D_{[a_1}\chi_{a_1\cdots a_p]}&=0\,.\label{LG-2}
		\end{align}
	\end{subequations}
	To solve the equations, 
	decompose $\blaa$ into exact and coexact parts
	\begin{equation}
	\blaa=\underbrace{\hat{\blaa}}_{\text{coexact}}+\underbrace{\di\beps}_{\text{exact}}\qquad\qquad \di\star\hat{\blaa}=0
	\end{equation}
Here we are using the Hodge decomposition theorem \cite{Folland1989} which states that a $p$-form on a compact $n$-manifold can be uniquely decomposed into exact, coexact and harmonic parts (see Appendix \ref{AppDiff} for details.) The only harmonic forms on an $n$-sphere are zero-forms and top forms. Since we are focusing on $p$-forms on $2p+2$ spheres, no harmonic froms appear in the decomposition.	
	\subsection{Coexact parameters}
	If the parameter is a coexact $p$-form, the field variation is
	\begin{subequations}
		\begin{align}
		\delta_{\hat{\laa}}A_{i_{1}\cdots i_{p+1}}&=(p+1)\partial_{[i_1}\hat{\laa}_{i_2\cdots i_{p+1}]}\\
		\delta_{\hat{\laa}}A_{\tau i_{1}\cdots i_{p}}&=\partial_\tau\hat{\laa}_{i_1\cdots i_{p}}\\
		\delta_{\hat{\laa}}A_{\rho i_{1}\cdots i_{p}}&=0
		\end{align}
	\end{subequations}
	From \eqref{Lorentz gauge split}, the coexact parameter is subject to
	\begin{equation}\label{laeq}
	(1-y^2)\hat{\laa}_{i_{1}\cdots i_{p}}^{\prime\prime}
	+\Delta\hat{\laa}_{i_{1}\cdots i_{p}}=0,\qquad y=\cos\tau.
	\end{equation}
	This is the same equation satisfied by $A_{\rho i_{1}\cdots i_{p}}$. 
	The conserved charge follows from the symplectic form \eqref{sympP}
	\begin{eqnarray}\label{pcochar}
	Q_{\hat{\laa}}=\frac{1}{p!}\oint \sqrt{h}\Big[\hat{\laa}_{i_1\cdots i_p}F^{\tau\rho i_1\cdots i_p}+\partial_\tau\hat{\laa}_{i_1\cdots i_{p}}A^{\rho i_{1}\cdots i_{p}}\Big].
	\end{eqnarray}
	This quantity is conserved as can be explicitly checked by taking a time derivative and using \eqref{laeq}.

	The physical significance of this quantity is better understood if  we consider parallel planar $p$-branes in $x^1$, $x^2, \cdots x^p$ directions and reduce the integration to the $(d-p-2)$-sphere surrounding the branes. Defining the orthogonal distance to the branes $h^2=x_mx^m$, and the associated hyperbolic coordinates $(\tilde{\rho},\tilde{\tau})$, one will find the following charges
	\begin{eqnarray}\label{pcochar11}
	Q_{\laa}=\oint_{S^{d-p-2}} \sqrt{\tilde{h}}\Big[\laa(\varphi^m)F^{\tilde{\tau}\tilde{\rho} 1\cdots p}+\partial_\tau\laa(\varphi^m)A^{\tilde{\rho} 1\cdots p}\Big]
	\end{eqnarray}
	This is the conserved charge of Maxwell theory in $d-p$ dimensions as expected: $p$-branes are seen as point particle from the orthogonal space.

	In principle, the general expression \eqref{pcochar} contains a larger set of charges since it is rotation-invariant and any alignment of the branes is admissible. In contrast, \eqref{pcochar11} is only applicable to branes in $x^1\cdots x^p$ directions.
	
	The field strength tensor for extended $p$-branes is not smooth on 4-sphere. For example, if $p=1$, the source will puncture the sphere at the two poles. This is an obstruction to the application of the Hodge decomposition theorem. For the same example, one has
	\begin{eqnarray}
	F_{tr\theta}=\frac{1}{r^{3}\sin^{3}\theta}
	\end{eqnarray}
	This 1-form on 4-sphere is both closed and co-closed (divergence-free), which means that it is a harmonic form, which contradicts the Hodge theorem.
	
	\subsection{Exact parameters}
	In this case, the gauge parameter consists only of $(p-1)$-forms $\epsilon$ and $\chi$, and the field variations are
	\begin{subequations}
		\begin{align}
		\delta_{\epsilon}A_{i_{1}\cdots i_{p+1}}&=0\\
		\delta_{\epsilon}A_{\tau i_{1}\cdots i_{p}}&=p\partial_\tau\partial_{[i_1}\epsilon_{i_1\cdots i_{p}]}\\
		\delta_{\epsilon}A_{\rho i_{1}\cdots i_{p}}&=p\partial_{[i_1}\chi_{i_2\cdots i_{p}]}
		\end{align}
	\end{subequations}
	$\epsilon$ and $\chi$ are related by the	Lorenz gauge. From equations \eqref{resideq} we have
	\begin{align}\label{chieq}
	(1-y^2)\chi_{i_{1}\cdots i_{p-1}}^{\prime\prime}
	+2y\chi_{i_{1}\cdots i_{p-1}}^{\prime}
	+\Delta\chi_{i_{1}\cdots i_{p-1}}&=0
	\end{align}
	\begin{align}
	\Delta\epsilon_{i_1\cdots i_{p-1}}^{\prime}&=\frac{2\chi_{i_1\cdots i_{p-1}}^\prime }{1-y^2}\,,\quad
	\epsilon_{i_1\cdots i_{p-1}}^{\prime\prime}=-\frac{2\chi_{i_1\cdots i_{p-1}}}{(1-y^2)^2}\,.
	\end{align}
	The surface charge for this variation is finite, given by
	\begin{equation}
	Q_{\epsilon}=-\frac{1}{p!}\oint\sqrt{h} \Big[\partial_\tau\di\epsilon_{i_1\cdots i_{p}} \A^{\rho i_1\cdots i_p}-\A^{\tau i_1\cdots i_p}\partial_{i_1}\chi_{i_2\cdots i_{p}}\Big].
	\end{equation}
	To this end, notice that by a Hodge decomposition of the potential as
	\begin{equation}
	\A_{\rho i_1\cdots i_p}=p\partial_{[i_1}\mathcal{C}_{i_2\cdots i_p]}+\hat{\A}_{\rho i_1\cdots i_p}
	\end{equation}
	one can write the charge in terms of $\mathcal{C}$ and $\chi$ only
	\begin{equation}\label{exact charge p}
	Q_{\epsilon}=\frac{2}{(p-1)!}\oint\sqrt{h} \Big[\partial_\tau\chi_{i_1\cdots i_{p-1}} \C^{ i_1\cdots i_{p-1}}-\partial_\tau\C_{ i_1\cdots i_{p-1}}\chi_{i_1\cdots i_{p-1}}\Big]
	\end{equation}
	Note that $\mathcal{C}_{i_1\cdots i_{p-1}}$ does not appear in the field strength tensor. Thus, the exact charge might be non-vanishing even in the absence of electromagnetic $(p+1)$-form fields.
	This however does not mean that the exact charges are conserved off-shell. 
	To see the role of the field equations in the conservation of exact charges, 	take all the free indices in \eqref{ap-coex-Field} on $(2p+2)$-sphere:
	\begin{align}\label{ap-co-Bianchi}
	F_{\rho i_1\cdots i_{p+1}}&=-(p+1)\partial_{[i_1}\hat{A}_{\rho i_2\cdots i_{p+1}]}\\
	F_{\rho\tau i_1\cdots i_p}&=-\partial_\tau \hat{A}_{\rho i_1\cdots i_p}-p\partial^{}_{[i_1}(\partial_\tau\mathcal{C}-A_{\rho\tau})_{i_1\cdots i_p]}.
	\end{align}
	Now, the equation of motion 
	\eqs{\label{ap-co-eom-rhotau}
		\mathcal{D}^{i_1}F_{\rho\tau i_1\cdots i_p
		}=0
	}
	implies that
	\begin{equation}\label{C def}
	A_{\rho\tau i_1\cdots i_p}= \partial_\tau\mathcal{C}_{i_1\cdots i_p}.
	\end{equation}
	This is important in proving conservation of the exact charges, since by \eqref{LG-2} one concludes that
	\begin{align}\label{Ceqm}
	(1-y^2)\mathcal{C}_{i_{1}\cdots i_{p-1}}^{\prime\prime}
	+2y\mathcal{C}_{i_{1}\cdots i_{p-1}}^{\prime}
	+\Delta\mathcal{C}_{i_{1}\cdots i_{p-1}}&=0.
	\end{align}
	%

	\subsection{Algebra of charges}\label{charge-algebra-sec}
	The expression for the charge may  be viewed as a functional over the phase space of form-field configurations. One can then compute algebra of charges (i.e. Poisson bracket of charges over the phase space). In the coexact sector,  these charges commute among themselves, since $\delta_\lambda \hat{A}_{\rho i}=0$ 
	\begin{align}
	\{ Q^{\text{\tiny coexact}}_{\laa_1},Q^{\text{\tiny coexact}}_{\laa_2}\}=-\delta^{}_{\laa_1} Q^{(\text{\tiny coexact})}_{\laa_2}=0\,.
	\end{align}
	
	In the exact sectors,   the gauge transformation on $\C$ (and $A_{\rho \tau}$) is non zero and acts on $\C$ as a shift
	\eqs{
		\C_{i_1\cdots i_{p-1}}\to \C_{i_1\cdots i_{p-1}}-\chi_{i_1\cdots i_{p-1}}\,.
	}
	We consequently find
	\begin{align}\label{exact-commutator}
	\{
	Q^{\text{\tiny exact}}_\chi,Q^{\text{\tiny exact}}_{\tilde{\chi}}\}
	&=-\delta_\chi Q^{(\text{\tiny exact})}_{\tilde{\chi}}\nn\\
	&=2\int_{S^4}\sqrt{\mathscr{G}}\Big[-\chi^{i_1\cdots i_{p-1}}\partial_\tau{\tilde{\chi}}_{i_1\cdots i_{p-1}}+{\tilde{\chi}}^{i_1\cdots i_{p-1}} \partial_\tau\chi_{i_1\cdots i_{p-1}}
	\Big]\,.
	\end{align}
	The charge algebra can hence be non-Abelian only if the $\rho$-component of the gauge parameters are non-zero. Moreover, the right-hand-side of\eqref{exact-commutator}, being independent of the gauge field $\A$, is a $c$-number over the phase space; i.e. a central term. In section \ref{2-form theo} we will explicitly compute the charges as well as their algebra for 2-form theory in six dimensions and find the corresponding central charge.

	\subsection{Magnetic charges}\label{sec-magnetic}
	
	For a $(p+1)$-form theory in $2p+4$ dimensions, the Hodge star operator maps the $(p+2)$-form field strength $\bF_{p+2}$ to its Hodge dual which is another $(p+2)$-form, 
	\begin{equation}
	\bF_{p+2}\to \star \bF_{p+2}.
	\end{equation}
	The source-free equations of motion $\di\star \bF=0$ which are relevant to the asymptotic region, will allow for a magnetic potential $\star \bF_{p+2}=\di{\tilde{\bA}}_{p+1}$. The action is invariant under `magnetic' gauge transformations ${\tilde{\bA}}_{p+1}\to{\tilde{\bA}}_{p+1}+\di {\tilde{\bLaa}}_p$, so we can ask about the conserved charges corresponding to this gauge symmetry. 
	
	The first question is whether the magnetic charges contain independent information about the fields. We have seen that the gauge potential can be decomposed into de Sitter differential forms:
	\begin{equation}\label{dsdecomposition}
	\A_{\nu_0\cdots \nu_{p}}:\qquad \left\{
	\begin{array}{ll}
	A_{\rho\,a_1\cdots a_p}& \ \text{de Sitter $p$-form}\,,\\
	A_{a_0\cdots a_p}& \ \text{de Sitter $(p+1)$-form}\,.
	\end{array}
	\right.
	\end{equation}
	We showed above that the conserved charges are built out of the de Sitter $p$-form $A_{\rho\,a_1\cdots a_p}$ and its corresponding field strength, independent of the de Sitter $(p+1)$-form $A_{a_0\cdots a_p}$ in \eqref{dsdecomposition}. Similarly, the magnetic charges involve only the $\tilde{A}_{a_1\cdots a_p\rho}$ components which are related to the electric gauge potentials by
	\begin{equation}
	(\di {\tilde A})_{a_0\cdots a_p\rho}=\frac{1}{(p+2)!}{\epsilon^{b_0\cdots b_{p+1}}}_{a_0\cdots a_p\rho}(\extd\A)_{b_0\cdots b_{p+1}}\,.
	\end{equation}
	Thus, the magnetic charges extract the information contained in $A_{a_0\cdots a_p}$ being missed by the electric charges.
	
	The expression for the magnetic charges and their conservation follows exactly alongside the discussions we had about electric charges, providing that the same boundary conditions as in  \eqref{bnyc23} are satisfied by the magnetic potential $\tilde{A}$.

	
	\section{$2$-form theory in six dimensions} \label{2-form theo}
	
	In this section we study the simplest yet non-trivial case with charges \eqref{exact charge p} associated to exact gauge transformation on, the $p=1$ case. The improved action is
	\begin{equation}\label{(6,1) improved action}
	S=-\frac{1}{12}\int_{\mathcal M}\sqrt{-g}\F_{\mu\nu\alpha}\F^{\mu\nu\alpha}-\int_{\Gamma}\sqrt{-h}\,A^{\rho a} A_{\rho a}\,,
	\end{equation}
	with the boundary conditions,
	\begin{equation}
	\A_{ \rho a}=\frac{A_{\rho a}}{\rho}+\ordr{-2}\,,\qquad \A_{ab}=A_{ab}+\ordr{-1}\,.\label{(1,6) gauge field b.c.}
	\end{equation}
	To leading order in $\rho$, the field equations and the Lorenz gauge condition are,
	\begin{subequations}
		\begin{align}
		D_cF^{cab}=0\,, \qquad &D^bF_{\rho ba}=0\,,\label{transvers eom}\\
		D_b A^{ba}+2A^{\rho a}=0
		\,,\qquad  &D^aA_{\rho a}=0\,,
		\label{radial-eom}
		\end{align}\label{eom16}
	\end{subequations}
	and we have  the $\rho$-component of the field 
	\begin{equation}\label{fabr16}
	F_{ab\rho}=2\partial_{[a}A_{b]\rho}\,,
	\end{equation} 
	which means  $F_{ab\rho}$ is a closed 2-form on $dS_5$ with the potential $A_{\rho a}$. 
	
	\paragraph{Exact, coexact and zero-mode conserved charges.}\label{6dcharges}
	
	The parameters of the boundary condition preserving gauge transformations
	at leading order in $\rho$
	generate the following transformations on the boundary 
	\eqs{
		\delta A_{a\rho}=\partial_a\laa_\rho\,,\qquad \delta A_{ab}=2\partial_{[a}\laa_{b]}\,.
	}
	They include  a de Sitter scalar $\lambda_\rho$ and a de Sitter vector $\laa_a$ which satisfy \eqref{resideq} in temporal gauge,
	\begin{subequations}\label{Lambda eq}
		\begin{align}
		&D^aD_{[a}\laa_{b]}=D_b\laa_\rho\,, \label{residual16}\\
		&\laa_\tau=0\,.\label{residual163}
		\end{align}
	\end{subequations}
	
	\paragraph{Coexact charges.} For constant $\tau$ slices and in the temporal gauge, the expression for coexact charges  \eqref{(6,1) coexact charge} simplifies to
	\begin{align}\label{(6,1) coexact charge}
	Q^{\text{\tiny coexact}}_{\laa}
	=-\int_{S^4}\sqrt{h}\Big(\partial_\tau\hat{\laa}_{i}\hat{A}^{\rho i} -\hat{\laa}^i\, \partial_\tau  {\hat{A}^\rho}_{i}\Big),
	\end{align}
	where in the first term above we used  \eqref{fabr16}.
	To compute the explicit expression of the charges we need to solve the equations for gauge potentials and gauge parameters. 
	The field equation \eqref{transvers eom} has an exact and a coexact part which are linearly independent and should be individually zero. For $\hat{A}_{\rho i}$ being coexact, it  simplifies as
	\begin{align}
	&(1-y^2)\hat{\laa}^{\pp}_i+\Delta \hat{\laa}_i
	=0\qquad \qquad
	y=\cos\tau\label{laa eq2}\,,\\
	&(1-y^2)\hat{A}^\pp_{\rho i}+\Delta \hat{A}_{\rho i}=0\,,\label{non-exact eq}
	\end{align}
	with solutions
	\eqs{
		\hat{A}_{\rho i}(y,\hat{x})=(1-y^2)^{\frac{1}{2}}\sum_{lm_\alpha}\omega_i^{lm_\alpha}(\hat{x})\Big[b^{(1)}_{lm_\alpha}P^1_{l+1}(y)+ b^{(2)}_{lm_\alpha}Q^1_{l+1}(y)
		\Big]\,\label{coexact solution}
		,}
	where $ \omega_i^{lm_\alpha}$ are coexact eigen 1-forms of the Laplace-Beltrami operator on $S^4$;
	\eq{\Delta_H\omega^{lm_\alpha}_i=[l(l+3)+2]\omega_i^{lm_\alpha}\qquad\text{with}\qquad l\geq 1\,.}{}

	In the coexact sector, both the gauge parameter $\hat{\laa}_i$ and the gauge field $\hat{A}_{i\rho}$ satisfy same equations \eqref{laa eq2} and \eqref{non-exact eq} with a general solution \eqref{coexact solution}. As in the coexact charges of the Maxwell theory in four dimensions, discussed in section \ref{Minkowski Maxwell}, not all terms in  \eqref{coexact solution} keep our boundary term \eqref{Bterm0} finite. 
	It turns out that  $Q^m_l$ solutions in \eqref{PQfunctions} are not square-integrable and make our boundary term \eqref{Bterm0} divergent and we discard them. 	Moreover, the finite contribution to the charge comes from the $Q^1_{l+1}(y)$ term in $\hat{\lambda}_i$:
	\begin{align}
	\hat{A}_{i\rho}(y,\hat{x})&=(1-y^2)^{\frac{1}{2}}\sum_{lm_\alpha}\omega_i^{lm_\alpha}(\hat{x})\Big[\frac{b_{lm_\alpha}}{\sqrt{(l+1)(l+2)}}P^1_{l+1}(y)\Big]\,,\\
	\hat{\laa}_i(y,\hat{x})&=(1-y^2)^{\frac{1}{2}}\sum_{lm_\alpha}\omega_i^{lm_\alpha}(\hat{x})\Big[\frac{\laa_{lm_\alpha}}{\sqrt{(l+1)(l+2)}}Q^1_{l+1}(y)\Big]\,,
	\label{laaB solution}
	\end{align}
	for $l\geq 1$.  This choice will determine the antipodal parity of the fields as explained below. Before delving into it, in order to motivate the antipodal boundary condition physically, note that the causal connection between points on $\I^+$ and $\I^-$ as boundaries of the de Sitter slices is made via null geodesics beginning on the sphere at $\I^-$ and reaching its antipode at $\I^+$ \cite{Strominger:2001pn}. This would verify the antipodal matching property of the field strength for generic $p$-form theories,
	\eqs{
		{F}_{\rho\tau i_1\cdots i_p}(\pi-\tau,-\hat{x})=
		F_{\rho\tau i_1\cdots i_p}(\tau,\hat{x})\,.\label{antipodality}} 
	This condition restricts the $p$-form $\hat{A}_{\rho i_1\cdots i_p}$ to its one branch of solutions similar to \eqref{coexact solution} and satisfying \eqref{non-exact eq} on $S^{2p+2}$. In particular, the time-dependence will be given  by $P^1_{l+p}(y)$ while the spacial-dependence on $S^{2p+2}$ is governed by $\omega^{lm_\alpha}_{(p)}(\hat{x})$ with the following parity transformations \cite{Folland1989}
	\begin{align}
	\label{paritytransformp}
	P^1_{l+p}(-y)&=(-1)^{l+p+1}P^1_{l+p}(y)\,,\nn\\
	\omega^{lm_\alpha}_{(p)}(-\hat{x})&=(-1)^{l+p}\omega^{lm_\alpha}_{(p)}(\hat{x}),\quad l\geq1\,.
	\end{align}
	Thus, $\hat{A}_{\rho i_1\cdots i_p}$ is odd under PT; implying that the field strength $\hat{F}_{\rho\tau i_1\cdots i_p}$ is even as in \eqref{antipodality}. Similarly one can confirm that the gauge parameters entering the coexact charges are also even.
	\eqs{\label{antip}
		Q^{+}_{{\laa}_+}[{\A}]=Q^{-}_{{\laa}_-}[{\A}]\,.
	}
	Now the expression of the charge in \eqref{(6,1) coexact charge} is the Wronskian of \eqref{non-exact eq} and by orthonormality of $\omega_i^{lm_\alpha}$ it yields
	\eqs{
		Q^{\text{\tiny coexact}}_\laa[\A]=\sum_{l\geq 1, m_\alpha} b_{lm_\alpha}\laa^\ast_{lm_\alpha}\,.
	}

	\paragraph{Exact charges.} These charges are associated with gauge transformations generated by the de Sitter scalar $\laa_\rho\equiv\chi$ and the de Sitter vector $\lambda_a$ which are related via solving \eqref{Lambda eq}.
	Expression for the charges \eqref{exact charge p} with the gauge parameters being exact on $S^4$ takes the form
	\begin{align}\label{(6,1) exact charge}
	Q^{\text{\tiny exact}}_{\laa}[\A]
	=2\int_{S^4}\sqrt{h}\,\Big( \partial_\tau\chi \C-\chi \partial_\tau\C\Big)\,,
	\end{align}
	The commutator of exact charges with the coexact ones are zero,
	\begin{align}
	\{ Q^{\text{\tiny coexact}}_\epsilon,Q^{\text{\tiny exact}}_\laa\}=0\,.
	\end{align}

	The more interesting part is the commutator of exact charges: the central charge for the exact charge sector turns out to be non-zero. The reason is simply that  the gauge transformation on $\C$ (and $A_{\rho \tau}$) is non zero and acts on $\C$ as a shift
	\eqs{
		\C\to \C-\chi.
	}
	We consequently find
	\begin{align}\label{exact-charge-commutator}
	\{
	Q^{\text{\tiny exact}}_\chi,Q^{\text{\tiny exact}}_{\tilde{\chi}}\}
	=-\delta_\chi Q^{(\text{\tiny exact})}_{\tilde{\chi}}
	=2\int_{S^4}\sqrt{\mathscr{G}}\Big[-\chi\partial_\tau{\tilde{\chi}}+{\tilde{\chi}} \partial_\tau\chi
	\Big]\,.
	\end{align}
	
	From \eqref{chieq} we have
	\begin{subequations}\label{laa-6d}
		\begin{align}
		(1-y^2)\chi^\pp+2y\chi^\p+\Delta\chi&=0\,,
		\label{laa eq4}
		\end{align}
	\end{subequations}

	Expanding in eigen-modes  $\mathcal{D}^2\chi=\sum_{l\geq0} l(l+3)\chi_l$   one gets,
	\begin{align}\label{chisol}
	\chi(y,\hat{x})=&\chi^{(1)}_0+\chi^{(2)}_0(\tfrac{y^3}{3}-y)\nn\\
	&\qquad+ (1-y^2)\sum_{l>0, m_\alpha}\big[\chi^{(1)}_{lm_\alpha} {\cal P}_{lm_\alpha}(y,\hat{x})+ \chi^{(2)}_{lm_\alpha}{\cal Q}_{lm_\alpha}(y,\hat{x})\big],
	\end{align}
	where ${\cal P}, {\cal Q}$ are the following functions,
	\begin{subequations}\label{PQfunctions}
		\begin{align}
		{\cal P}_{lm_\alpha}(y,\hat{x})&\equiv\sqrt{\frac{2(l-1)!}{(l+3)!}}Y_{lm_\alpha}(\hat{x})P^2_{l+1}(y)\,,\\
		{\cal Q}_{lm_\alpha}(y,\hat{x})&\equiv
		\sqrt{\frac{2(l-1)!}{(l+3)!}}Y_{lm_\alpha}(\hat{x})Q^2_{l+1}(y)\label{normalized-modes}\,, 
		\end{align}
	\end{subequations}
	with $l\geq 1$. These functions on $dS_5$ are normalized as:
	\eq{
		\int \di y\di^4\hat{x}\sqrt{-h}(1-y^2)^2 {\cal P}_{lm_\alpha}(y,\hat{x}) {\cal P}_{l'm'_\alpha}(y,\hat{x})=\de_{ll}\delta_{m_\alpha,m'_\alpha},
	}{normalization-PQ}

	In contrast to the coexact part,   for exact gauge fields, the boundary term \eqref{Bterm0} is a total derivative, thus we need not disallow one of the branches allowed by equations of motion.\footnote{The boundary term $S_b$ in this case describes a massless scalar on $dS_{5}$ which can be regularized by holographic renormalization means \cite{deHaro:2000vlm}.}
	On the other hand, unlike the coexact case, both $\lambda^{(1)}_{lm_\alpha}$ and $\lambda^{(2)}_{lm_\alpha}$ modes in \eqref{chisol} can contribute to exact charges and we hence have two sets of exact charges, which will conveniently be denoted by $Q^{(a)}_{lm_\alpha}, a=1,2$.  In this respect, the exact charges are different than the coexact and zero-mode charges. However, one class of the parameters $\lambda^{(a)}_{lm_\alpha}$, say $a=1$ leads to non-zero charges only if $\mathcal{C}$ belongs to the opposite class, which has opposite behavior under PT. 
	As a result, the exact charges exhibit antipodal matching property too.

	The more interesting part is, however, the algebra of the exact charge sector, for which we need to evaluate \eqref{exact-charge-commutator}, by the normalized Legendre functions \eqref{normalized-modes} and \eqref{normalization-PQ}
	\be\label{charge-algebra}
	\{
	Q^{(a)}_{lm_\alpha},Q^{(b)}_{ l^\prime m^\prime_\alpha}\}=4\de _{ll^\prime}\de _{m_\alpha,-m_\alpha^\prime}
	\epsilon^{ab},\qquad l\geq 1, \quad a,b=1,2,
	\ee
	where $\epsilon^{ab}$ is the anti-symmetric symbol. As mentioned $\lambda^{(1)}_{lm_\alpha}, \lambda^{(2)}_{lm_\alpha}$ behave oppositely under parity. Nonetheless, the expression of commutator of exact charges \eqref{exact-charge-commutator} involves a time derivative and hence the expression receive a non-zero contribution for $Q^{(1)}_{lm_\alpha}, Q^{(2)}_{lm_\alpha}$ charges.

	\paragraph{Zero-mode charges.}
	We  discussed above exact gauge parameters with $\laa_\rho\neq 0$. It remains to consider exact gauge parameters with $\laa_\rho=0$. For this class, $(\di\laa)_{ij}=0$ on the sphere, so they leave the gauge parameter invariant; they are \emph{exact symmetries} \cite{Hajian:2015xlp} and the charge reduces to 
	\be\label{zero-mode-6d}Q^{\text{\tiny zero-mode}}_\laa=\int_{S^3}\star \bF.
	\ee
	These charges are non-vanishing only in the presence of sources that pierce the celestial sphere, like infinite strings. Zero-mode charges obviously commute with all charges in the theory.

	\section{Discussion}
	Let us summarize this chapter with a few comments
	\begin{itemize}
		\item 
		There are three classes of charges for the $(p+1)$-form theory in $2p+4$ dimensions. The zero-mode charges are specified by time-independent $(p-1)$-forms (the $\epsilon$) on $S^{2p+2}$.  The exact and coexact charges are respectively specified by exact and coexact 1-forms on the $S^{2p+2}$. There is one set of coexact charges but two sets of exact charges. The zero-mode and coexact charges commute with all other charges and only the exact charges of a different kind do not commute. Their commutator is given in \eqref{charge-algebra} which is an infinite-dimensional Heisenberg algebra. In deriving the algebra \eqref{charge-algebra} we assumed $\lambda$ is independent of the gauge field $\A$. One may construct other algebras through a quadratic combination of these charges associated with linearly field-dependent gauge parameters \cite{Grumiller:2019fmp,Afshar:2016wfy,Afshar:2016kjj}. 

		\item The coexact and zero-mode charges have analogs in the usual $4d$ Maxwell theory, but the exact charges are new features.
		As a comment on the physical meaning of these exact charges, we note that the same expression as in \eqref{(6,1) exact charge} appears for electric conserved charges in the 6d Maxwell theory with the same boundary conditions given in \eqref{bnyc1}. Fields that appear in exact charges are flat connections \eqref{C def}. For the same reason, we expect that these charges can not be excited by soft radiation and be measured by memory effects.

		\item $p$-form zero-mode  charges are known as higher form global symmetries \cite{Lake:2018dqm,Hofman:2018lfz,Gaiotto:2014kfa}. For a $p$-form global symmetry, charged operators have $p$ space-time dimensions, like Wilson lines and  surface defects, and  charged excitations have $p$ spatial dimensions, like strings, membranes, etc. For $p=0$,  the  symmetry operator is the integration of Noether current  on $(d-1)$-dimensional surfaces. Similarly, $p$-form global symmetries are associated with $(d-p-1)$-form currents integrated on a hypersurface $M^{(d-p-1)}$. For two elements $g, g^\prime$ of $p$-form global symmetry group, there exist topological operators $U(M^{(d-p-1)})
		$ such that
		\begin{eqnarray}
		U_g(M^{(d-p-1)})U_{g^\p}(M^{(d-p-1)})=U_{gg^\p}(M^{(d-p-1)})\,.
		\end{eqnarray}
		The simplest examples are electric and magnetic 1-form symmetries in four dimensional $U(1)$ gauge theory
		
		\begin{align}
		U^E(M^{(2)})_{g=e^{i\alpha}}&=\exp\Big(\frac{i\alpha}{2\pi}\int_{M^{(2)}}\star\bF\Big)\\
		U^M(M^{(2)})_{g=e^{i\alpha}}&=\exp\Big(\frac{i\alpha}{2\pi}\int_{M^{(2)}}\bF\Big)\,.
		\end{align}
		The symmetry group is abelian in $p>0$ case \cite{Gaiotto:2014kfa,Henneaux:1986ht,Bekaert:2000qx}. Higher-form symmetries are powerful tools in studying non-perturbative dualities:  despite gauge symmetries in different formulations need not match, global symmetries must match. Symmetry breaking and anomalies in higher form symmetries have also been studied.
	\end{itemize}

	\chapter{String Memory Effect}\label{memory chapter}
	This chapter is devoted to the memory effect of 2-form soft radiation. 2-forms couple to strings so in principle there are both open and closed string memory effects, the former being explained in the next chapter. The special feature of the 2-form memory effect is both the test object (string) and the way it is influenced: memory is encoded in the internal states of the string probe, not on its position and momentum. This is the first example of internal memory.
	The 2-form memory exists in the whole tower of the string state. 
	
	It is worth mentioning that there is no reference here to any string theory. As far as memory effect is concerned, there is no preferred string length and the probe can have large classical values. However, even for the strings at the Planck scale, one can detect the memory on massless states: the graviton, the NSNS two-form, and the dilaton.
	
	In section \ref{closed memory}, we consider a closed string in flat spacetime located at the radiation zone of a scattering process, which involves 2-form soft radiation and we calculate the response of the string to the 2-form soft radiation. This is essentially a  worldsheet calculation that involves the 2-form background as a gauge interaction and is valid for any target space dimension.  Section \ref{quantum closed} displays the same calculation in light cone quantization of the worldsheet theory. In the low energy limit, the theory contains a graviton, a 2-form gauge field, and a dilaton. We show in section \ref{effective} how these particles can convert into each other under a soft 2-form radiation. The next couple of sections discuss memory effect on open string probes.
	
	\section{The worldsheet action}
	The history of a 1-dimensional object, a string, defines a 2-dimensional manifold in spacetime, which we call a \emph{worldsheet}. Given the worldsheet coordinate $(\tau,\sigma)$ one needs to determine the embedding functions $X^{\mu}(\tau,\sigma)$ to locate the string. The evolution of the embedding functions can be postulated to arise from the Polyakov action

	\begin{equation}\label{polyakov action}
	S_{P}=-\frac{1}{4\pi\alpha^\p}\int d\tau d\sigma \sqrt{-\gamma}\gamma^{ab}\partial_aX^\mu\partial_b X_\mu\,,
	\end{equation}
	in which the dynamical fields are both $X^\mu$ and $\gamma_{ab}$.  Equations of motion for metric constrain matter fields by making the energy-momentum tensor vanish
	\begin{eqnarray}\label{Tconst}
	T_{ab}=\partial_aX^\mu\partial_b X_\mu-\frac{1}{2}\gamma_{ab}\partial^c X^\mu\partial_c X_\mu=0
	\end{eqnarray}
	The trace ${T^a}_a$ is identically vanishing since the action is identically invariant under Weyl transformations of the metric. As a result, \eqref{Tconst} consists of two constraints and there are $d-2$ independent solutions among $X^\mu(\sigma,\tau)$.
	
	Define two null embedding  coordinates
	\begin{align}
	X^-=X^0-X^1\qquad\qquad X^+=X^0+X^1\,.
	\end{align}
	Fix the worldsheet time coordinate to light-cone gauge $X^-=\ell_s\tau$. Also, set the conformal gauge choice $\gamma_{ab}=e^{\omega}\eta_{ab}$ for the worldsheet metric. These make three gauge choices corresponding to three local symmetries of \eqref{polyakov action}.
	Variations with respect to the worldsheet metric $\gamma_{ab}$ and the target space coordinates $X^\mu$ give the following constraint and equations of motion,
	\begin{subequations}
		\begin{align}
		&-\ell_s^2-2\ell_s\partial_\pm X^++(\partial_\pm X_\ii)^2=0\,\label{constraint},\\
		&   \partial_+\partial_-X^\ii=0
		\end{align}
	\end{subequations}
	
	where the left/right worldsheet derivatives are defined as
	\begin{equation}
	\partial_+=\partial_\tau+\partial_\sigma \qquad
	\partial_-=\partial_\tau-\partial_\sigma \,.
	\end{equation} 
	The free string equations are
	\begin{equation}\label{Y equation}
	\partial_-\partial_+ X^\ii=0\,,
	\end{equation}
	whose general solution is
	\begin{equation}\label{gensol}
	X^i(\tau,\sigma)=X^\ii_{L}(\tau-\sigma)+X^\ii_{R}(\tau+\sigma)\,.
	\end{equation}
	For a closed string $(\sigma\sim \sigma+2\pi)$ one  imposes the periodic boundary conditions $X^\ii_{L,R}(x)=X^\ii_{L,R}(x+2\pi)$. Thus, embedding functions are bosonic fields on the worldsheet.
	The general solution \eqref{gensol} subject to these boundary conditions, takes the following form,
	\begin{equation}\label{closed mode exp}
	X^\ii(\tau,\sigma)=x_0^\ii+\alpha^\p p^\ii\tau+i\sqrt{\frac{\alpha^\p}{2}}\sum_{n\neq 0}\frac{1}{n}\left(\alpha^\ii_n e^{-in(\tau-\sigma)}+
	\tilde{\alpha}^\ii_n e^{-in(\tau+\sigma)}
	\right)\,.\end{equation}
	
	For an open string $\sigma\in [0,\pi]$ one  first needs to impose boundary conditions on string endpoints
	\begin{eqnarray}
	\delta X_\ii \partial_\sigma{X}^\ii=0\qquad\qquad \sigma=0,\pi\,.
	\end{eqnarray}
	There are usually two obvious choices:  either \emph{the Dirichlet condition} $\delta X^\ii=0$ or \emph{the Neumann condition} $\partial_\sigma X^\ii=0$. There can also be mixed conditions which we encounter when considering the interaction with 2-forms.
	
	\section{2-form memory on closed string}\label{closed memory}
	Let us couple the free string action  to a background 2-form field $\B_{\mu\nu}(X)$ \begin{equation}\label{Sint}
	S_{\text{int.}}=-\frac{1}{4\pi\alpha^\p}\int_{\mathcal{W}}\B
	=-\frac{1}{4\pi\alpha^\p}\int_{\mathcal{W}} d\tau d\sigma \epsilon^{ab}\partial_aX^\mu\partial_b X^\nu \B_{\mu\nu}\,.
	\end{equation}
	The $\B$-term is a topological term (it is independent of the worldsheet metric $\gamma^{ab}$) and does not contribute to the  constraint \eqref{constraint} which determines the non-zero-mode part of the $r$-coordinate. In  the radial gauge $\B_{r\mu}=0$ for the background 2-form field we have (at the leading order in $r$)
	\begin{equation}
	\partial_+\partial_-X^\ii=-\ell_s\partial_u{\B}^{\ii\jj} X^{\p \jj}\,,\label{closed eom}
	\end{equation}
	In our analysis, we neglect variation of $\B$ around the string; we assume  $\partial_\sigma\B=0$, that is, the string length is much smaller than variation of $\B$.
	In order to solve \eqref{closed eom} in presence of small $\B$-field,
	consider the ansatz
	\begin{equation}\label{X def Y}
	X^\ii(\tau,\sigma)\equiv [\delta^{\ii\jj}+\frac{1}{2}\B^{\ii\jj}(u)]X^\jj_L+     [\delta^{\ii\jj}-\frac{1}{2}\B^{\ii\jj}(u)]X^\jj_{R}\,,
	\end{equation}
	which satisfies
	\begin{equation}
	\partial_+\partial_-X^\ii=-\ell_s \partial_u{\B}^{\ii\jj}X'^\jj+\frac{\ell_s^2}{2}\partial^2_u{\B}^{\ii\jj}(X^\jj_{L}-X^\jj_{R})+\ord{\B^2}\,.
	\end{equation}
	Assuming that  $\B$ is \emph{soft}, \emph{i.e.} its frequency is much smaller than the string frequency $\ell_s^{-1}$, the second term is negligible and $X^i$ satisfy equations of motion \eqref{closed eom}. Therefore, the closed string solution to \eqref{closed eom} (with periodic boundary conditions) for a slowly varying background $\ell_s\partial_u\B\ll \B$ and neglecting $\ord{\B^2}$ terms is
	\begin{align}\label{mode exp closed}
	\hspace*{-0.3cm}  X^\ii=x_0^\ii+\alpha^\p p^\ii\tau
	+i\sqrt{\frac{\alpha^\p}{2}}\sum_{n\neq 0}\frac{1}{n}&\Big[(\delta^{\ii\jj}+\frac{1}{2}\B^{\ii\jj}(u))\alpha^\jj_n e^{-in(\tau-\sigma)}\nn\\
	& \quad\quad+(\delta^{\ii\jj}-\frac{1}{2}\B^{\ii\jj}(u))\tilde{\alpha}^\jj_n e^{-in(\tau+\sigma)}\Big]
	\end{align}
	where $\alpha^\ii_n$ and $\tilde{\alpha}^\ii_n$ are the oscillator modes of $X^\ii_L$ and $X^\ii_R$ in \eqref{closed mode exp}, respectively.

	\paragraph{Classical closed string memory effect.} The memory effect is the imprint of the background field on the string, found by comparing the state of the string at far future and far past.
	Defining
	\begin{equation}\label{Theta def}
	\Delta\B_{ij}=\B_{ij}(u\to+\infty)-\B_{ij}(u\to-\infty)\,,
	\end{equation}
	the closed string memory effect follows from \eqref{mode exp closed}:
	\begin{subequations}
		\begin{align}\label{classical memory closed}
		\alpha^\ii_n(u\to+\infty)&=(\delta^{\ii\jj}+\frac{1}{2}\Delta\B^{\ii\jj})\alpha^\jj_n(u\to-\infty)\\
		\tilde{\alpha}^\ii_n(u\to+\infty)&=(\delta^{\ii\jj}-\frac{1}{2}\Delta\B^{\ii\jj})\tilde{\alpha}^\jj_n(u\to-\infty)\\
		p^\ii(u\to+\infty)&= p^\ii(u\to-\infty)\label{closedmemory}
		\end{align}
	\end{subequations}
	plus terms of $\ord{\B^2}$ and $\ord{\ell_s\partial_u\B}$ which are negligible.
	The center of mass motion is unaltered, while oscillation modes are rotated, with a relative minus sign between left- and right-moving modes.

	\section{Quantum closed string memory}\label{quantum closed}
	To study the quantum closed string memory, we start from \eqref{classical memory closed}-\eqref{closedmemory} and quantize the system\footnote{Quantum operators are not hatted in this section.}. We employ a Hamiltonian approach, perform the calculations in the light-cone gauge, and treat the soft background as a perturbation. Using proposition \ref{canonic}, we first introduce a canonical transformation that turns the initial and the final Hamiltonian into similar forms.
	The light-cone Hamiltonian for the action \eqref{polyakov action} is
	\begin{equation} 
	H= \frac{\ell_s^2}{4}\int_0^{2\pi} d\sigma\left[(\tilde{P}_\ii-\frac{2}{\ell_s^2}X^{\p \jj}\B_{\ii\jj})(\tilde{P}_\ii-\frac{2}{\ell_s^2}X^{\p \kk}\B_{\ii\kk}) +\frac{4}{\ell_s^4}( X^{\p \ii} X^{\p \ii})\right]
	\end{equation}
	in which $\tilde{P}_\ii=\frac{\partial L}{\partial \dot{X}^\ii}=\frac{2}{\ell_s}\left(\dot{X}_\ii+X'^j\B_{\ii\jj}\right)$ is the canonical momentum of $X^\ii$. Consider the canonical transformation on the momenta
	\begin{equation}
	P_\ii=\tilde{P}_\ii-\frac{2}{\ell_s^2} X^{\prime \jj}\B_{\ii\jj}\,,
	\end{equation}
	and leave the coordinates  unaltered. The transformed Hamiltonian $K$ is
	\begin{equation}\label{Classical K}
	K= \frac{\ell_s^2}{4}\int_0^{2\pi} d\sigma\left[P^\ii P^\ii +\frac{4}{\ell_s^4}  X^{\p \ii} X^{\p \ii}+\frac{4}{\ell_s^4}X^\ii X^{\p \jj} \dot{\B}_{\ii\jj}\right]\,.
	\end{equation}
	The generating function for the transformation is
	\begin{equation}
	G=X^\ii(P_\ii+\frac{1}{\ell_s^2}X^{\p \jj}\B_{\ii\jj})\,.
	\end{equation}
	
	To quantize, as usual, we promote the phase space coordinates to quantum operators and impose the commutation relations
	\eq{
		[X^\ii(\sigma),P^\jj(\sigma^\p)]=i\delta^{\ii\jj}\delta(\sigma-\sigma^\p)\,,\qquad [\bar{r},p_{\bar{r}}]=i\,,
	}{}
	where $\bar{r}=\int d\sigma r(\sigma)$\,.  The free string mode expansion \eqref{closed mode exp} then yields the following brackets
	\begin{subequations}
		\begin{align}\label{closed-string-brackets}
		[\alpha^\ii_m,\alpha^\jj_{n}]&=m\delta^{\ii\jj}\delta_{m+n},\\
		[\tilde{\alpha}^\ii_m,\tilde{\alpha}^\jj_{n}]&=m\delta^{\ii\jj}\delta_{m+n}\,,\\
		[x_0^\ii,p^\jj]&=\delta^{\ii\jj}i\,.
		\end{align}
	\end{subequations}
	
	The Hamiltonian is\footnote{One could perform the canonical transformation \emph{after} quantization, which would be a unitary transformation on Hilbert space and operators.
		Consider the following unitary transformation
		\eq{
			U[\B]=\exp{\left(-\frac{i}{\ell_s^2}\int_0^{2\pi}d\sigma\,
				X^\ii X^{\p \jj}\B_{\ii\jj}
				\right)}\,.\nonumber
		}{unitary}
		with its action on momentum operator
		\begin{equation}
		P^\ii\to UP^\ii U^{\dagger}=P^\ii+\frac{2}{\ell_s^2}X^{\p \jj}\B_{\ii\jj}-\frac{1}{\ell_s^2 }X^\jj\partial_\sigma\B_{\ii\jj}\nonumber
		\end{equation}
		with the transformed Hamiltonian 
		\begin{equation}\label{KKH}
		K=UHU^\dagger+i\dot{U}U^\dagger.\nonumber
		\end{equation}
		Assuming  the string length is much smaller than the variation of $\B$, $\partial_\sigma\B\simeq 0$, $K$ 
		takes the same form as \eqref{Classical K}, but now as a quantum operator. }
	\begin{equation}\label{quantum hamiltonian}
	K=K_0+\sum_{n\neq 0}\frac{i}{2n}(\alpha^\ii_{-n}\alpha^\jj_{n}-\tilde{\alpha}^\ii_{-n}\tilde{\alpha}^\jj_{n}+2\alpha^\ii_n\tilde{\alpha}^\jj_{n})\dot{\B}_{\ii\jj}
	\end{equation}
	where $K_0=\frac{1}{2}\sum_{n}\left(\alpha_n^\ii\alpha_{-n}^\ii+\tilde{\alpha}_n^\ii\tilde{\alpha}_{-n}^\ii\right)$ is the unperturbed Hamiltonian with  $\alpha_0^\ii=\tilde{\alpha}_0^\ii=\sqrt{\alpha^\p/2}\,p^\ii$. 
	In our analysis we will drop the zero point energy as we implicitly assume that our $X$ modes are a part of a superstring theory where the zero point energy cancels out by the worldsheet superpartners.
	
	For a slowly-varying and small $\B$-field background (\emph{i.e.} the adiabatic evolution, in which the perturbation varies much slower than the energy gap among states), the transition amplitude between states of different energy is vanishing. As a result,  the very last term in \eqref{quantum hamiltonian}, which does not commute with the free Hamiltonian $K_0$  gives no contribution to the evolution of states (Recall that the adiabatic evolution can give rise to transitions only in degenerate states). In addiiton, the first couple of terms are identified as the left and right components of the angular momentum operators. Using the notation of \cite{Blumenhagen:2013fgp}, the angular momentum operator is defined as
	\begin{equation}
	J^{\mu\nu}=2\int_0^{2\pi}d\sigma X^{[\mu}P^{\nu]} =\ell^{\mu\nu}+E^{\mu\nu}+\tilde{E}^{\mu\nu}\,,\nonumber
	\end{equation}
	and in terms of Fourier coefficients, $\ell^{\mu\nu}=2x_0^{[\mu}p^{\nu]}$ and
	\begin{equation}
	E^{\ii\jj}\equiv\sum_{n\neq 0}\frac{i}{n}\alpha^\ii_n\alpha^\jj_{-n}\,,\qquad
	\tilde{E}^{\ii\jj}\equiv\sum_{n\neq 0}\frac{i}{n}\tilde{\alpha}^\ii_n\tilde{\alpha}^\jj_{-n}.
	\end{equation}
	These operators commute with the Hamiltonian. The Hamiltonian is then easily integrated to give the  time evolution operator
	\begin{equation}
	U(u_2,u_1)=\exp\left(-\frac{i}{4}(\B_{\ii\jj}(u_2)-\B_{\ii\jj}(u_1)(E^{\ii\jj}-\tilde{E}^{\ii\jj})\right)\exp\left(-iK_0(u_2-u_1)\right)\,.
	\end{equation}
	The $S$ operator which maps $in$ and $out$ states is thus identified as
	\begin{equation}\label{S-matrix-closed-string}
	S=\exp\left(-\frac{i}{4}\Delta\B_{\ii\jj}(E^{\ii\jj}-\tilde{E}^{\ii\jj})\right)\,,
	\end{equation}
	in which $\Delta\B_{\ii\jj}$ is defined in \eqref{Theta def}. The 2-form memory effect on a closed string is then given by the Heisenberg-picture  evolution of the operators,
	\begin{equation}\label{quantum closed memory}
	\begin{split}
	S^{-1}\alpha^k_m\,S&=+\frac{1}{2}\Delta\B_{\ii\kk}\alpha_m^\kk +\ord{\B^2}\,,\cr
	S^{-1}\tilde{\alpha}^\kk_m\,S&=-\frac{1}{2}\Delta\B_{\ii\kk}\tilde{\alpha}_m^\kk +\ord{\B^2}\,,
	\end{split}
	\end{equation}
	consistent with the classical result \eqref{classical memory closed}.
	
	Eqs.\eqref{quantum closed memory} are our main result of this section. To explore it further, consider the general massless state in bosonic string theory
	\begin{equation}
	\zeta_{\ii\jj}\alpha^{\ii}_{-1}\tilde{\alpha}_{-1}^\jj|0;k\rangle\,.
	\end{equation}
	The memory effect on this state is the transition in its polarization tensor according to \eqref{quantum closed memory}:
	\begin{equation}
	\zeta_{\ii\jj}\to \left(\delta^{\ii\mathsf{m}}+\frac{1}{2}\Delta\B^{\ii\mathsf{m}}\right) \zeta_{\mathsf{m}\mathsf{n}}\left(\delta^{\jj\mathsf{n}}-\frac{1}{2}\Delta\B^{\jj \mathsf{n}}\right)\,,
	\end{equation}
	where as usual the dilaton, graviton and $b$-field states are respectively associated with trace, symmetric-traceless and antisymmetric parts of polarization tensor $\zeta_{\ii\jj}$:
	\begin{subequations}
		\begin{align}
		\varphi&\equiv\frac{1}{\sqrt{d-2}} \delta^{\ii\jj}\zeta_{\ii\jj}\,,\\
		b_{\ii\jj}&\equiv\frac{1}{2}(\zeta_{\ii\jj}-\zeta_{\jj\ii})\,,\\
		h_{\ii\jj}&\equiv\frac{1}{2}(\zeta_{\ii\jj}+\zeta_{\jj\ii})-\frac{1}{\sqrt{d-2}}\delta_{\ii\jj}\varphi\,.
		\end{align}
	\end{subequations}
	
	Variation of different components are
	\begin{subequations}\label{string-theory-conversion-amplutides}
		\begin{align}
		\Delta\varphi&=-\frac{1}{\sqrt{d-2}}\Delta\B^{\mathsf{m}\mathsf{n}}b_{\mathsf{m}\mathsf{n}}\,,\\
		\Delta b_{\ii\jj}&=\frac{\varphi}{\sqrt{d-2}}\Delta\B_{\ii\jj}-\Delta\B_{\mathsf{n}[\ii}h_{\jj]\mathsf{n}}\,,\\
		\Delta h_{\ii\jj}&=\frac{1}{d-2}\delta_{\ii\jj}\Delta\B^{\mathsf{m}\mathsf{n}}b_{\mathsf{m}\mathsf{n}}+\Delta\B_{\mathsf{n}(\ii}b_{\jj)\mathsf{n}}\,.
		\end{align}
	\end{subequations}
	Variations of $\varphi$ and $h_{\ii\jj}$ depend exclusively on the initial 2-form state $b_{\ii\jj}$. On the other hand, variation of the 2-form field comes from  other components, \emph{i.e.} $\varphi$ and $h_{\ii\jj}$. This is of course quite expected from 
	group theory viewpoint; the contracted product of an antisymmetric tensor and symmetric one is an antisymmetric tensor and the contracted product of two antisymmetric tensors is a symmetric one. That is, the passage of a soft $2$-form converts the massless $b$ state to dilaton or graviton and vice versa, without changing its momentum or energy.
	
	
	\section{Effective field theory analysis}\label{effective}
	In this section, we will provide a field-theoretic explanation of the closed string memory effect. We write the effective field theory for the graviton $h_{\mu\nu}$,   the dilaton $\varphi$, and the Kalb-Ramond 2-form field $b_{\mu\nu}$, on a slowly varying 2-form background. We calculate the transition amplitudes of previous sections by reading the vertices in Feynman rules.
	
	\begin{figure}
		\centering
		\begin{tikzpicture}
		\usetikzlibrary{snakes}
		\draw[snake=zigzag,thick](0,0)--(1.8,0);
		\node [thick] at (2,0){\large$\otimes$};
		\draw[very thick,dashed] (2.2,0)--(4,0);
		\node [] at (0,.5) {$b_{\mu\nu}(p)$};
		\node [] at (4,.5) {$\varphi(p)$};
		\node [] at (2,.5) {$\Delta\B_{\mu\nu}$};

		\begin{scope}[shift={(6,0)}]
		\draw[snake=zigzag,thick](0,0)--(1.8,0);
		\node [thick,opacity=1] at (2,0){\large$\otimes$};
		\draw[thick,snake=snake] (2.2,0)--(4,0);
		\node [] at (0,.5) {$b_{\mu\alpha}(p)$};
		\node [] at (4,.5) {${h_{\nu}}^\alpha(p)$};
		\node [] at (2,.5) {$\Delta\B^{\mu\nu}$};

		\end{scope}
		\end{tikzpicture}
		\caption{Kinetic mixing vertices. }
		\label{Fig-2}
	\end{figure}
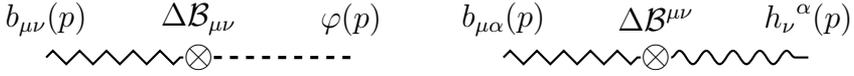
	
	The effective action of NSNS sector of string theory in string frame is \cite{Johnson:2003gi}
	\begin{align}
	S=\frac{1}{2\kappa^2}\int d^dx\sqrt{-G}e^{-2\Phi}&\left[R+4\nabla_\mu\Phi\nabla^\mu\Phi-\frac{1}{12}\Hh_{\mu\nu\laa}\Hh^{\mu\nu\laa}
	\right]
	\end{align}
	where $\kappa^2=8\pi G_N$. For each field we consider a perturbation on top of a background field
	\begin{equation}
	G_{\mu\nu}=\bar{G}_{\mu\nu}+h_{\mu\nu}\,,\qquad \B_{\mu\nu}=\bar{\B}_{\mu\nu}+b_{\mu\nu}\,,\qquad \Phi=\bar{\Phi}+\varphi
	\end{equation}
	We assume that background fields satisfy their own classical equations of motion, thus, only terms of second and higher order in perturbation fields appear in the action. The second-order terms are kinetic terms. We assume that $\bar{G}_{\mu\nu}=\eta_{\mu\nu}$ and $\bar{\Phi}=0$
	and neglect second order terms in background $\B$-field. The second order action in Einstein frame
	\footnote{
		The Einstein frame action is
		\begin{align}
		S=\frac{1}{2\kappa^2}\int d^Dx\sqrt{-G}&\left[R-\frac{4}{d-2}\nabla_\mu\Phi\nabla^\mu\Phi-\frac{1}{12}e^{-8\Phi/(d-2)}\Hh_{\mu\nu\laa}\Hh^{\mu\nu\laa}
		\right]\,.\nonumber
		\end{align}
	} becomes
	\begin{equation}
	S^{(2)}=\frac{1}{2\kappa^2}\int d^dx\left[\mathcal{L}^{(2)}_h+\mathcal{L}^{(2)}_b+\mathcal{L}^{(2)}_\varphi\right]
	\end{equation}
	where
	\begin{subequations}
		\begin{align}    \mathcal{L}^{(2)}_h&=-\frac{1}{2}(\nabla_\mu h_{\nu\rho})^2+\nabla_\mu h_{\nu\rho}\nabla^\nu h^{\mu\rho}-\nabla_\mu h\nabla_\rho h^{\mu\rho}+\frac{1}{2}(\nabla_\mu h)^2\\
		\mathcal{L}^{(2)}_b&=-\frac{1}{12}(\di b)_{\mu\nu\laa}(\di b)^{\mu\nu\laa}\\
		\mathcal{L}^{(2)}_\varphi&=-\frac{4}{d-2}\nabla_\mu\varphi\nabla^\mu\varphi
		\end{align}
	\end{subequations}
	There are also a couple of kinetic mixing terms (interaction with background $\B$-field)  at second order due to the background $\Hh$-field, as depicted in Figure \ref{Fig-2}
	\begin{align}
	\mathcal{L}^{(2)}_{b\varphi}&=\frac{4}{3(d-2)}\varphi\bar{\Hh}_{\mu\nu\laa}(\di b)^{\mu\nu\laa}\\
	\mathcal{L}^{(2)}_{bh}&=-\frac{1}{6}\left(3h^{\alpha\beta}\bar{\Hh}_{\alpha\nu\laa}{(\di b)_{\beta}}^{\nu\laa}+\frac{1}{2}h\bar{\Hh}_{\mu\nu\laa}(\di b)^{\mu\nu\laa}\right)
	\end{align}

	\paragraph{Gauge fixed action; TT gauge.}We impose transversality condition both on the graviton and the 2-form field:
	\begin{equation}\label{transverse}
	\nabla^\mu h_{\mu\nu}=0\,,\qquad \nabla^\mu b_{\mu\nu}=0\,.
	\end{equation}
	In addition, the temporal components can be set to zero by using the residual gauge symmetry while the metric perturbation is set to traceless:
	\begin{equation}\label{traceless}
	\eta^{\mu\nu}h_{\mu\nu}=0\,,\qquad h_{u\alpha}=0\,, \qquad b_{u\alpha}=0\,.
	\end{equation}
	Equations \eqref{transverse} and \eqref{traceless} comprise $2d$ conditions on graviton  and $2d-3$ conditions on the 2-form field to give the right degrees of freedom. 
	
	We implement canonical quantization of the fields on constant-$u$ surfaces. Although $u$ is a timelike coordinate, constant-$u$ surfaces, where outgoing wavefronts lie, are null hyperplanes. The canonical momenta are the following
	\begin{align}
	\Pi_{(\varphi)}(\x)&=\frac{\partial\mathcal{L}}{\partial\dot{\varphi}}=\frac{4}{\kappa^2(d-2)}\partial_r\varphi\\
	\Pi^{ab}_{(b)}(\x)&=\frac{\partial\mathcal{L}}{\partial\dot{b}_{ab}}=-\frac{1}{2\kappa^2}H^{uab}\\
	\Pi^{ab}_{(h)}(\x)&=\frac{\partial\mathcal{L}}{\partial\dot{h}_{ab}}=\frac{1}{2\kappa^2}\nabla_rh^{ab}\,.
	\end{align}
	The interaction-picture free fields are\footnote{The completeness relations are
		\begin{align}
		\sum_{s}\zeta^s_{ab}(\pii)\zeta^{s\ast}_{cd}(\pii)&=\delta_{a[c}\delta_{d]b}-2\pii_{[a}\delta_{b][d}\pii_{c]}/|\pii|^2\,,\nonumber\\
		\sum_{s}\xi_{ab}(\pii,s)\xi^{s\ast}_{cd}(\pii,s)&=\delta_{a(c}\delta_{d)b}-2\pii_{(a}\delta_{b)(d}\pii_{c)}/|\pii|^2\,.\nonumber
		\end{align}
	}
	\begin{subequations}
		\begin{align}
		\varphi(x)&=\frac{\kappa}{2}\sqrt{d-2}\int \di^{d-1}p\frac{1}{\sqrt{2p_r}}\left(\boldsymbol{\varphi}_{\pii}e^{ip\cdot x}+\boldsymbol{\varphi}^\dagger_{\pii}e^{-ip\cdot x}\right),\\
		b_{ab}(x)&=\kappa\sum_s\int \frac{\di^{d-1}p}{\sqrt{2p_r}}\left(\zeta_{ab}(\pii,s)\,\mathbf{b}^{s}_{\pii}\,e^{ip\cdot x}+\zeta^{\ast}_{ab}(\pii,s)\,\mathbf{b}^{s\dagger}_{\pii}\,e^{-ip\cdot x}\right)\,,\\
		h_{ab}(x)&=\kappa\sum_s\int \frac{\di^{d-1}p}{\sqrt{2p_r}}\left(\xi_{ab}(\pii,s)\,\mathbf{h}^{s}_{\pii}\,e^{ip\cdot x}+\xi^{\ast}_{ab}(\pii,s)\,\mathbf{h}^{s\dagger}_{\pii}\,e^{-ip\cdot x}\right)\,.
		\end{align}
	\end{subequations}
	
	The quantum operators satisfy the commutation relations below
	\begin{subequations}
		\begin{align}
		[\boldsymbol{\varphi}_\pii,\boldsymbol{\varphi}^\dagger_{\textbf{q}}]&=\delta^{d-1}(\pii-\textbf{q})\,\\
		[\mathbf{b}^r_\pii,\mathbf{b}^{s\dagger}_{\textbf{q}}]&=\delta^{rs}\,\delta^{d-1}(\pii-\textbf{q})\,,\qquad s=1,\cdots(d-2)(d-3)/2\\
		[\mathbf{h}^r_\pii,\mathbf{h}^{s\dagger}_{\textbf{q}}]&=\delta^{rs}\,\delta^{d-1}(\pii-\textbf{q})\,,\qquad s=1,\cdots,d(d-3)/2\,,
		\end{align}
	\end{subequations}
	and the  free Hamiltonian in terms of quantum operators becomes
	\begin{equation}
	H=\int d^{d-1}p\,\omega_\pii\left(\boldsymbol{\varphi}^\dagger_\pii \boldsymbol{\varphi}_\pii+\sum_s \mathbf{b}^{s\dagger}_\pii \mathbf{b}^{s}_\pii+\sum_s\mathbf{h}^{s\dagger}_\pii \mathbf{h}^{s}_\pii\right)\,,
	\end{equation}
	where
	\begin{equation}  
	\omega_\pii=p_u=\frac{p_r^2+p_i^2}{2p_r}\,.
	\end{equation}
	
	\paragraph{Two-form--dilaton conversion amplitude.}
	
	The time evolution operator (interaction Hamiltonian) at first order in $\B$ is
	\begin{equation}
	-i\int du H_I(u)=-\frac{2i}{3(d-2)\kappa^2}\int d^d x\varphi(x)\bar{\Hh}_{\mu\nu\laa}(x)(\di b)^{\mu\nu\laa}(x)\,.
	\end{equation}
	By a  Fourier transform on the spatial dimensions and using free field expansions we have
	\begin{align}\label{b phi vertex}
	&  -i\int du H_I(u)=\frac{3}{\sqrt{d-2}}\int du \bar{\Hh}_{u\ii\jj}(u)\nonumber\\
	&\qquad\quad\times\sum_{s}\int \frac{d^{d-1}p}{2p_r}\left(p^{[u}\zeta^{\ii\jj]\ast}(\pii,s)\boldsymbol{\varphi}_{\pii}\mathbf{b}^{s\dagger}_\pii\,-p^{[u}\zeta^{\ii\jj]}(\pii,s)\boldsymbol{\varphi}^\dagger_{\pii}\mathbf{b}^{s}_\pii\,\right).
	\end{align}
	The first integral picks out the soft mode of the background field
	\begin{equation}
	\int du\bar{\Hh}_{u\ii\jj}=\tilde{\Hh}_{u\ii\jj}(\omega\to 0)=\Delta\B_{\ii\jj}\,.
	\end{equation}
	The transition amplitude is hence proportional to the intensity of the soft 2-photon. 
	
	The simplest example the matrix element $\mathcal{M}(b_{\mu\nu}\to\varphi)$, for an in-going 2-form particle with momentum $p _\mu=(p,2p,0,\cdots,0)$ and polarization tensor $\zeta_{\ii\jj}(\pii,s)$ as the \emph{probe in state}, and a scalar particle of the same momentum as the \emph{probe out state}. The amplitude according to \eqref{b phi vertex} is
	\begin{equation}
	\mathcal{M}(b_{\mu\nu}\to\varphi)=\frac{-1}{\sqrt{d-2}}\,\Delta \B_{\ii\jj}\zeta^{ \ii\jj}(\pii,s)\,.
	\end{equation}

	\paragraph{Two-form--graviton conversion amplitude.}
	The time evolution operator (interaction Hamiltonian) is
	\begin{align}\label{b h vertex}
	-i\int du H_I(u)=\frac{i}{4\kappa^2}\int d^dx h^{\alpha\beta}(x)\bar{\Hh}_{\alpha\nu\laa}(x){(\di b)_{\beta}}^{\nu\laa}(x)\,,
	\end{align}
	and in terms of quantum operators it equals to
	\begin{align}
	\frac{3}{4} \int du\bar{\Hh}_{u\ii\jj}(u)&\sum_{s,w}\int \frac{d^{d-1}p}{2p_r}\nn\\
	&\times\left({\xi_\beta}^{[u}(\pii,w)p^{\ii}{\zeta^{\jj]\beta\ast}}(\pii,s)\, \mathbf{h}^w_{\pii}\mathbf{b}^{s\dagger}_\pii\,-c.c.\right).
	\end{align}
	The amplitude for an incoming graviton of momentum $p$, to convert to a two-form $b$ is
	\begin{equation}
	\mathcal{M}(h_{\mu\nu}\to b_{\mu\nu})=-\Delta \B_{\ii\jj}\xi^{\mathsf{m}[\ii}(\pii,s)\zeta^{\jj]\mathsf{m}\ast}(\pii,w).
	\end{equation}
	The above of course matches our string theory computation and the result \eqref{string-theory-conversion-amplutides}.

	\section{The 2-form memory on an open string} \label{sec:42}
	In the previous section, we showed how the closed string memory effect was a consequence of the invariance of the worldsheet, or the corresponding low energy effective action under $\Lambda$-gauge transformations, $\B\to \B+d\Lambda$. For the open string case and due to the presence of worldsheet endpoints the $\Lambda$ gauge symmetry should be modified by the addition of the boundary  1-form gauge field \cite{Witten:1995im, SheikhJabbari:1999ba}.	The action \eqref{polyakov action} augmented by the boundary term
	\begin{equation}
	S_b=\frac{2}{\ell_s^2}\int d\tau\dot{X}^I\A_I\,,
	\end{equation}
	where $I,J=1,\cdots,p$ denote the spatial directions 
	along the D$_p$-brane and $\A_I$ is the U(1) gauge field on it, leads to the total open string world-sheet action. For open strings $\sigma\in [0,\pi]$.
	Variation of this action, and then fixing the light-cone gauge yields equations of motion \eqref{closed eom} and a boundary term,
	\begin{align}
	-\frac{2}{\ell_s^2}\int d\tau &\big(\gamma^{\sigma b}(\delta X^I\partial_b X_I)+\epsilon^{\sigma b}\delta X^I(\partial_b X^J \B_{IJ}+\B_{Iu})+\delta X^I\frac{d}{d\tau}\A_I\big)
	\end{align}
	$(b=\tau,\sigma)$ which results in the mixed (Neumann and Dirichlet) boundary conditions for directions along the brane \cite{SheikhJabbari:1997yi, SheikhJabbari:1999xd}
	\begin{equation}\label{Neumann b.c.}
	X^\p_I+\F_{IJ}\dot{X}^J+\ell_s\F_{Iu}=0\,.
	\end{equation}
	
	The boundary action introduces a coupling of the boundary $\partial\Sigma$ of the open string to the gauge field on the D-brane which is invariant under the new $\Lambda$-gauge transformations,
	\be
	\delta\B_{IJ}=2\partial_{[I}\Laa_{J]},\qquad \delta \A_I=\Laa_I\,.
	\ee  
	In other words, the combination
	\be\label{calF}
	{\cal F}\equiv\B-d\A
	\ee
	is invariant under  $\Lambda$-gauge transformation as well as under $\lambda$-gauge transformation, $\A\to \A+d\lambda$.
	
	As in the closed string case,  in the light-cone gauge $u=\ell_s\tau$ by fixing the radial gauge $\B_{r\mu}=0$, we find
	\begin{align}\label{mode exp general}
	X^I&=x_0^I+2\alpha^\p p^I\tau+w^I \sigma\nn\\&+i\sqrt{\frac{\alpha^\p}{2}}\sum_{n\neq 0}\frac{1}{n}\left(\Big[\delta^{IJ}+\frac{\B^{IJ}(u)}{2}\Big]\alpha^J_n e^{-in(\tau-\sigma)}+
	\Big[\delta^{IJ}-\frac{\B^{IJ}(u)}{2}\Big]\tilde{\alpha}^J_n e^{-in(\tau+\sigma)}\right)
	\end{align}
	Next, we impose the boundary condition \eqref{Neumann b.c.}:\footnote{Here we considered the case with both ends of the open string with the same boundary conditions. In principle one can consider cases where the two ends have different Neumann, Dirichlet or mixed boundary conditions, e.g. as in \cite{SheikhJabbari:1997yi}.}
	\begin{equation}\label{mode neumann}
	(1-\F)^{IJ}\left(1+\frac{\B}{2}\right)_{JL}\alpha^L_n=
	(1+\F)^{IJ}\left(1-\frac{\B}{2}\right)_{JL}\tilde{\alpha}^L_n\,,
	\end{equation}
	Since $\B_{IJ}$ and $\F_{IJ}$ have time dependence, \eqref{mode neumann} is in general not consistent with constant $\alpha^i_n$ and $\tilde{\alpha}^i_n$ (which is required by equations of motion). In a slowly varying and small 2-form $\B_{IJ}$ background, this can, however, be remedied if
	\begin{equation}\label{F-B-Lambda}
	\F^{IJ}(u)-\frac{1}{2}\B^{IJ}(u)=C^{IJ}\,,
	\end{equation}
	where $C$ is a constant 2-form. We can further choose  $C=0$ by making an appropriate $\Lambda$-gauge transformation on $\B$ such that, the boundary condition \eqref{Neumann b.c.} yields
	\begin{equation}\label{alpha-tilde-alpha-open}
	\alpha_n^I=\tilde{\alpha}_n^I\,.
	\ee
	Eq.\eqref{Neumann b.c.} also fixes the zero mode part, and the mode expansion becomes
	\begin{align}\label{open-mode-expansion}
	\hspace*{-4mm}X^I(\tau,\sigma)=x_0^I&+2\alpha^\p p^I\tau-2\alpha^\p\F^{IJ}p^J\sigma-2\alpha^\p\ell_s \F_{Iu}\sigma\nn\\
	&+\sqrt{2\alpha^\p}\sum_{n\neq 0}\frac{e^{-in\tau}}{n}\left(i\cos n\sigma\delta^{IJ}-\sin n\sigma\F^{IJ}\right)\alpha^J_n\,,
	\end{align}
	where the $\ord{\B^2}$ and higher order terms are neglected. 
	
	The analysis above summarized in  equations \eqref{alpha-tilde-alpha-open} and \eqref{open-mode-expansion} shows that the effects of adiabatically time-varying $\B$-field on an open string is completely different than on closed string. Specifically, the mode expansion coefficients $\alpha_n^i$ and the Hamiltonian of the system in the appropriate canonical frame are both time-independent. 
	This is essentially the same as the open string mode expansion in a constant $\B$-field background \cite{SheikhJabbari:1997yi, SheikhJabbari:1999vm, Ardalan:1999av, SheikhJabbari:1999ba,Chu:1998qz,Chu:1999gi}. The effects of the time-variation of the $\B$-field may be seen in the center of mass motion of the string 
	\be
	{\bar X}^I(u)=\frac{1}{\pi}\int_0^{\pi} d\sigma\ X^I(\tau,\sigma)=x_0^I+2\alpha^\p p^I\tau-\pi\alpha^\p\F^{IJ}p^J-\pi\alpha^\p\ell_s \F_{Iu},
	\ee
	in the last two terms. We will discuss this further in the following subsection.
	
	\subsection*{Quantum treatment}
	
	Having the mode expansion we can readily quantize the open string by imposing \cite{Ardalan:1999av, SheikhJabbari:1999vm, Chu:1998qz,Chu:1999gi, Seiberg:1999vs}
	\be
	[x^I_0, x^J_0]=i\pi\alpha'\F^{IJ},\qquad [x^I_0, p^J]=i\delta^{IJ},\qquad [\alpha^I_n, \alpha^J_m]=n\delta^{IJ} \delta_{m+n,0}.
	\ee
	As we see the effects of the adiabatically changing background $\B$-field has appeared only in the noncommutativity of the $x_0^i$ coordinates. Note that $x_0^i$ are basically the coordinates of the D$_p$-brane the open string endpoint is attached to.  Since the Hamiltonian is not affected by the background $\B$-field in the open string case we do not have a memory effect in the usual sense. Nonetheless, open strings in the adiabatically changing $\B$-field background behave like an electric dipole \cite{SheikhJabbari:1999vm}, whose dipole moment is changing in time:
	\begin{equation}
	\Delta d^I\equiv  d^I (u\to+\infty)- d^I (u\to-\infty)=2\pi\alpha^\p\Delta\B^{IJ}p^J,
	\end{equation}
	where $d_I(u)=\langle X^I(\sigma=\pi, u)-X^I(\sigma=0, u)\rangle$. This result can be also understood from the effective field theory of the open strings, which is the Born-Infled theory residing on the brane. In the presence of an adiabatically changing $\B$-field we are dealing with a noncommutative gauge theory \cite{SheikhJabbari:1998ac,Douglas:1997fm} with a slowly varying noncommutativity parameter (see also \cite{Bachas:2002jg} for a related analysis). As it is known in the noncommutative field theories the kinetic term is not affected by noncommutativity, e.g. see \cite{Micu:2000xj} and the effects of noncommutativity appear only in interaction terms which are not captured in the usual memory effect described in previous sections. 
	
	\section{D-brane probes and boundary states}
	D-branes are a  part of the spectrum of the string theory. They appear by requiring T-duality in theories of open strings \cite{Polchinski:1996fm,Polchinski:1996na} by the fact that under T-duality, Neumann and Dirichlet boundary conditions are interchanged.
	Besides this open string description \cite{Polchinski:1995mt}, it is known that D-branes can be represented as a coherent (bound) state of closed strings they can emit. The amplitude of the closed string emission is given through the boundary state \cite{DiVecchia:1997vef},  for a review of constructing these states  see \cite{DiVecchia:1999mal,DiVecchia:1999fje} and references therein. 
	
	The open string satisfies the boundary condition \eqref{Neumann b.c.} at its end along the D-brane world-volume and along the transverse directions we have $X^i(\sigma=0)=x^i_0$ for $i=p+1,\cdots,d-1$.
	Boundary state for D$_p$-branes in a constant $\B$-field background has also been worked out and studied \cite{DiVecchia:1999uf,Arfaei:1999jt}. One may generalize the analysis of \cite{Arfaei:1999jt} to adiabatically changing $\B$-field background, which should satisfy\footnote{Here we prefer covariance and do not to go to the light-cone gauge.}
	\be
	\left(\partial_\tau X_\mu+\F_{\mu\nu} \partial_\sigma X^\nu\right)_{\tau=\tau_0}|\mathscr{B}(\tau_0)\rangle=0\,.
	\ee
	We note that the boundary state $|\mathscr{B}(\tau_0)\rangle$ should satisfy the above condition at the given arbitrary time $\tau_0$, which in the light-cone gauge $\tau_0=u_0/\ell_s$. That is, $|\mathscr{B}(\tau_0)\rangle$ is giving the amplitude of closed string  emissions from a D-brane at time $\tau_0$. In a static background like the cases analyzed in \cite{DiVecchia:1999uf, Arfaei:1999jt}, the $\tau_0$ dependence of the boundary state  appears simply through $e^{in\tau_0}$ dependence of the string oscillator modes,  while in our case there is an extra dependence due to the slowly varying background $\B$ field.
	Since the analysis is  similar to the static case where $\F$ is constant, we skip the details of computation and quote the final result:
	\begin{align}\label{boundary-state}
	|\mathscr{B}(\tau_0)\rangle&=\exp\left(-\sum_{n=1}\frac{e^{2in\tau_0}}{n}\alpha^{\mu}_{-n}D_{\mu\nu}(\tau_0)\tilde{\alpha}^{\nu}_{-n}\right)|0\rangle\,,\\ D_{\mu\nu}&\equiv(\mathcal{Q}_{rs}(\tau_0),-\delta_{AB})\,.\nonumber
	\end{align}
	$r,s$ label all directions parallel to the brane, while those normal to the brane are labelled by $A,B$. The matrix $\mathcal{Q}(\tau_0)$ is defined as
	\begin{align}
	\mathcal{Q}(\tau_0)&=(1-\F(\tau_0)+\B(\tau_0)/2)^{-1}(1+\F(\tau_0)-\B(\tau_0)/2)\nonumber\\
	&=1+2\F(\tau_0)-\B(\tau_0)+\cdots\,,
	\end{align}
	where the braces denote higher orders in $\B$ or $\F$. Having the D$_p$-brane boundary state we can use it as the probe to explore the memory effect associated with passage of a $\B$-field. As the first example, let us compute the action of the closed string memory $S$-matrix \eqref{S-matrix-closed-string} on this state:
	\begin{align}
	S|\mathscr{B}\rangle&=\Bigg\{1+\frac{1}{2}\sum_{n=1}^{\infty}\frac{e^{2in\tau_0}}{n}
	\alpha^r_{-n} \tilde{\alpha}^s_{-n}
	\left(
	\Delta \B_{rs}-\Delta \B_{sr}-2\Delta \B_{rs}
	\right)+\cdots\Bigg\}|\mathscr{B}\rangle,
	\end{align}
	where braces denote terms second or higher order in $\B, \F$. We therefore have
	\eq{
		S|\mathscr{B}\rangle=|\mathscr{B}\rangle+\ord{\B^2}\,.
	}{}
	The boundary state is unchanged at first order in background fields. This result may be understood as follows. 
	The D$_p$-brane in a slowly-varying $\B$-field background is a non-marginal bound state of D$_p$ and lower dimensional D$_{(p-2n)}$-branes where $n$ rank of the $\B$ field along the brane, much like the constant $\B$-field case \cite{SheikhJabbari:1997yi}. The  mass density of the brane is $\frac{1}{g_s}\sqrt{\det{(1+\F)}}\simeq \frac{1}{g_s}(1+{\cal O}(\F^2))$ where $g_s$ is the string coupling. This is compatible with the softness of the passing $B$-field wave. The ($p-2n$)-form RR charge density carried by the bound state is proportional to $\F^n$. Therefore, in our approximation only  $n=1$ case is remaining. The change in this D$_{(p-2)}$-brane charge density is then proportional to $\Delta\B$. We note, however, that to see this RR charge density from the boundary state one needs to go beyond the bosonic sector discussed above and to consider the superstring case and include fermionic degrees of freedom. Moreover, using the boundary state \eqref{boundary-state} one can study the scattering of two such D$_p$-branes off each other, where the $|in\rangle$ and $|out\rangle$ boundary states differ in their value of the $\F$ field. 
	
	\section{2-form soft theorem}
	Calculations in previous sections verified how a  transformation $\Delta\B_{\ii\jj}$ influences various probes, which presumed the fact that such a transformation exists. In gravity and electromagnetism, non-zero gauge transformation between $u=+\infty$ and $u=-\infty$ relies on the $\frac{1}{\omega}$ poles of the soft photon amplitudes, the \emph{leading soft theorems}. The relevant question is whether there is a pole in 2-form soft theorems or not.
	
	Soft theorem for 2-form field has been studied in the same footing as graviton and dilaton in bosonic string theory \cite{DiVecchia:2015oba,DiVecchia:2015srk, DiVecchia:2017gfi}. To quote the results, the leading factor is the following
	\begin{eqnarray}
	M_{n+1}=\epsilon_{\mu}\tilde{\epsilon}_{\mu}\sum_{i=1}^n\frac{p_i^\mu p_i^\nu}{p_i\cdot q}M_n(p_i)+\mathcal{O}(q^0)
	\end{eqnarray}
	where $\epsilon$ and $\tilde{\epsilon}$ are Yang-Mills polarization vectors in a formulation which treats massless states of string theory as double-copied Yang-Mills states. The pole obviously drops out for an anti-symmetric polarization tensor for the 2-form field, and also for the dilaton, since the hard particles are massless $p_i^2=0$.  Hence, two form theory does not have a leading soft pole, at least through the  interaction vertices of bosonic string theory. Of course this does not imply that asymptotic symmetries and memories in this case must be trivial. One can associate these quantities to \emph{subleading} terms in the soft expansion. The $\mathcal{O}(q^0)$ term in 2-form soft theorem in bosonic string theory is
	\begin{equation}
	M_{n+1}=\epsilon_{\mu}\tilde{\epsilon}_{\mu}\sum_{i=1}^n\Bigg(\frac{p_i^{[\mu} (S_i-\tilde{S}_i)^{\nu]\sigma}q_\sigma}{p_i\cdot q}-\frac{1}{2}(S_i-\tilde{S}_i)^{\nu\sigma}\Bigg)M_n(p_i)  \,,
	\end{equation}
	where
	\begin{equation}
	S_{i\mu\nu}=i\big(\epsilon_{i\mu}\frac{\partial}{\partial\epsilon_i^\nu}-
	\epsilon_{i\nu}\frac{\partial}{\partial\epsilon_i^\mu}\big).
	\end{equation}
	One expects that this expression can be Fourier-transformed, as in case of QED to produce a pure gauge transformation on some component of the asymptotic expansion of the gauge field. Our general calculation of the 2-form memory effect suggests how such memories may be observed.

	\section{Discussion}
	In order to study the string memory effect, we revisited the problem of strings in a slowly varying NSNS 2-form $\B$-field background. It is known that the $\B$-field affects closed and open strings in different ways and hence we considered these two cases separately. For the closed string case the string memory effect is encoded in the transition of the massless states of the closed string into each other.  Thus, in the string memory effect, the probes themselves can be massless states, in contrast with  other usual memories in which the probe is a massive state.\footnote{Note that in the internal string memory effect the probe can also be one of the massive modes of string excitations.}
	
	In the open string, the constant part of the $\B$-field affects string dynamics through the boundary conditions. It is known that the endpoints of open strings attached to a D-brane in the $\B$-field background parameterize a noncommutative space \cite{Ardalan:1999av, SheikhJabbari:1999vm, Chu:1998qz,Chu:1999gi, Seiberg:1999vs}. The open string memory effect is then encoded in the change in the noncommutativity of the open string endpoints.  This memory can hence be observed through the effective noncommutative field theory residing on the brane where the open string endpoints attach. Alternatively, one may use the D-branes as probes of the background time-varying $\B$-field, using boundary state formulation for D-branes \cite{DiVecchia:1997vef, Arfaei:1999jt}. In this system, the change in the $\B$-field is encoded in the mass density or other Ramond-Ramond (RR) charges of the D-brane probe. The similar problem of strings in a gravitational background pulse was considered in \cite{Bachas:2002jg} and the possibility of a permanent shift in brane separation was discussed, which resembles point particle gravitational memory effects. 
	
	\chapter{Concluding remarks and outlook}

	Here we conclude with a brief summary of the results presented in this thesis, remark future directions, and discuss some open problems.
	
	After a review of the concepts of asymptotic symmetry and memory effect in chapter \ref{2 chapter}, we discussed the asymptotic structure of Maxwell theory in generic dimensions both in Minkowski and anti-de Sitter backgrounds (chapter \ref{max chapter}). Afterward, we considered the more interesting $(p+1)$-form gauge theories in  \ref{p chapter}. In all cases, we presented the boundary conditions in the hyperbolic coordinate system. This approach although does obscure the physical intuition behind the temporal and radial directions, has its own virtue of making Lorentz symmetry manifest and covering the whole spatial infinity region. Furthermore, taking the limit to future and past null infinities is simple in this coordinate system which facilitates proving the consistency of symmetry algebras appearing in the spatial and the null infinity approaches. To mention a couple of evidence,  one refers to the studies of electromagnetism and gravity in three and four dimensions \cite{Campiglia:2017mua, Troessaert:2017jcm,Compere:2017knf, Esmaeili:2019hom}. Of course, one must keep in mind that boundary conditions prescribed at spatial and null infinity do not possess the same physical content, since radiative modes are absent in the former.
	
	There is another kind of hyperbolic slicing of Minkowski space which covers \emph{inside} the light-cone. This coordinate system is convenient when considering the phase space of free massive particles at early or late times. Moreover, this version makes it possible to use AdS/CFT techniques in flat space holography \cite{Cheung:2016iub,Ball:2019atb} cause the constant-time slices are Euclidean-AdS (i.e. hyperbolic space $\mathbb{H}_{d-1}$). We did not elaborate on this `Milne patch' as our main focus has been on \emph{the spatial infinity point of view} and physical gauge transformations at the spatial boundary.
	
	We showed how an infinite-dimensional asymptotic symmetry group can be realized in higher dimensional Maxwell theory by pure gauge fluctuations at the boundary. Despite the fact that no radiation flux exists at spatial infinity, conservation of surface charges is not guaranteed and in general non-trivial, as we discussed in chapters \ref{max chapter} and \ref{p chapter}. The asymptotic Lorenz gauge is needed as a boundary equation of motion to make the charges conserved and the action principle well-defined. The need for a gauge fixing while adding no boundary degrees of freedom differentiates this procedure from the Hamiltonian approach \cite{Henneaux:2018gfi}. In any case, two copies of angle-dependent $U(1)$ algebras are found unless additional conditions are imposed.

	The antipodal matching condition of gauge fields in Minkowski space is established as the standard boundary condition which maps asymptotic symmetry generators in Hilbert spaces at future and past null infinity in four dimensions and reduces the asymptotic symmetry algebra to one copy of angle-dependent $U(1)$ algebra. There are different motivations for these boundary conditions as we discussed in section \ref{antip sec}. In quantum theory, this condition restricts how quantum states are `dressed'.  In the conventional Fock space of quantum field theories, matter states are not gauge-invariant. Recovering gauge-invariance demands \emph{dressing} matter states by soft hair of gauge bosons \cite{Giddings:2019hjc,Giddings:2019wmj,Gabai:2016kuf,Mirbabayi:2016axw,Javadinazhed:2018mle}. As a matter of fact, there exist physically reasonable dressings in classical Maxwell theory that violate the antipodal matching condition \cite{Giddings:2019ofz}. Nevertheless, these dressings involve electromagnetic fields with infinite boost charges and they may identify superselection sectors of the theory.

	In chapter \ref{p chapter}, we focused on $(p+1)$-form theories. We showed the presence of non-abelian exact charges even in pure gauge configurations in specific dimensions, which was confirmed in four dimensions \cite{Henneaux:2018mgn}. $(p+1)$-form theories can have flexible Coulombic boundary conditions depending on the sources being compact or not.  The existence of exact charges in generic dimensions and with different boundary conditions is an interesting statement to prove. Coexact charges, on the other hand, are non-vanishing only in backgrounds with a non-trivial electromagnetic field. In particular, our chosen boundary conditions allow arbitrary brane alignments which are coded in the zero-mode charges of the theory.
	
		Our calculation of $p$-form soft charges especially in the exact sector can bear relevance to black hole physics, as $p$-brane solutions in supergravity generically have non-zero $p$-form flux through the horizon. The fact that generic matter states falling into a black hole can supertranslate the near-horizon metric \cite{Hawking:2016sgy,Hawking:2016msc} and cause memory effect \cite{Rahman:2019bmk,Donnay:2018ckb} has led to the idea that the soft hair store (at least part of) the infalling information and release it during black hole evaporation.   There are also various soft hair proposals which reproduce Bekenstein-Hawking entropy by the near-horizon soft modes, see e.g. \cite{Strominger:1997eq, Guica:2008mu, Afshar:2016uax,Sen:2007qy}.  Since $p$-form zero-mode charges contribute to $p$-brane mechanics \cite{Compere:2007vx}, one expects that $p$-form soft hair is treated on equal footing with gravitational soft hair.
	
	In chapter \ref{memory chapter}, we provided a systematic definition of the memory effect  and proposed an observational setup for 2-form memory on string-like probes. The memory imprinted on closed strings amounts to opposite rotation in left- and right-moving sectors and results in non-vanishing transition amplitude among internal modes of the string with the same mass. Soft 2-form radiation on open strings rotates Dirichlet and Neumann boundary conditions into each other and changes the non-commutativity parameter on the brane.
	
	Although memory effect has been proposed in numerous examples of gauge theories and gravity as an observational feature of soft modes, the hypothesis that they are actually observable in physically reasonable conditions is disputed \cite{Bousso:2017rsx,Bousso:2017dny,Bousso:2017xyo}. The argument is tenable: soft modes defined as strict zero-frequency modes of radiation field take an infinite time to observe–by definition. Moreover, since they cost arbitrarily low energies, they can not be measured even \emph{approximately} by long enough exposure. For any time interval $t\in(-T,T)$  of observation–no matter how long–soft modes can change arbitrarily at $t>T$, still with infinitesimal energy cost. These considerations call for more accurate definitions for observable supertranslations induced by scattering of massive celestial bodies.   In addition, objections similar to those against the observability of the memory effect have been raised against the relevance of soft modes to the black hole information paradox as well \cite{Mirbabayi:2016axw}.

		The IR triangle (i.e. memory/surface charge/soft theorem)  which generically implies an infinite-dimensional asymptotic symmetry group  fundamentally belongs to asymptotically flat theories. However, one may reasonably investigate parallel ideas in other backgrounds as well.  For example, although Maxwell fields have different asymptotics in anti-de Sitter space,  a similar procedure to that of Minkowski space yields an infinite-dimensional asymptotic symmetry group. A clear understanding of this problem, in particular for the gauge fields appearing in supergravity can bring new information about the dual field theories.
	
	As a technical point, we mention that our discussion lies in \emph{gauge fixing approach} to asymptotic symmetries. In the \emph{geometric approach}, \emph{a la} Penrose \cite{Penrose:1965am,Geroch:1972up},  the spacetime is mapped to a compactified manifold, whose boundaries are at a finite distance. Therefore, the dynamical fields as well as the metric can be defined unambiguously \emph{on the boundary}. The  boundary conditions are defined by fixing data on the boundary of the compactified spacetime. The asymptotic symmetries in this method are defined as the gauge transformations which preserve the boundary data, modulo those that vanish on the boundary. This approach is manifestly gauge-invariant and geometric. However, the study of the asymptotic fields in the gauge-fixing approach is  more tractable. In practice, one starts the analysis in a specific coordinate system  and translates the results into geometric terms \emph{a posetriori}.
	
	Otherwise, from a conceptual point of view, one might try to regard gauge symmetry as a redundancy altogether, by demanding gauge invariance of the symplectic form. Generically, the symplectic form of a gauge theory on an \emph{open manifold}  is not gauge-invariant (this is of course the root of surface charges).
  In order to describe physics inside an open subregion in a gauge-invariant manner, one needs to postulate \emph{edge modes}   on the boundary with  a compensating  gauge transformation. These edge modes are part of the definition of a subregion in a gauge theory and they are associated with canonical charges\footnote{See \cite{Speranza:2017gxd} for a comprehensive review and  \cite{Riello:2019tad,Gomes:2019xto,Gomes:2018dxs,Gomes:2018shn}   for a geometric approach.}. This approach  technically differs from the usual asymptotic symmetry method presented here and generically yields larger symmetry algebras  \cite{Balachandran:1994up,Geiller:2019bti,Donnelly:2016auv,Donnelly:2016rvo}. However, it makes clear the conceptual fact that surface degrees of freedom are essential in the understanding of gauge theories in open spaces and they are associated with physical phenomena.

	\appendix
	
	\chapter{Basics of Hodge theory}\label{AppDiff}
	Here we define some basic elements and state important theorems relevant to our physical discussion \cite{de2007hodge,Warner1983,Folland1989}.
	Let $V$ be an $d$-dimensional real vector space.   $V_{r,s}$, \emph{the tensor space of type $(r,s)$ } is the vector space
	\begin{equation}
	V\otimes\cdots\otimes V\otimes V^\ast\otimes\cdots\otimes V^\ast
	\end{equation}
	and the direct sum
	\begin{equation}
	T(V)=\sum_{r,s\geq 0} V_{r,s}
	\end{equation}
	where $V_{0,0}=\mathbb{R}$ is called \emph{the tensor algebra of }$V$, and its elements are called \emph{tensors}.
	$T(V)$ is a non-commutative, associative, graded algebra under $\otimes$ such that the product of two tensors in $V_{r,s}$ and $V_{r^\prime,s^\prime}$ is in $V_{r+r^\prime,s+s^\prime}$. 
	
	Let $C(V)$ denote the subalgebra $\sum_{k=0}^{\infty}V_{k,0}$ of $T(V)$ and $I(V)$ be an ideal of $C(V)$ generated by the set of elements  of the form $v\otimes v$ for $v\in V$, and
	\begin{equation}
	I_{k}(V)=I(V)\cap V_{k,0}.
	\end{equation}
	The exterior algebra $\Lambda(V)$ of $V$ is the graded algebra $C(V)/I(V)$. If we set
	\begin{equation}
	\Lambda_k(V)=V_{k,0}/I_k(V)\qquad (k\geq 2),\qquad \Lambda_0(V)=\mathbb{R}\qquad \Lambda_1(V)=V,
	\end{equation}
	then
	\begin{equation}
	\Lambda(V)=\sum_{k=0}^\infty\Lambda_k(V).
	\end{equation}
	The multiplication in $\Lambda(V)$ is called \emph{wedge} or \emph{exterior product} and is denoted by $\wedge$.
	
	If  $u\in\Lambda_k(V)$ and  $v\in\Lambda_l(V)$ then $u\wedge v\in\Lambda_{k+l}(V)$ and $u\wedge v=(-1)^{kl}v\wedge u$. It also follows that
	\begin{equation}
	\Lambda_d(V)\cong \mathbb{R},\qquad  \Lambda_{d+j}(V)=0 \quad (j>0).
	\end{equation}
	In addition,
	\begin{equation}
	\text{dim}\Lambda(V)=2^d,\qquad\qquad  \text{dim}\Lambda_k(V)=\binom{d}{k}.
	\end{equation}

	\begin{defin}
		[Exterior bundle] Let $M$ be a differentiable manifold. Define \emph{exterior $k$-bundle} over $M$ by
		\begin{equation}
		\Lambda^\ast(M)=\bigcup_{m\in M}\Lambda_k(M^\ast)
		\end{equation}
	\end{defin}

	\begin{defin}
		[$p$-forms] A $C^\infty$ mapping of $M$  into $\Lambda^\ast_p(M)$ whose composition with the canonical projection is the identity map is  called a differential $p$-form on $M$. We denote by $E^p(M)$ the set of all smooth $p$-forms on $M$.
	\end{defin}
	\begin{defin}
		[Exterior differentiation] There exists a unique anti-derivation $\di: E^\bullet(M)\to E^\bullet(M)$ of degree $+1$ such that $\di^2=0$ and whenever $f\in C^\infty(M)=E^0(M)$, $df$ is the differential of $f$.
	\end{defin}
	
	[Closed/Exact] A $p$-form $u$ is said to be closed if $du=0$ and is said to be exact if there exists $v\in E^{p-1}(M)$ such that $dv=u$.
	\begin{defin}
		[de Rham cohomology] The quotient space of the real vector space of closed $p$-forms modulo the subspace of exact $p$-forms is called the $p$-th de Rham cohomology group of $M$.
	\end{defin}
	
	\begin{theo}
		[Poincar\'e Lemma] A closed $p$-form $u$, $p\geq 1$ on  $M$ is locally exact, i.e for every $m\in M$ there exists an open neighbourhood $U$ of $m$ and $v\in E^{p-1}(U)$ such that
		\begin{equation}
		u=dv.
		\end{equation}
	\end{theo}

	\begin{defin}
		
		[The $\star$ operator] If $V$ is an inner product space, the Hodge $\star$ operator is the unique liniear isomorphism
		\begin{equation}
		\star:\Lambda(V)\to\Lambda(V)
		\end{equation}
		such that
		\begin{equation}
		\star:\Lambda^p(V)\to\Lambda^{n-p}(V)
		\end{equation}
		and
		\begin{equation}
		u\wedge\star v=\langle u,v\rangle dV,\qquad\qquad \forall u,v\in\Lambda(V), \forall p\,.
		\end{equation}
	\end{defin}
	For an oriented Riemannian Manifold, we have $\star$ defined on $\Lambda(M^\star_m)$ for each $m$, and it is smooth. We thus have
	\begin{equation}
	\star: E^p(M)\to E^{n-p}(M)
	\end{equation}
	as a map between differential forms on $M$, and it satisfies
	\eqs{\label{hodgesquare2}
		\star\star=(-1)^{p(n-p)}.
	}
	The co-differential operator $\di^\dagger$ acting on $p$-forms is defined by\footnote{Both \eqref{hodgesquare2} and \eqref{codif2} acquire one more minus sign when the signature is Lorentzian.},
	\eqs{\label{codif2}
		\di^\dagger=(-1)^{n(p+1)+1}\star\di\star}
	and acts as
	\eqs{
		(\di^\dagger\omega)_{B_2\cdots B_p}=-\mathcal{D}^{B_1}\omega_{B_1\cdots B_p}\,.
	}
	The Laplace-Beltrami operator $\Delta:E^{p}(M)\to E^{p}(M)$ is defined by
	\eqs{
		\Delta=\di\di^\dagger+\di^\dagger\di\,.
	}
	One can define an inner product on the space of $p$-forms:
	\eqs{
		\langle\alpha,\beta\rangle=\int_\M\alpha\wedge\star\beta\,.\label{inner product2}
	}
	It follows that $\di^\dagger$ is the adjoint of $\di$
	\eqs{
		\langle\di\alpha,\beta\rangle=\langle\alpha,\di^\dagger\beta\rangle\,,
	}
	while $\Delta$ is self-adjoint.
	\begin{pro}
		$\Delta\alpha=0$ if and only if  $\di\alpha=0$ and $\did\alpha=0$.
	\end{pro}
	It follows that the only harmonic functions $\Delta f=0$ on a compact connected, oriented, Riemannian manifold are the constant functions. 
	
	\begin{defin}[Harmonic forms] Define the space of \emph{harmonic $p$-forms} as
		\begin{equation}
		\mathbf{H}^p(M):=Ker(\Delta)
		\end{equation}
		The space of harmonic forms depends on the metric. A form may be harmonic with respect to one metric and fail to be harmonic with respect to another metric
	\end{defin}
	\begin{theo}[The Hodge decomposition theorem]
		For each integer $p$ with $0\leq p \leq n$, $\mathbf{H}^p$ is finite-dimensional, and we have the following orthogonal direct sum decomposition of  the space $E^p(M)$ of smooth $p$-forms on $M$:
		
		\begin{align}
		E^p(M)&=\Delta(E^p)\oplus\mathbf{H}^p\nonumber\\
		&=\di\did(E^p)\oplus \did\di (E^p)\oplus \mathbf{H}^p\nonumber\\
		&=\di(E^{p-1})\oplus\did(E^{p+1})\oplus \mathbf{H}^p
		\end{align}
		Consequently, the equation $\Delta\omega=\alpha$ has a solution $\omega\in E^{p}(M)$ if and only if the $p$-form $\alpha$ is orthogonal to the space of harmonic $p$-forms.
	\end{theo}
	
	The number of harmonic $p$-forms on $\M$ is equal to the dimension of the $p$-th de Rham cohomology group $H_{\text{dR}}^p(\M)$.

	\begin{theo}
		[Hodge isomorphism theorem] Let $(M,g)$ be a compact oriented Riemannian manifold. There is an isomorphism depending only on on the metric $g$:
		\begin{equation}
		\mathbf{H}^p(M)\cong  H^{n-p}_{dR}(M)
		\end{equation}
	\end{theo}

	\begin{theo}[Poincar\'e duality theorem] Let $M$ be a compact oriented  manifold. The pairing
		\begin{equation}
		H^p_{dR}(M)\times  H^{n-p}_{dR}(M)\to\mathbb{R},\qquad (u,v)\to\int_M u\wedge v
		\end{equation}
		is non-degenerate, i.e. it induces an isomorphism (the Poincare\'e duality isomorphism)
		\begin{equation}
		H^p_{dR}(M)\cong  H^{n-p}_{dR}(M)
		\end{equation}
	\end{theo}

	A corollary of Poincar\'e duality theorem is that for an oriented compact manifold, $H^n_{dR}\cong\mathbb{R}$.
	
	\chapter{The Rindler Patch}\label{rindler}
	\paragraph{De Sitter slicing of AdS.} 
	The $d$-dimensional anti-de Sitter space is defined as the hyperboloid
	\begin{equation}
	-X_{\mone}^2+X_\mu X^\mu=-\ell^2,\qquad a=0,\cdots,d-1
	\end{equation}
	embedded in $(d+1)$-dimensional Minkowski space with signature $(--+\cdots+)$. The isometries are generated by
	\begin{equation}
	iL_{AB}=X_A\partial_B-X_B\partial_A\qquad A,B=-1,0,\cdots,d-1
	\end{equation}
	which form a $(d-1)$-dimensional conformal algebra $SO(d-1,2)$,
	\begin{equation}
	-i[L_{AB},L_{CD}]=\eta_{AC}L_{BD}+\eta_{BD}L_{AC}-\eta_{AD}L_{BC}-\eta_{BC}L_{AD}\,.
	\end{equation}
	A Lorentz subgroup $SO(d-1,1)$  can be identified with generators $L_{\mu\nu}$ which preserves $X_\mu X^\mu$, so that  the algebra can be cast into the form
	\begin{subequations}
		\begin{align}
		i[L_{\mone \mu},L_{\mone \nu}]&=L_{\mu\nu},\\
		-i[L_{\mu\nu},L_{\mone \alpha}]&=\eta_{\mu\alpha}L_{\mone \nu}-\eta_{\nu\alpha} L_{\mone \mu},\\
		-i[L_{\mu\nu},L_{\alpha\beta}]&=\eta_{\mu\alpha}L_{\nu\beta}+\eta_{\nu\beta}L_{\mu\alpha}-\eta_{\mu\beta}L_{\nu\beta}-\eta_{\nu\alpha}L_{\mu\beta}.
		\end{align}
	\end{subequations}
	We will call $L_{-1 \mu}$ the ``AdS-translation'' generators since they reduce to translation vectors near the origin (i.e. in the large $\ell$ flat space limit).  They are
	\begin{equation}
	iL_{\mone \mu}=-\sqrt{\ell^2+X_\nu X^\nu}\ \partial_\mu-X_\mu\partial_{\mone },\qquad X_{\mone }>0.
	\end{equation}

	We would like to set a coordinate system which makes the Lorentz  $SO(d-1,1)$ symmetry generated by $L_{\mu\nu}$ manifest and also allows for taking the flat space limit in a simple way. That is achieved by the parametrization
	\begin{equation}
	X^\nu X_\nu\equiv -(X^0)^2+X^iX_i=\rho^2>0,\qquad X_{\mone }^2=\ell^2+\rho^2\,.
	\end{equation}
	The AdS$_d$ metric becomes\footnote{
		The global AdS coordinates are related to embedding coordinates by
		\begin{equation}\label{glo hyper}
		X^0=\sqrt{\ell^2+r^2}\sin t\qquad X^{\mone}=\sqrt{\ell^2+r^2}\cos t\qquad r^2=X^i X_i\nn
		\end{equation}
		so that the metric is
		\begin{equation}
		ds^2=-(\ell^2+r^2)dt^2+\frac{dr^2}{1+r^2/\ell^2}+r^2d\Omega_{d-2}^2
		\end{equation}}
	\begin{equation}\label{adsmetric}
	ds^2=\frac{\ell^2 d\rho^2}{\rho^2+\ell^2}+\rho^2h_{ab}dx^a dx^b,\qquad a,b=0,1,\cdots,d-2\,,
	\end{equation}
	where $h_{ab}$ is the $(d-1)$-dimensional Lorentzian metric on unit radius de Sitter space. A convenient choice of coordinate system is \cite{Compere:2017knf}
	\begin{equation}\label{dS-metric}
	h_{ab}dx^a dx^b=\frac{1}{\sin^2\tau}\Big(-d\tau^2+ d\Omega^2_{d-2}\Big),\
	\end{equation}
	where
	\begin{equation}\label{hyp emb}
	X^0=\rho\cot\tau,\qquad \sqrt{{X}^i{X}_i}=\frac{\rho}{\sin\tau}\,.
	\end{equation}
	
	By the choice of radial coordinate $\rho$, the induced boundary metric will be dS$_{d-1}$, a positively curved, Lorentzian maximally symmetric space. This property facilitates the study of isometries on the AdS$_d$ boundary. Moreover, these coordinates are suitable for taking the flat space $\ell\to\infty$ limit. We should, however, point out that while the dS slicing is a convenient choice, as mentioned above, a similar charge analysis with the same general results may be carried out adopting any other coordinate system on AdS.
	
	As depicted in Figure \ref{fig1} \cite{Esmaeili:2019mbw}, there is an AdS PT (parity times time reversal) transformation, $X^A\to -X^A$, which also acts at the AdS boundary as 
	$\tau\to -\tau$ and an antipodal map on the $S^{d-2}$. Under this map, the radial coordinate $\rho$ does not change. This PT-symmetry may be combined with the AdS isometries to form an $O(d-1,2)$ symmetry group.
	
	\begin{figure}
		\centering
		\begin{tikzpicture}[scale=.4]
		\draw[very thick, reverse directed] (-3,5)--(-3,-5);
		\draw[very thick,  reverse directed] (3,5)--(3,-5);
		\draw[dashed] (-3,3)--(3,-3);
		\draw[dashed] (-3,-3)--(3,3);
		\draw [purple, thick] (0,3) ellipse (3cm and .3cm);
		\draw [purple, thick] (0,-3) ellipse (3cm and .3cm);
		
		\draw [blue, reverse directed] (3,3)..controls (0,0)..(3,-3);
		\draw [blue, reverse directed] (3,3)..controls (0.75,0)..(3,-3);
		\draw [blue, reverse directed] (3,3)..controls (1.5,0)..(3,-3);
		\draw [blue, reverse directed] (3,3)..controls (2,0)..(3,-3);
		\draw [blue, reverse directed] (3,3)..controls (2.5,0)..(3,-3);
		\draw [blue, reverse directed] (3,3)..controls (2.8,0)..(3,-3);
		\draw [blue, reverse directed] (-3,3)..controls (-0,0)..(-3,-3);
		\draw [blue, reverse directed] (-3,3)..controls (-0.75,0)..(-3,-3);
		\draw [blue, reverse directed] (-3,3)..controls (-1.5,0)..(-3,-3);
		\draw [blue, reverse directed] (-3,3)..controls (-2,0)..(-3,-3);
		\draw [blue, reverse directed] (-3,3)..controls (-2.5,0)..(-3,-3);
		\draw [blue, reverse directed] (-3,3)..controls (-2.8,0)..(-3,-3);
		%
		%
		\node[] at( 3.5,4.5) {$\mathscr{I}$};
		\node[] at( -3.5,4.5) {$\mathscr{I}$};
		\node[rotate=-90] at( 3.5,0) {$\rho=\infty$};
		%
		\node[] at( 5,3) {$\tau=\pi/2$};
		\node[] at( 5,-3) {$\tau=-\pi/2$};
		\draw [ purple] (0,0)..controls (2,.8)..(3,.8);
		\draw [ purple] (0,0)..controls (2,-.8)..(3,-.8);
		\draw [ purple] (0,0)..controls (2.1,1.4)..(3,1.4);
		\draw [ purple] (0,0)..controls (2.1,-1.4)..(3,-1.4);
		\draw [ purple] (0,0)..controls (2.3,1.8)..(3,1.8);
		\draw [ purple] (0,0)..controls (2.3,-1.8)..(3,-1.8);
		\draw [ purple] (0,0)..controls (2.4,2.1)..(3,2.1);
		\draw [ purple] (0,0)..controls (2.4,-2.1)..(3,-2.1);
		\draw [ purple] (0,0)..controls (-2.1,1.2)..(-3,1.2);
		\draw [ purple] (0,0)..controls (-2.1,-1.2)..(-3,-1.2);
		\draw [ purple] (0,0)--(3,0);

		\node[purple] at( -3.8,1.2) {{$\Sigma_{\tau}$}};
		\node[purple] at( -3.8,-1.2) {${\Sigma_{-\tau}}$};
		\end{tikzpicture}
		\caption{The patch of AdS$_d$ covered by hyperbolic coordinates. Blue curves are constant $\rho$ hypersurfaces, preserved by $SO(d-1,1)$ Lorentz transformations and the red lines represent codimension one constant time $\tau$ slices, in particular $\Sigma_{\tau},\Sigma_{-\tau}$ are two surfaces at $\tau$ and $-\tau$. Note that in our coordinate system all constant time slices for finite $\tau$ pass through $\rho=0$. The arrows on the boundary and on blue curves show the flow of time $\tau$. In this coordinate system, AdS conformal boundary is a maximally symmetric manifold (dS$_{d-1}$ at $\rho\to\infty$), and  it is conformally invariant under AdS-translations.}
		\label{fig1}
	\end{figure}
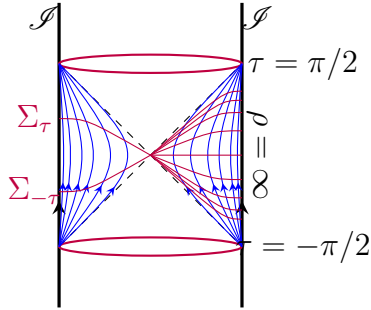

	\paragraph{Isometries in the de Sitter slicing.} 
	To the Lorentz generators  $L_{ab}=X_a\partial_b-X_b\partial_a$ one can associate differential 1-forms on the ambient Minkowski space
	\begin{equation}
	L^{\mu\nu}=X^{\mu}\di X^\nu-X^\nu \di X^\mu\,.
	\end{equation}
	In the hyperbolic coordinates adapted above they become
	\begin{align}
	L^{\mu\nu}&=(X^{\mu}\bar{X}^\nu-X^\nu \bar{X}^\mu)\di\rho +\rho^2(\bar{X}^{\mu}\di\bar{X}^\nu-\bar{X}^\nu \di\bar{X}^\mu)=2\rho^2\bar{X}^{[\mu} \bar{M}^{\nu]}_a \di x^a
	\end{align}
	where 
	\begin{equation}
	\bar{X}^\mu=\frac{X^\mu}{\rho},\qquad\qquad\bar{M}^{\mu}_a =\frac{\partial \bar{X}^\mu}{\partial x^a}\,.
	\end{equation}
	Similarly, for the ``AdS-translation'' generators $L_{\mone \mu}=X_{\mone}\partial_\mu-X_\mu\partial_{\mone}$, the corresponding 1-forms are
	\begin{align}\label{transl form}
	L^{\mone \mu}&=-X_{\mone} \di X^\mu+X^\mu\di X_{\mone}
	\end{align}
	In hyperbolic coordinates they become
	\begin{align}
	L^{\mone \mu}&=(-X_{\mone} \frac{{X}^\mu}{\rho}+X^\mu \frac{\rho}{X_{\mone}})\di \rho- X_{\mone}\ \rho \bar{M}_a^\mu \di x^a\cr
	&=-\frac{\ell^2}{\rho X_{\mone}}X^\mu\di \rho - \rho X_{\mone}\bar{M}_a^\mu \di x^a
	\end{align}
	The associated vectors using the AdS inverse metric, are
	\begin{align}
	\text{{Lorentz:}}\, \xi_L\quad&\qquad L^{\mu\nu}= 2\bar{X}^{[\mu} \bar{M}^{\nu]}_a h^{ab}\partial_b\label{xi-L} \\
	\text{AdS-{Translation:}}\, \xi_T \quad&\qquad L^{\mone \mu}= -X_{\mone}\bar{X}^\mu\ \partial_\rho - \frac{X_{\mone}}{\rho} \bar{M}_a^\mu h^{ab}\partial_b\label{xi-T}\,.
	\end{align}
	
	From \eqref{transl form}, if $\rho\ll\ell$, we have $L^{\mone \mu}\approx \ell \di X^\mu$. Hence, $L_{\mone \mu}$ are translation generators near the origin.
	On the other hand if  $\rho\gg\ell$, i.e. near the boundary of AdS, the AdS-translation generators are
	\begin{equation}
	L^{\mone \mu}\approx  \bar{X}^\mu\rho\partial_\rho + \bar{M}_a^\mu h^{ab}\partial_b\,.
	\end{equation}
	Infinitesimal AdS-translations act on $\rho$ as $\rho\to \rho(1+\epsilon \bar{X}^\mu)$. One can then observe that these
	AdS-translations are conformal Killing vectors of the boundary metric in hyperbolic slicing $h_{ab}$.

	Note that for Lorentz transformations $\xi_L^\rho=0$ and we also have, 
	\begin{align}
	D^{}_\mu\xi^L_\nu+D^{}_\nu\xi^L_\mu=0,\qquad D^{}_\nu\xi_L^\nu=0.
	\end{align}

	AdS-translations satisfy the following relations,
	\begin{align}\label{xi-T-iden}
	D^{}_a\xi^T_b+D^{}_b\xi^T_a=2\Gamma_{ab}^\rho \xi^T_\rho=-2\rho\xi_T^\rho h_{ab},\qquad\quad
	D_b\xi_T^b=\frac{1}{\rho}\xi_T^\rho(1-d).
	\end{align}
	In the $\rho\gg l$ limit only the leading terms are relevant and the expression for the AdS-translation vector simplifies to
	\begin{align}
	\xi_T=(\xi_T^\rho,\xi_T^a)=(\bar{X}^a\rho,\bar{M}^{a}_b h^{ab}).
	\end{align}
	It is convenient to define, 
	\begin{equation}
	\bar{\xi}^\rho_T=\lim_{\rho\to\infty} \frac{{\xi}^\rho_T}{\rho},\qquad\qquad 
	\bar{\xi}^b_T=\lim_{\rho\to\infty}{\xi}^b_T.
	\end{equation}
	A useful implication of Killing equation is the following asymptotic relation
	\begin{equation}\label{6.8}
	\bar{\xi}_T^a=D^a\bar{\xi}^\rho_T\,.
	\end{equation}

	\chapter{More on three dimensional Maxwell theory}
	Here we compare our results  in the three-dimensional maxwell theory with a previous study \cite{Barnich:2015jua} in Bondi coordinates $(u,r,\varphi)$. In that work, the Einstein-Maxwell theory was considered. We will switch the gravity sector off to compare the Maxwell sector. The boundary conditions are
	\begin{equation}
	A_r=0\qquad A_u=\ord{\ln\frac{r}{r_0}}\qquad A_\varphi=\ord{\ln\frac{r}{r_0}}.
	\end{equation}
	The leading terms in the solution space were parametrized as
	\begin{align}
	A_\varphi&=\alpha(u,\varphi)\ln\frac{r}{r_0}+A^0_\varphi(u,\varphi)+\ord{r^{-1}}\\
	A_u&=-\lambda\ln\frac{r}{r_0}+A^0_u(u,\varphi)+\ord{r^{-1}}
	\end{align}
	where $\alpha(u,\varphi)=-\omega(\varphi)-u\lambda^\prime(\varphi)$ and $\dot{A}^0_\varphi=-\lambda^\prime(\varphi)+(A^0_u)^\prime$. Here, the function $\omega(\varphi)$ is an ``integration constant'', and $A^0_u(u,\varphi)$ is the electromagnetic news. 
	The field strength tensor at leading order is (\text{notation of \cite{Barnich:2015jua}})
	\begin{equation}
	F_{u\varphi }=-\lambda^\prime(\varphi)\qquad F_{\varphi r}=\frac{\omega(\varphi)+u\lambda^\prime(\varphi)}{r}\qquad F_{ur}=\frac{\lambda(\varphi)}{r}\label{Barn bc}
	\end{equation}

	($\laa$ in notation of \cite{Barnich:2015jua} is a physical component of the gauge field while in our notation it is the gauge parameter not appearing in the field strength) The surface charges in \cite{Barnich:2015jua} consist of gravitational and gauge parts. By setting the super-rotation and supertranslation generators to zero, and in absence of electromagnetic news function $A^0_u=0$, the charge is integrable and conserved and is given by
	\begin{equation}\label{barn ch}
	Q_E=\frac{1}{4\pi G}\int d\varphi\lambda (\varphi)\bar{E}(\varphi)\qquad\text{(Maxwell sector only)}
	\end{equation}
	where $\bar{E}(\varphi)$ is the leading term of the gauge parameter in this specific case. Clearly, this non-gravitational subgroup of the asymptotic symmetry group is $U(1)$ at every angle.

	In our analysis, the dual scalar field behaves as $\Phi\sim\ord{1}$ in $\rho$-expansion. According to \eqref{duality}, the leading order $\Phi^{(0)}$ is even under antipodal condition, and the general solution is given in \eqref{even}, which has the following behaviour near future null infinity  at $\tau=0$
	\begin{equation}
	\Phi^{(0)}(\tau,\varphi)=p_0 +\sum_{n\neq 0} p_n e^{in\varphi}\cos n\tau\,=\bar{\Phi}^{(0)}-\bar{\Phi}^{(0)\p\p}\frac{u}{r}+\cdots
	\end{equation}
	where we have used $\tau^2\approx-2u/r$, and defined
	\begin{equation}
	\bar{\Phi}^{(0)}\equiv \sum_{n} p_n e^{in\varphi} 
	\end{equation}
	(note that the superscript $(n)$ shows the $\rho$-expansion at spatial infinity) .
	For the subleading order $\Phi^{(1)}$, we take it to be odd (like $\A_\rho^{(0)}$) under antipodal map. In this case, the solution is given in \eqref{odd}, and at $\Ti\to 0$ it behaves as
	\begin{equation}
	\Phi^{(1)}(\tau,\varphi)=q_0 \tau+\sum_{n\neq 0} \frac{q_n}{n} e^{in\varphi}\sin n\tau=\frac{\bar{\Phi}^{(1)}}{r}+\cdots
	\end{equation}
	where we have defined	
	\begin{equation}
	\bar{\Phi}^{(1)}\equiv \sum_{n} q_n e^{in\varphi} \,.
	\end{equation}
	
	The dual field strength at leading order will be
	\begin{subequations}
		\begin{align}
		\F_{ur}={\epsilon^\varphi}_{ur}\partial_\varphi\bar{\Phi}&=\frac{\bar{\Phi}^{(0)\p}}{r}+\cdots\\
		\F_{u\varphi}=\epsilon_{ru\varphi}(\partial_r\Phi
		-\partial_u\Phi)&=-\Phi^{(0)\p\p}+\cdots\\
		\F_{\varphi r}=\epsilon_{ur\varphi}\partial_r\Phi&=\frac{-\Phi^{(1)}+u\Phi^{(0)\p\p}}{r}+\cdots
		\end{align}
	\end{subequations}
	These are exactly the leading null infinity behaviour \eqref{Barn bc} of Ref. \cite{Barnich:2015jua}, for $\laa(\varphi)=\Phi^{(0)\p}$ and $\omega(\varphi)=-\Phi^{(1)}$.
	
	There only remains to check the agreement of conserved charges. 
	Close to the future null infinity  at $\tau=0$, the fields behave as follows
	\begin{subequations}
		\begin{align}
		\A_{\rho}^{(0)}&=\bar{\A}_{\rho}^{(0)}(\varphi)\tau+\ord{\tau^3}\qquad\text{gauge field}\\
		\laa&=\bar{ \laa}(\varphi)+\ord{\tau^2}\qquad\quad\text{gauge parameter}
		\end{align}
	\end{subequations}

	At null infinity, only the second term of the charge \eqref{3d charge} remains non-vanishing,
	\begin{equation}
	Q_\laa=\int_{S^1}d\varphi\bar{\laa}\bar{\A}_{\rho}^{(0)}=\int_{S^1}d \varphi\bar{\laa}\Phi^{(0)\p}
	\end{equation}
	where we have used $\partial_\tau\bar{\A}_{\rho}^{(0)}=\Phi^{(0)\p}$. This expression is in agreement with \eqref{barn ch} by substitution $\bar{\laa}\to \bar{E}$ and $\bar{\Phi}^{(0)\p}\to\laa$. In conclusion, the spatial infinity power-law boundary condition \eqref{3d falaf} advocated here reproduces the leading order field strength behavior at null infinity, prescribed in \cite{Barnich:2015jua}. Finally, we mention that \cite{Barnich:2015jua} takes a larger set of initial data (and hence solution space) than considered here, by including logarithmic terms in the asymptotic expansion. This difference is however does not affect to the asymptotic symmetry group.

	\bibliographystyle{fullsort}
	
	\bibliography{Bib}
	
\end{document}